\documentclass[11pt]{article}
\usepackage{amssymb}
\usepackage{amsmath}
\usepackage{amscd}
\usepackage{latexsym}
\usepackage[all]{xy}
\xyoption{curve}
\xyoption{line}
\usepackage{graphicx}
\usepackage{graphics}
\usepackage{color}
\catcode`@=11
\renewcommand{\section}
{\@startsection{section}{1}{0pt}{\medskipamount}{\medskipamount}{\large\bf}}
\makeatletter\renewcommand{\subsection}
{\@startsection{subsection}{2}{\z@}{-3.25ex plus -1ex minus -.2ex}
{1.5ex plus .2ex}{\it }}

\numberwithin{equation}{section}
\catcode`@=12

\def\th{\theta}

\def\de{\delta}

\def\cgstack#1#2{{{\scriptstyle #1}\atop{\scriptstyle #2}}}

\def\pa{\partial}

\def\beq{\begin{equation}}
\def\eeq{\end{equation}}
\def\bea{\begin{eqnarray}}
\def\eea{\end{eqnarray}}

\newcommand{\im}{\,\mathrm{i}\,}
\newcommand{\diff}{\mathrm{d}}

\newcommand{\R}{{\mathbb{R}}}

\newcommand{\C}{{\mathbb{C}}}
\newcommand{\Z}{{\mathbb{Z}}}

\newcommand{\F}{{\mathbb{F}}}
\newcommand{\CP}{{\C P}}

\newcommand{\Idd}{\mathbf{1}}
\newcommand{\Hcal}{{\cal H}}
\newcommand{\Ecal}{{\cal E}}

\newcommand{\Pcal}{{\cal P}}
\newcommand{\Vcal}{{\cal V}}
\newcommand{\Ical}{{\cal I}}
\newcommand{\Tcal}{{\cal T}}
\newcommand{\Qcal}{{\cal Q}}
\newcommand{\Ccal}{{\cal C}}
\newcommand{\fh}{\hat{f}}

\newcommand{\zh}{\hat{z}}
\newcommand{\zbh}{\hat{\bar{z}}}
\newcommand{\yb}{{\bar{y}}}
\newcommand{\Yb}{{\bar{Y}}}
\newcommand{\zb}{{\bar{z}}}
\newcommand{\ab}{{\bar{a}}}
\newcommand{\ub}{{\bar{u}}}
\newcommand{\vb}{{\bar{v}}}
\newcommand{\jb}{{\bar{\jmath}}}
\newcommand{\ib}{{\bar{\imath}}}
\newcommand{\betab}{{\bar{\beta}}}
\newcommand{\gammab}{{\bar{\gamma}}}
\newcommand{\bb}{{\bar{b}}}

\newcommand{\ca}{{\cal{A}}}
\newcommand{\cf}{{\cal{F}}}

\newcommand{\Imm}{{\rm im}}
\newcommand{\K}{{\rm K}}
\newcommand{\HQ}{{\rm H}}

\newcommand{\ch}{{\mathrm{ch}}}
\newcommand{\rep}{{\rm R}}

\newcommand{\Lcal}{{\cal L}}
\newcommand{\Pf}{{\rm Pf}}
\newcommand{\vol}{{\rm vol}}
\newcommand{\Tr}{{\rm Tr}}

\newcommand{\tr}{{\rm tr}}

\newcommand{\su}{{{\rm SU}(2)}}
\newcommand{\suL}{{{\rm su}(2)}}
\newcommand{\sut}{{{\rm SU}(3)}}
\newcommand{\uo}{{{\rm U}(1)}}
\newcommand{\uoL}{{{\rm u}(1)}}

\newcommand{\urm}{{{\rm U}}}
\newcommand{\urmL}{{{\rm u}}}
\newcommand{\ut}{{{\rm U}(3)}}
\newcommand{\utwo}{{{\rm U}(2)}}

\newcommand{\utwoL}{{{\rm u}(2)}}
\newcommand{\Sp}{{\rm S}}
\newcommand{\slc}{{{\rm SL}(2,\C)}}

\newcommand{\sltc}{{{\rm SL}(3,\C)}}
\newcommand{\sltcL}{{{\rm sl}(3,\C)}}
\newcommand{\gltwoc}{{{\rm GL}(2,\C)}}

\newcommand{\quiver}{{\sf Q}}
\newcommand{\rel}{{\sf R}}

\newcommand{\mphi}{{{\mbf\phi}}}

\newcommand{\mmu}{{{\mbf\mu}}}

\newcommand{\mbf}[1]{{\boldsymbol {#1} }}
\newcommand{\xr}{\xrightarrow}
\newcommand{\noverq}{{\stackrel{\scriptstyle n}{\scriptstyle q}}}

\newcommand{\noverqpmt}{{\stackrel{\scriptstyle n}{\scriptstyle
      q\pm2}}}

\newcommand{\noverqpt}{{\stackrel{\scriptstyle n}{\scriptstyle q+2}}}
\newcommand{\noverqmt}{{\stackrel{\scriptstyle n}{\scriptstyle q-2}}}
 
\newcommand{\npoverqpmo}{{\stackrel{\scriptstyle n+1}{\scriptstyle
      q\pm1}}} 
\newcommand{\nmoverqpmo}{{\stackrel{\scriptstyle n-1}{\scriptstyle
      q\pm1}}} 
\newcommand{\npoverqmo}{{\stackrel{\scriptstyle n+1}{\scriptstyle
      q-1}}} 
\newcommand{\nmoverqmo}{{\stackrel{\scriptstyle n-1}{\scriptstyle
      q-1}}} 
\newcommand{\npoverqpo}{{\stackrel{\scriptstyle n+1}{\scriptstyle
      q+1}}} 
\newcommand{\nmoverqpo}{{\stackrel{\scriptstyle n-1}{\scriptstyle
      q+1}}} 
\newcommand{\npmoverqpo}{{\stackrel{\scriptstyle n\pm1}{\scriptstyle
      q+1}}} 
\newcommand{\npmoverqmo}{{\stackrel{\scriptstyle n\pm1}{\scriptstyle
      q-1}}}

\def\Dirac{{D\!\!\!\!/\,}} 

\def\chern{{\rm Ch}}
\def\Hom{{\rm Hom}}
\def\Ext{{\rm Ext}}
\def\End{{\rm End}}
\def\ind{{\rm index}}

\def\>{\rangle}
\def\<{\langle}
\def\+{\dagger}
\def\={\ =\ }

\allowdisplaybreaks

\begin{document}

\begin{titlepage}
\setcounter{page}{0}
\begin{flushright}
ITP--UH--06/08\\
CERN--PH--TH--2008--054\\
HWM--08--5\\
EMPG--08--10\\
\end{flushright}

\vskip 1.8cm

\begin{center}

{\Large\bf SU(3)-Equivariant Quiver Gauge Theories
  \\[10pt] and Nonabelian Vortices}

\vspace{15mm}

{\large Olaf Lechtenfeld${}^{1}$}, \ \ {\large Alexander D. Popov${}^{1,2}$}
\ \ and \ \ {\large Richard J. Szabo${}^3$}
\\[5mm]
\noindent ${}^1${\em Institut f\"ur Theoretische Physik,
Universit\"at Hannover \\
Appelstra\ss{}e 2, 30167 Hannover, Germany }
\\[5mm]
\noindent ${}^2${\em Bogoliubov Laboratory of Theoretical Physics, JINR\\
141980 Dubna, Moscow Region, Russia}
\\[5mm]
\noindent ${}^3${\em Department of Mathematics and Maxwell Institute
  for Mathematical Sciences\\ Heriot-Watt University,
Colin Maclaurin Building, Riccarton, Edinburgh EH14 4AS, U.K.}
\\[5mm]
{Email: {\tt lechtenf, popov @itp.uni-hannover.de , R.J.Szabo@ma.hw.ac.uk}}

\vspace{15mm}

\begin{abstract}
\noindent
We consider $\sut$-equivariant dimensional reduction of Yang-Mills theory on
K\"ahler manifolds of the form $M\times\sut/H$, with $H=\su\times\uo$
or $H=\uo\times\uo$. The induced rank two quiver gauge theories on $M$
are worked out in detail for representations of $H$ which descend from
a generic irreducible $\sut$-representation. The reduction of the
Donaldson-Uhlenbeck-Yau equations on these spaces induces nonabelian
quiver vortex equations on $M$, which we write down explicitly. When
$M$ is a noncommutative deformation of the space $\C^d$, we construct
explicit BPS and non-BPS solutions of finite energy for all cases. We
compute their topological charges in three different ways and propose
a novel interpretation of the configurations as states of
D-branes. Our methods and results generalize from $\sut$ to any
compact Lie group.

\end{abstract}

\end{center}
\end{titlepage}

\tableofcontents

\newpage

\section{Introduction and summary \label{Intro}}

\noindent
BPS type equations for gauge theories in higher dimensions were
proposed long ago~\cite{Corrigan2} as generalizations of the
self-duality equations in four dimensions. For nonabelian gauge theory
on a K\"ahler manifold the most natural BPS condition lies in the
Donaldson-Uhlenbeck-Yau equations~\cite{DUY1}. These equations arise
in compactifications of superstring theory down to four-dimensional
Minkowski spacetime as the condition for at least one unbroken
supersymmetry. In this paper we will study the structure of solutions
to these equations on product manifolds of the form $M_q\times G/H$,
where $G$ is the Lie group $\sut$ and $H$ is a closed subgroup of
$G$. 

Building on earlier work~\cite{LPS}, nonabelian gauge theory on the
manifold $M_q\times S^2$, the product of a real
$q$-dimensional manifold $M_q$ with lorentzian or riemannian
nondegenerate metric and a two-sphere $S^2$, was considered
in~\cite{PS1}. The $\su$-equivariant dimensional reduction induces a
Yang-Mills-Higgs theory on $M_q$ which is a quiver gauge theory of
rank one. For K\"ahler manifolds $M_q$, with $q=2d$, the proper
reduction of the Yang-Mills equations on $M_{2d}\times S^2$ induces
quiver gauge theory equations on $M_{2d}$, and quiver vortex equations
on $M_{2d}$ in the BPS sector~\cite{Garcia1}. The Seiberg-Witten
monopole equations~\cite{Witten} for $d=2$ and the ordinary vortex
equations~\cite{Jaffe} for $d=1$ are particular instances of quiver
vortex equations. It has also been shown that the ordinary vortex
equations are integrable when $M_2$ is a compact Riemann surface of
genus $g>1$~\cite{group3}. In that case $M_2\times S^2$ is a
gravitational instanton, and the vortex equations are the
compatibility conditions of two linear equations (Lax pair) so that
the standard methods of integrable systems can be applied to the
construction of their solutions. Explicit $\su$-equivariant monopole,
dyon and monopole-antimonopole pair solutions of the Yang-Mills
equations on $M_2\times S^2$ were also obtained in~\cite{group4} for
$M_2$ of lorentzian signature $(-+)$ and with the topology of $\R^2$
or $\R\times S^1$.

While the criteria for existence of solutions to all of these BPS
quiver vortex equations are by now
well-understood~\cite{A-CG-P1,A-CG-P2}, in practice it is usually
quite difficult to write down explicit solutions of them. The
construction of exact solutions is facilitated when $M_{2d}$ is the
noncommutative space $\R_\theta^{2d}$. In this instance both BPS and
non-BPS solutions were obtained for various cases
in~\cite{LPS,PS1},\cite{group1}--\cite{LPS3}. In this paper we will
extend these constructions by $\sut$-equivariant dimensional reduction
over the coset spaces $\sut/H$, where $H=\su\times\uo$ or
$H=\uo\times\uo$. In the former case the coset space is the complex
projective plane $\C P^2$. We will find that many aspects of the
induced rank two quiver gauge theory on $M_{2d}$ in this case are
qualitatively similar to that obtained from the symmetric 
space $\CP^1\times\CP^1$~\cite{LPS2}. However, the technical aspects
are much more involved and some new features emerge from the
nonabelian $\su$ instanton degrees of freedom which now reside at the
vertices of the quiver. In the latter case the coset space is the
six-dimensional homogeneous manifold $Q_3$, and the qualitative
features of the rank two quiver gauge theories are much
different in this case, even though the quiver vertex degrees of
freedom involve two $\uo$ monopole charges as in the
$\CP^1\times\CP^1$ case~\cite{LPS2}. This space has appeared before in
a variety of different physical contexts, such as in connection with
the dynamics of M-theory on a manifold of G$_2$ holonomy which
develops a conical singularity~\cite{AW1}, and for the dimensional
reduction of ten-dimensional supersymmetric gauge theories over
six-dimensional coset spaces in much the same spirit as our more
general reductions~\cite{Lopes}. In these latter applications the
usage of the non-symmetric space $Q_3$ induces a four-dimensional
field theory with softly broken $\mathcal{N}=1$ supersymmetry. We
expect that our constructions will have direct applications in these
classes of models, though this is left for future work. 

The outline of this paper is as follows. Throughout we present
detailed constructions and examples of the pertinent quiver gauge
theories. Although most of our analyses are model dependent, our
techniques and results apply to reductions over more general coset
spaces $G/H$, where $G$ is a compact Lie group and $H$ is a closed
subgroup of $G$. In Section~\ref{quiverreps} we give explicit
constructions of the quivers and their representations which will
underlie the gauge theories considered in this paper. In
Section~\ref{quiverbuns} we carry out the $\sut$-equivariant
dimensional reductions and explicitly construct the fields of the
quiver gauge theory. In Section~\ref{NCinst} we study the BPS
equations of the quiver gauge theory which describe nonabelian quiver
vortices. In Section~\ref{NCquivvort} we construct explicit BPS and
non-BPS solutions of finite energy on the noncommutative deformation
of $M_{2d}=\C^d$. Finally, in Section~\ref{Dcharges} we compute
topological charges of our noncommutative instanton solutions from
various points of view, and use the constructions to propose a novel
interpretation of the configurations as states of D-branes. 

\bigskip

\section{Homogeneous bundles and quiver
  representations\label{quiverreps}}

\noindent
In this section we will give purely algebraic constructions of the
quivers that will play a role throughout this paper, following the
general formalism developed in~\cite{A-CG-P1}. They are based
on the representation theory of the Lie group $G=\sut$, and
are naturally associated to homogeneous vector bundles whose fibres
transform under irreducible representations of $\sut$. We
will begin with the simplest instance of the fundamental
representation as illustration of the method. Then we will give the
general construction, and illustrate the formalism with
various other explicit examples.

\subsection{Fundamental representations\label{fundquiverreps}}

We are interested in the geometry of coset spaces of the form
$G/H$, where $H$ is a closed subgroup of $G=\sut$. Given a
finite-dimensional representation $\underline{V}$ of $H$, the
corresponding induced, homogeneous hermitean vector bundle over
$G/H$ is given by the fibred product
\beq
\mathcal{V}=G\times_H\,\underline{V} \ .
\label{indhermbungen}\eeq
Every $G$-equivariant bundle of finite rank over $G/H$, with
respect to the standard transitive action of $G$ on the homogeneous
space, is of the form (\ref{indhermbungen}). If $\underline{V}$ is
irreducible, then $H$ is the structure group of the associated
principal bundle. The category of such homogeneous bundles is
equivalent to the category of finite-dimensional representations of a
certain quiver with relations, whose structure is determined entirely
by the subgroup $H$. In contrast to the treatment of~\cite{A-CG-P1},
we restrict to those representations
$\underline{V}$ which descend from some irreducible representation of
$\sut$ by restriction to $H$. We will now give an elementary
construction of this quiver representation in the simplest case where
$\underline{V}=\underline{C}^{1,0}\big|_H$ is the restriction of the
three-dimensional fundamental representation of~$\sut$.

The Dynkin diagram for $\sut$ consists of a pair of roots
$\alpha_1,\alpha_2$. The complete set $\Delta$ of non-null roots is
$\pm\,\alpha_1,\pm\,\alpha_2,\pm\,(\alpha_1+\alpha_2)$, with the inner
products $(\alpha_1,\alpha_1)=(\alpha_2,\alpha_2)=1$ and
$(\alpha_1,\alpha_2)=-\frac12$ so that
$(\alpha_1+\alpha_2,\alpha_1+\alpha_2)=1$. For the system
$\Delta^+$ of positive roots we
take $\alpha_1=(1,0)$, $\alpha_2=\frac12\,(-1,\sqrt3\,)$ and
$\alpha_1+\alpha_2=\frac12\,(1,\sqrt3\,)$. Let $e_{ij}$ be the matrix
units obeying $e_{ij}\,e_{kl}=\delta_{jk}\,e_{il}$, and abbreviate
$e_i:=e_{ii}$. The generators of $\sut$ for the Cartan-Weyl basis in
the $3\times3$ fundamental representation are then given by the
Chevalley generators
\beq
E_{\alpha_1}\=e_{12} \ , \qquad E_{\alpha_2}\=e_{23} \ , \qquad
\mbox{and} \qquad
E_{\alpha_1+\alpha_2}~:=~[E_{\alpha_1},E_{\alpha_2}]\=e_{13}
\label{fundgens1}\eeq
along with
\beq
E_{-\alpha_1}\=E_{\alpha_1}^\dag\=e_{21} \ , \qquad
E_{-\alpha_2}\=E_{\alpha_2}^\dag\=e_{32} \ , \qquad \mbox{and} \qquad
E_{-\alpha_1-\alpha_2}\=E_{\alpha_1+\alpha_2}^\dag\=e_{31} \ .
\label{fundgens2}\eeq
We take 
\beq
H_{\alpha_1}\=e_1-e_2 \qquad \mbox{and} \qquad
H_{\alpha_2}\=e_1+e_2-2e_3
\label{fundgenalso}\eeq
as the generators of the Cartan subalgebra $\uoL\oplus\uoL$. The
commutation relations are
\bea
[H_{\alpha_1},E_{\pm\,\alpha_1}]\=\pm\,2E_{\pm\,\alpha_1} \qquad
&\mbox{and}& \qquad [H_{\alpha_2},E_{\pm\,\alpha_1}]\=0 \ , \nonumber
\\[4pt] [H_{\alpha_1},E_{\pm\,\alpha_2}]\=
\mp\,E_{\pm\,\alpha_2} \qquad &\mbox{and}& \qquad
[H_{\alpha_2},E_{\pm\,\alpha_2}]\=\pm\,3
E_{\pm\,\alpha_2}
\label{HEEcomms}\eea
along with
\beq
[E_{\alpha_1},E_{-\alpha_1}]\=H_{\alpha_1} \ , \quad
[E_{\alpha_2},E_{-\alpha_2}]\=\mbox{$\frac12$}\,(H_{\alpha_2}-
H_{\alpha_1}) \ , \quad
[E_{\alpha_1+\alpha_2},E_{-\alpha_1-\alpha_2}]\=\mbox{$\frac12$}\,
(H_{\alpha_1}+H_{\alpha_2}) \ ,
\label{EEHcomms}\eeq
the Lie brackets
\beq
[H_{\alpha_1},E_{\pm\,(\alpha_1+\alpha_2)}]\=\pm\,
E_{\pm\,(\alpha_1+\alpha_2)} \qquad \mbox{and} \qquad [H_{\alpha_2},
E_{\pm\,(\alpha_1+\alpha_2)}]
\=\pm\,3E_{\pm\,(\alpha_1+\alpha_2)} \ ,
\label{HEEplus}\eeq
and
\beq
[E_{\pm\,\alpha_1},E_{\pm\,\alpha_2}]\=E_{\pm\,(\alpha_1+\alpha_2)} \
, \quad [E_{\pm\,\alpha_1},E_{\mp\,(\alpha_1+\alpha_2)}]\=\mp\,
E_{\mp\,\alpha_2} \ , \quad [E_{\pm\,\alpha_2},
E_{\mp\,(\alpha_1+\alpha_2)}]\=\pm\,E_{\mp\,\alpha_1} \ .
\label{EEEcomms}\eeq
All other Lie brackets of the generators vanish.

The fundamental weights are
$\mu_{\alpha_1}=\frac12\,\big(1,\frac1{\sqrt3}\,\big)$ and
$\mu_{\alpha_2}=\big(0,\frac1{\sqrt3}\,\big)$, with
$\mu_{\alpha_1}$ the highest weight of the defining representation
$\underline{C}^{1,0}$. The corresponding weight diagram is
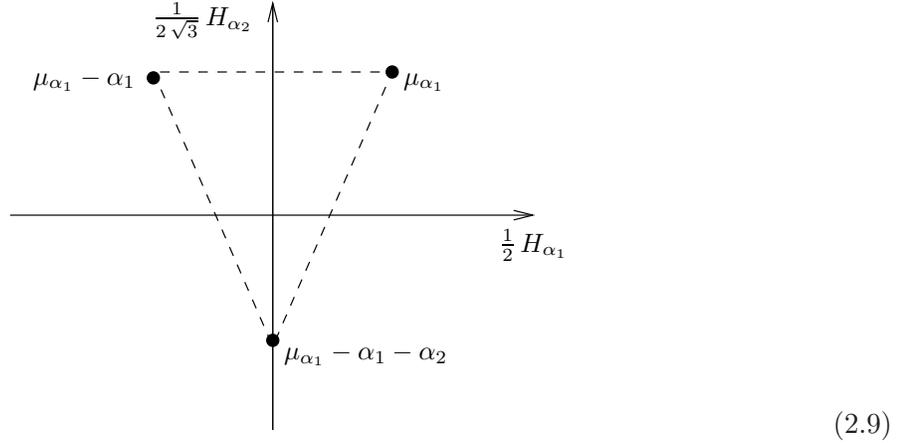
\begin{equation}
\input{SU3_weight_fund.pstex_t}
\label{fundweightdiag}\end{equation}
and the Young tableau consists of a single box. There are two
homogeneous spaces of interest that we now analyse in turn.

\bigskip

\noindent
{\bf Symmetric $\mbf{\underline{C}^{1,0}}$ quiver. \ } Our first
example is the four-dimensional complex projective plane
\beq
\C P^2=\sut\,\big/\,\Sp\big(\utwo\times\uo\big)
\label{CP2coset}\eeq
which is a symmetric space whose isometry group is
isomorphic to $\sut$. We can use the weight diagram
(\ref{fundweightdiag}) to decompose the fundamental representation of
$\sut$ as a representation of the subgroup
$H=\Sp(\utwo\times\uo)\cong\su\times\uo$ (locally) to get
\beq
\underline{C}^{1,0}\big|_{\su\times\uo}=\underline{(1,1)}~\oplus~
\underline{(0,-2)} \ ,
\label{fundSU2U1decomp}\eeq
since the restriction of the $\sut$ operators $E_{\pm\,\alpha_1}$ to
$\su$ shifts vertices along the horizontal directions of the weight
diagram. The first integer $n=2I$ in the pairs on the right-hand side
labels twice the isospin $I$ (the eigenvalue of $H_{\alpha_1}$) such that
$(n+1)$ is the dimension of the irreducible $\su$ representation and the
second integer $m=3Y$ labels three times the hypercharge $Y$ which is
the $\uo$ magnetic charge (the eigenvalue of $H_{\alpha_2}$). We use
the two pairs of integers in (\ref{fundSU2U1decomp}) to label vertices
in a directed graph. The arrow between the two vertices comes from
increasing the $H_{\alpha_1}$ and $H_{\alpha_2}$ eigenvalues using the
raising operators $E_{\alpha_2}$ and $E_{\alpha_1+\alpha_2}$, via the
commutation relations  (\ref{HEEcomms}) and (\ref{HEEplus}). This
gives the elementary quiver diagram
\begin{equation}
\input{SU3_quiverCP2_fund.pstex_t}
\label{fundCP2quiver}\end{equation}
associated to the symmetric space $\C P^2$ for the fundamental
representation of $\sut$, where $\Phi_{0,-2}^+$ is the $H$-equivariant
intertwiner between the two $H$-modules induced by the action of the
$\sut$ raising operators $E_{\alpha_2}$ and $E_{\alpha_1+\alpha_2}$. 
Later on this quiver will be
associated to a holomorphic triple representing the basic unstable
brane-antibrane pair. A completely analogous construction applies
to the conjugate representation $\underline{C}^{0,1}$, with highest
weight $\mu_{\alpha_2}$, whose weight diagram is the counterclockwise
rotation of (\ref{fundweightdiag}) through angle $\frac\pi3$ about the
origin. The Young diagram consists of a single column of two boxes,
and the analog of the decomposition (\ref{fundSU2U1decomp}) is
\beq
\underline{C}^{0,1}\big|_{\uo\times\su}=\underline{(1,-1)}~\oplus~
\underline{(0,2)} \ .
\label{C01SU2U1decomp}\eeq

\bigskip

\noindent
{\bf Non-symmetric $\mbf{\underline{C}^{1,0}}$ quiver. \ } Our second
example is the six-dimensional reducible flag manifold $\F(1,2;3)\cong
Q_3$ given by
\beq
Q_3=\sut\,\big/\,\uo\times\uo
\label{Q3coset}\eeq
which is homogeneous but not symmetric. Now we decompose the
fundamental representation of $\sut$ as a representation of the
maximal torus $H=T=\uo\times\uo$ of $\sut$. This can be achieved
starting from (\ref{fundSU2U1decomp}) by decomposing the fundamental
representation of $\su$ as a $\uo$ representation to get
$\mbf2\big|_{\uo}=\underline{(1)}\oplus\underline{(-1)}$, where
$\underline{(q)}$ denotes the irreducible one-dimensional
representation of $\uo$ with magnetic charge $q\in\Z$ which is twice
the third component of isospin of an irreducible $\sut$
representation. It follows that
\beq
\underline{C}^{1,0}\big|_{\uo\times\uo}=\underline{(1,1)_1}~\oplus~
\underline{(-1,1)_1}~\oplus~\underline{(0,-2)_0} \ ,
\label{fundU1U1decomp}\eeq
where now the pairs label the $\uo$ charges $(q,m)_n$, $m=3Y$, of the
torus $T$, and the subscripts label the original $\sut$ isospin integer
$n$ which keeps track of multiplicities of states in the weight diagram. 
The three vertices are now joined by the actions of the
raising operators $E_{\alpha_1+\alpha_2}$, $E_{\alpha_1}$ and
$E_{\alpha_2}$ through the commutation relations (\ref{HEEcomms}) and
(\ref{HEEplus}), giving the quiver diagram
\begin{equation}
\input{SU3_quiverQ3_fund.pstex_t}
\label{fundQ3quiver}\end{equation}
Compatibility with the commutation relations of the $\sut$ raising
operators implies the relation
\beq
{}^0\Phi_{0,-2}^+-{}^1\Phi_{-1,1}^0\,{}^0\Phi_{0,-2}^-=0 \ .
\label{fundQ3quiverrel}\eeq

\subsection{General constructions\label{AlgQuiverGen}}

We now turn to the general construction of the quiver diagram
associated with the homogeneous bundle (\ref{indhermbungen}). As
complex algebraic varieties, the homogeneous space $G/H$ is
diffeomorphic to the flag manifold $G^\C/P$, where $P$ is a parabolic
subgroup of $G^\C=\sltc$. Given an irreducible holomorphic
representation $\underline{V}\,_\mu$ of $P$, corresponding to an
integral dominant weight $\mu\in\Lambda_P^+$, the induced,
irreducible, holomorphic homogeneous vector bundle over $G^\C/P$ is
$\mathcal{O}_\mu:=G^\C\times_P\,\underline{V}\,_\mu$. Let $L\subset
P$ be the universal complexification of the subgroup $H$. Then there
is a Levi decomposition $P=U\ltimes L$. By Engel's theorem, a
holomorphic module $\underline{V}\,_\mu$ is an irreducible
representation of $P$ if and only if its restriction
$\underline{V}\,_\mu\big|_L$  is an irreducible representation of the
reductive Levi subgroup $L$ and the action of $U$ on
$\underline{V}\,_\mu$ is trivial. Then
$\mathcal{V}_\mu=G\times_H\,\underline{V}\,_\mu$ is the smooth,
$G$-equivariant hermitean vector bundle underlying
$\mathcal{O}_\mu$. The quiver is associated to an appropriate basis of
weights of $L$ which specifies the vertices. The action of $U$,
extending the $L$-action to a $P$-action, specifies the arrows
satisfying relations which invoke the commutation relations of the Lie
algebra $\mathfrak{u}$ corresponding to $U$. Homogeneous bundles
with the same slope will typically lie along diagonal lines of the
quiver diagram~\cite{A-CG-P1,A-CG-P2}. As we will see later on, the
relations of the quiver correspond to integrability of the Dolbeault
operator on the induced homogeneous vector bundle over $G^\C/P$, which
will arise dynamically as BPS equations for a quiver gauge theory.

For each pair of non-negative integers $(k,l)$ there is an irreducible
holomorphic representation $\underline{C}^{k,l}$ of $\sltc$ of
dimension 
\beq
d^{k,l}~:=~\dim\big(\,\underline{C}^{k,l}\,\big)
\=\mbox{$\frac12$}\,(k+l+2)\,(k+1)\,(l+1)
\label{dimkl}\eeq
and highest weight $\mu=k\,\mu_{\alpha_1}+l\,\mu_{\alpha_2}$. The
restriction $\underline{C}^{k,l}\big|_L$ then determines a
finite-dimensional representation of this quiver, and conversely. Let
us illustrate the construction explicitly on our two coset spaces of
interest. 

\bigskip

\noindent
{\bf Symmetric $\mbf{\underline{C}^{k,l}}$ quiver. \ } The complex
projective plane can be modelled locally as the Grassmann manifold
${\rm Gr}(2,3)={\rm GL}(3,\C)/K\cong\CP^2$, where $K$ is the
stability subgroup of block upper triangular matrices preserving the
two-dimensional lowest weight subspace of the complex vector space
$\underline{C}^{1,0}\cong\C^3=\C^2\oplus\C$. Thus in this case $P$ is
the parabolic subgroup of block upper triangular matrices in
$\sltc$. The Levi subgroup $L$ is the group of unit determinant
matrices in $\gltwoc\times\C_*$, where $\C_*:=\C\setminus\{0\}$. The
commutative algebra $\mathfrak{u}$ corresponding to $U$ is generated
by $E_{\alpha_2}$ and $E_{\alpha_1+\alpha_2}$. The set of vertices
$\quiver_0$ of the associated quiver consists of the weight vectors
$(n,m)$ of $\utwo\cong\su\times\uo$ (locally) in $\sltc$ with respect
to the basis $(H_{\alpha_1},H_{\alpha_2})$. From the commutation
relations (\ref{HEEcomms}) and (\ref{HEEplus}) it follows that the
arrow set $\quiver_1$ of the quiver is determined by the action of $U$
on the weights as
\beq
(n,m)~\longmapsto~(n\pm1,m+3) \ ,
\label{CP2arrows}\eeq
depending on which particular weight vectors $(n,m)$ the raising
operators $E_{\alpha_1+\alpha_2}$ and $E_{\alpha_2}$ act on. Since
$\mathfrak{u}$ is an abelian algebra, the arrows
(\ref{CP2arrows}) generate commutative quiver diagrams with quadratic
holomorphic relations $\rel$ around any elementary square of the
quiver. These graphs are formally the same as the rectangular lattice
quiver diagrams associated to the symmetric space $\C P^1\times\C
P^1$~\cite{A-CG-P1,LPS2}, which are generically a product of two
chains.

For a fixed pair of non-negative integers $(k,l)$, the irreducible
$\sut$ representation $\underline{C}^{k,l}$ determines a {\it finite}
quiver vertex set $\quiver_0(k,l)$ as follows. The quiver diagram can
be obtained by collapsing the ``horizontal'' $\su$
representations to single nodes in the weight diagram for
$\underline{C}^{k,l}$. Algebraically, the integer $k$ is the
number of fundamental representations $\underline{C}^{1,0}$ and $l$
the number of conjugate representations $\underline{C}^{0,1}$
appearing in the tensor product construction of
$\underline{C}^{k,l}$. This means that the Young diagram of
$\underline{C}^{k,l}$ contains $k+l$ boxes in its first row and $l$
boxes in its second row. To determine $\quiver_0(k,l)$, we decompose
$\underline{C}^{k,l}$ as a representation of $\su\times\uo$. From the
weight diagram for $\underline{C}^{k,l}$ including multiplicities, one
can extract the $\su\times\uo$ representation content for each
row. For example, with $k>l$ one finds
\bea
m&=&k+2l \qquad , \qquad \ \ \quad n\=k \nonumber \\
m&=&k+2l-3 \qquad , \qquad n\=k+1 \ , \ k-1 \nonumber \\
m&=&k+2l-6 \qquad , \qquad n\=k+2 \ , \ k \ , \ k-2 \nonumber \\
& & \qquad\qquad\qquad\qquad \vdots \nonumber \\
m&=&k-l \qquad , \qquad  \ \ \ ~ \quad n\=k+l \ , \ k+l-2 \ , \ k+l-4 \
, \ \ldots \ , \ k-l \nonumber \\
m&=&k-l-3 \qquad , \qquad ~ \ n\=k+l-1 \ , \ k+l-3 \ , \ k+l-5 \ , \
\ldots \ , \ k-l-1 \nonumber \\
& & \qquad\qquad\qquad\qquad \vdots \nonumber \\
m&=&-2k+2l \qquad , \qquad \ \ n\=2l \ , \ 2l-2 \ , \ 2l-4 \ , \
\ldots \ , \ 2 \ , \ 0 \nonumber \\
m&=&-2k+2l-3 \qquad , \ \ \ n\=2l-1 \ , \ 2l-3 \ , \ 2l-5 \ , \
\ldots \ , \ 3 \ , \ 1 \nonumber \\
& & \qquad\qquad\qquad\qquad \vdots \nonumber \\
m&=&-2k-l+3 \qquad , \quad  \ \ n\=l+1 \ , \ l-1 \nonumber \\
m&=&-2k-l \qquad , \qquad \quad \ n\=l \ .
\label{Q3genvertices}\eea
The set of quiver vertices $(n,m)$ in this case generates a
rectangular lattice tilted through angle $\frac\pi4$ with corners
located at $(k,k+2l)$, $(k+l,k-l)$, $(0,-2k+2l)$ and $(l,-2k-l)$. The
cases $k\leq l$ are treated similarly.
Conversely, given any finite quiver associated to the symmetric space
$\C P^2$, the ranges of the vertices determine a corresponding
irreducible representation $\underline{C}^{k,l}$ of $\sut$. The
integers $(n,m)$ have the same even/odd parity.

\bigskip

\noindent
{\bf Non-symmetric $\mbf{\underline{C}^{k,l}}$ quiver. \ } In this
instance $P$ is the Borel subgroup of (purely) upper triangular
matrices in $\sltc$. The Levi subgroup is
$L=(\C_*)^2\subset\sltc$. The algebra $\mathfrak{u}$ is now generated
by $E_{\alpha_1}$, $E_{\alpha_2}$ and $E_{\alpha_1+\alpha_2}$,
which is the non-abelian three-dimensional Heisenberg algebra with
central element $E_{\alpha_1+\alpha_2}$. The set of vertices $(q,m)_n$
is now precisely the weight lattice $\Lambda\cong\Z^2$ of $\sltc$, and
from the commutation relations (\ref{HEEcomms}) and (\ref{HEEplus})
it follows that the action of $U$ on the weight vectors is given by
\bea
E_{\alpha_1}\,:\,(q,m)_n&\longmapsto&(q+2,m)_n \ , \nonumber\\[4pt]
E_{\alpha_2}\,:\,(q,m)_n&\longmapsto&(q-1,m+3)_{n\pm1} \ ,
\nonumber\\[4pt] E_{\alpha_1+\alpha_2}\,:\,(q,m)_n&
\longmapsto&(q+1,m+3)_{n\pm1} \ .
\label{Q3arrows}\eea
The arrows of the quiver thus translate weight vectors by the set of
positive roots. Moreover, the Heisenberg commutation relations induce
linear terms in the relations expressing commutativity of the
corresponding quiver diagrams. In contrast to the symmetric quivers
above, in this case there can be multiple arrows emanating between two
vertices due to degenerate weight vectors $(q,m)_n$ and $(q,m)_{n'}$
with $n\neq n'$.

Given an irreducible representation $\underline{C}^{k,l}$ of $\sut$,
the quiver diagram is now simply the weight diagram for
$\underline{C}^{k,l}$. Thus the non-symmetric vertex set
$\quiver_0(k,l)$ is generated by lattice points $(q,m)_n$ with
\beq
q\=-n \ , \ -n+2 \ , \ \dots \ , \ n-2 \ , \ n \ .
\label{qrange}\eeq
Note that the symmetric vertex $(n,m)$, representing a full isospin
multiplet, can be obtained by collapsing the
non-symmetric quiver vertices $(q,m)_n$ to $(n,m)$. Given any
finite quiver associated to the homogeneous space $Q_3$, the ranges of
the vertices determine a corresponding representation
of~$\sut$. Again, the integers $(q,m)_n$ have the same parity.

\subsection{Examples}

We conclude this section by giving some further explicit examples of
quiver representations corresponding to irreducible $\sut$
representations, as illustration of the general
algorithm of Section~\ref{AlgQuiverGen} above. From these examples the
generic features of the symmetric and non-symmetric space quivers will
become apparent.

\bigskip

\noindent
{\bf Symmetric $\mbf{\underline{C}^{2,0}}$ quiver. \ } Let us consider
the six-dimensional representation $\underline{C}^{2,0}$ of $\sut$. In
this case the highest weight is $2\mu_{\alpha_1}$ and the
corresponding weight diagram is
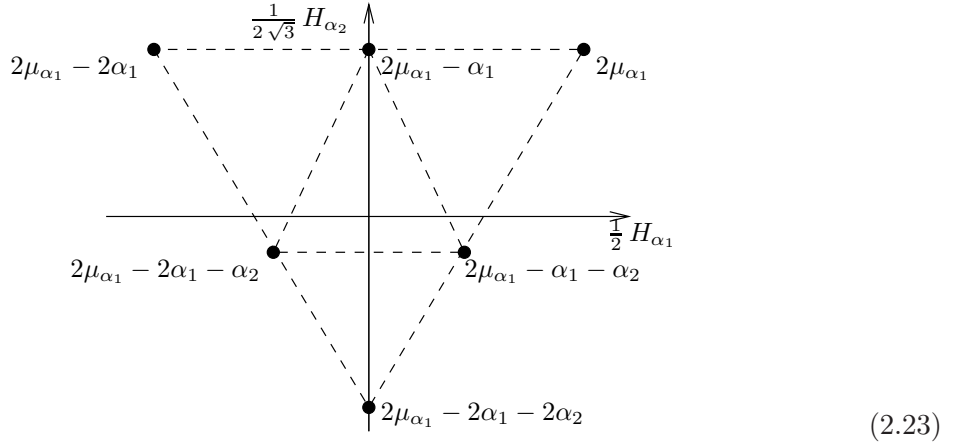
\begin{equation}
\input{SU3_weight_six.pstex_t}
\label{sixweightdiag}\end{equation}
The Young diagram consists of a single row of two boxes, and from the
weight diagram (\ref{sixweightdiag}) one can work out the
decomposition of $\underline{C}^{2,0}$ as a representation of
$\su\times\uo$ to get
\beq
\underline{C}^{2,0}\big|_{\su\times\uo}=\underline{(2,2)}~\oplus~
\underline{(1,-1)}~\oplus~\underline{(0,-4)} \ .
\label{C20CP2decomp}\eeq
The corresponding quiver diagram is
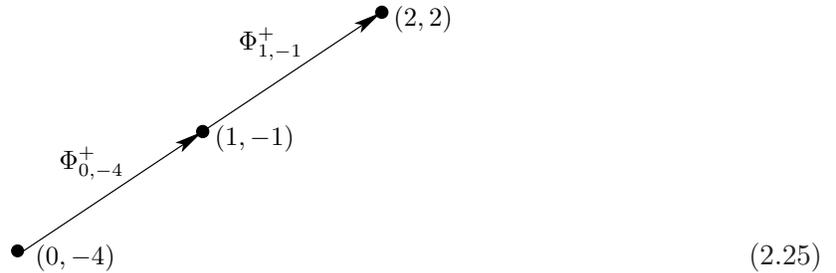
\begin{equation}
\input{SU3_quiverCP2_six.pstex_t}
\label{sixCP2quiver}\end{equation}
which will be represented in Section~\ref{quiverbuns} by a holomorphic
chain.

\bigskip

\noindent
{\bf Symmetric $\mbf{\underline{C}^{1,1}}$ quiver. \ } The final
example of a quiver associated to the symmetric space $\C P^2$
corresponds to the eight-dimensional adjoint representation
$\underline{C}^{1,1}$. The highest weight is given by
$\mu:=\mu_{\alpha_1}+\mu_{\alpha_2}=\frac12\,(1,\sqrt3\,)$, and the
weight diagram is
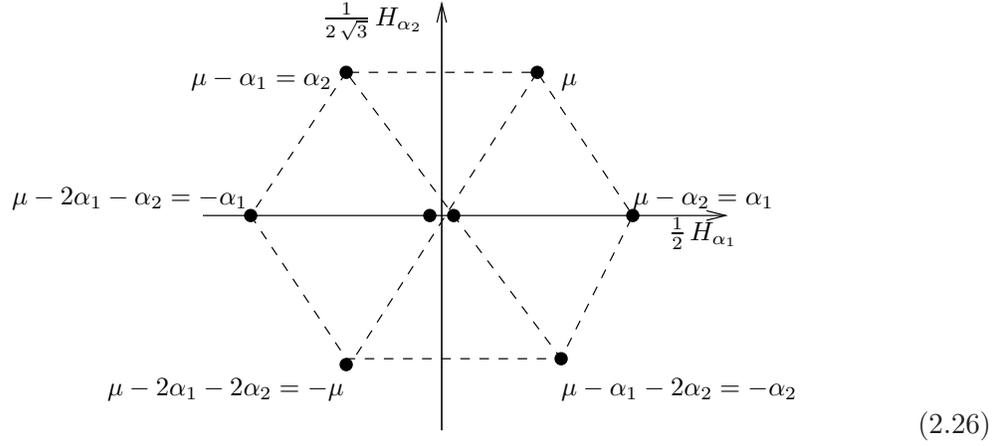
\begin{equation}
\input{SU3_weight_ad.pstex_t}
\label{adweightdiag}\end{equation}
The corresponding decomposition of $\underline{C}^{1,1}$ is given by
\beq
\underline{C}^{1,1}\big|_{\su\times\uo}=\underline{(1,-3)}~\oplus~
\underline{(2,0)}~\oplus~\underline{(0,0)}~\oplus~\underline{(1,3)} \
.
\label{C11CP2decomp}\eeq
The quiver diagram is
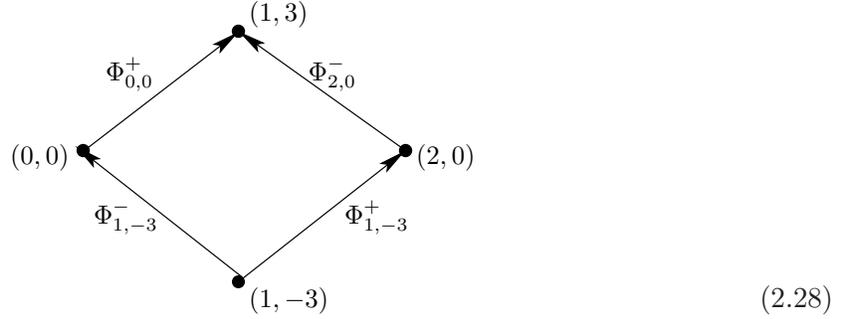
\begin{equation}
\input{SU3_quiverCP2_ad.pstex_t}
\label{adCP2quiver}\end{equation}
whose commutativity is expressed through the holomorphic relation
\beq
\Phi_{2,0}^-\,\Phi_{1,-3}^+-\Phi_{0,0}^+\,\Phi_{1,-3}^-=0 \ .
\label{adCP2rel}\eeq

\bigskip

\noindent
{\bf Non-symmetric $\mbf{\underline{C}^{2,0}}$ quiver. \ } Next we
turn to the quiver associated to the non-symmetric
space $Q_3$ which corresponds to the representation
$\underline{C}^{2,0}$ of $\sut$. Decomposing the $\su$-modules
in (\ref{C20CP2decomp}) as irreducible $\uo$ representations, the
decomposition of $\underline{C}^{2,0}$ as a representation of the
maximal torus $T$ is given by
\beq
\underline{C}^{2,0}\big|_{\uo\times\uo}=\underline{(2,2)_2}~\oplus~
\underline{(0,2)_2}~\oplus~\underline{(-2,2)_2}~\oplus~
\underline{(1,-1)_1}~
\oplus~\underline{(-1,-1)_1}~\oplus~\underline{(0,-4)_0} \ .
\label{C20Q3decomp}\eeq
The corresponding quiver diagram coincides with the weight diagram
(\ref{sixweightdiag}) and can be presented as
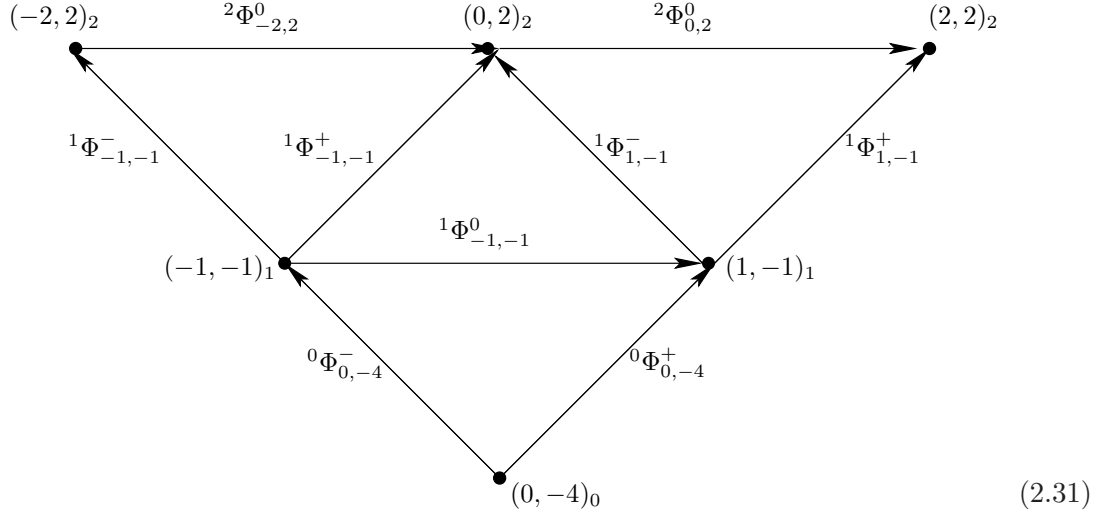
\begin{equation}
\input{SU3_quiverQ3_ad.pstex_t}
\label{sixQ3quiver}\end{equation}
Compatibility of the maps in (\ref{sixQ3quiver}) implies the set of
holomorphic relations
\bea
{}^1\Phi_{1,-1}^+-{}^2\Phi_{0,2}^0\,{}^1\Phi_{1,-1}^-&=&0 \ , 
\nonumber\\[4pt]
{}^2\Phi_{0,2}^0\,{}^1\Phi_{-1,-1}^+-{}^1\Phi_{1,-1}^+\,{}^1
\Phi_{-1,-1}^0&=&0 \ , \nonumber\\[4pt]
{}^1\Phi_{-1,-1}^++{}^1\Phi_{1,-1}^-\,{}^1\Phi_{-1,-1}^0-2~{}^2
\Phi_{-2,2}^0\,{}^1\Phi_{-1,-1}^-&=&0 \ , \nonumber\\[4pt]
{}^1\Phi_{-1,-1}^+\,{}^0\Phi_{0,-4}^--{}^1\Phi_{1,-1}^-\,{}^0
\Phi_{0,-4}^+&=&0 \ , \nonumber\\[4pt]
{}^0\Phi_{0,-4}^+-{}^1\Phi_{-1,-1}^0\,{}^0\Phi_{0,-4}^-&=&0 \ .
\label{sixQ3rels}\eea

\bigskip

\noindent
{\bf Non-symmetric $\mbf{\underline{C}^{1,1}}$ quiver. \ } Our final
example of this section is the quiver associated to $Q_3$ which is
built on the adjoint representation $\underline{C}^{1,1}$ of
$\sut$. Using (\ref{C11CP2decomp}) one has the decomposition
\bea
\underline{C}^{1,1}\big|_{\uo\times\uo}&=&\underline{(1,3)_1}~\oplus~
\underline{(-1,3)_1}~\oplus~\underline{(2,0)_2}~\oplus~
\underline{(0,0)_2}\nonumber\\ && 
\oplus~\underline{(-2,0)_2}~\oplus~\underline{(0,0)_0}~\oplus~
\underline{(1,-3)_1}~\oplus~\underline{(-1,-3)_1}
\label{C11Q3decomp}\eea
and the corresponding quiver diagram coincides with the weight diagram
(\ref{adweightdiag}) presented as
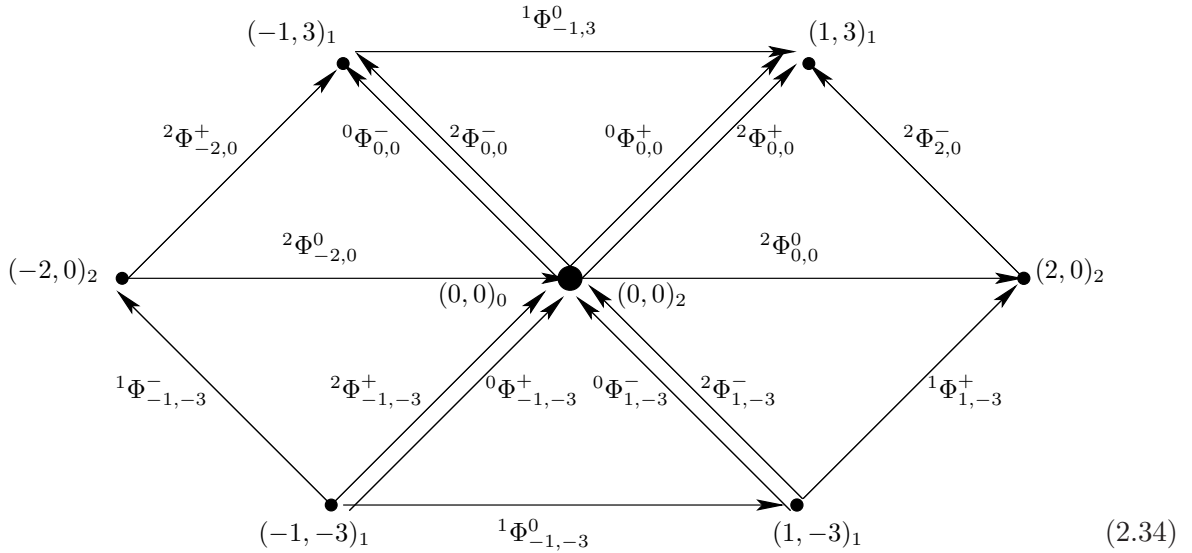
\begin{equation}
\input{SU3_quiverQ3_six.pstex_t}
\label{adQ3quiver}\end{equation}
The four extra arrows arise from the doubly degenerate weight vector
$(0,0)_0,(0,0)_2$ occuring in (\ref{C11Q3decomp}). We omit the rather 
lengthy list of holomorphic relations expressing compatibility of the
equivariant $T$-module morphisms in (\ref{adQ3quiver}).

\bigskip

\section{Equivariant gauge theories and quiver
  bundles\label{quiverbuns}}

\noindent
In this section we will consider Yang-Mills theory with
$G$-equivariant gauge fields on manifolds of the form
\beq
X~:=~M_D\times G\big/H\=G\times_HM_D \ ,
\label{M2ntimesSU3H}\eeq
where $M_D$ is a manifold of dimension $D$ and $G=\sut$ acts
trivially on $M_D$. We will reduce the gauge theory on
(\ref{M2ntimesSU3H}) by compensating the isometries of $G/H$ with
gauge transformations, such that the Lie derivative with respect to a
Killing vector field is given by an infinitesimal gauge transformation
on $X$. This unifies gauge and Higgs fields in the
higher-dimensional gauge theory, and reduces to a quiver gauge theory
on $M_D$. The twisted reduction will be accomplished by uniquely
extending the homogeneous vector bundles (\ref{indhermbungen}) by
$H$-equivariant bundles $E\to M_D$, providing a representation of
the quivers with relations of Section~\ref{quiverreps} in the category
of complex equivariant vector bundles over $M_D$. Such a 
representation is
called a quiver bundle. As previously, we will illustrate the idea
behind the construction by considering first the quiver bundles
associated to the fundamental representation of $\sut$. Then we give
the general construction of the $G$-equivariant gauge connections,
and present various other explicit examples to demonstrate generic
features of the quiver gauge theory. While the equivariant reduction
of gauge connections has been carried out in generality
in~\cite{A-CG-P1}, the construction there relies on various formal
isomorphisms. In the following we carry out the coset space
dimensional reduction explicitly by exploiting the geometry of the
homogeneous spaces involved and a suitable basis for the irreducible
representations of $\sut$.

\subsection{Fundamental representations\label{Quivbunfund}}

We begin by constructing the quiver bundles associated to the
fundamental representation $\underline{C}^{1,0}$ of $\sut$, for the
two coset spaces considered in Section~\ref{quiverreps}. In each case
we begin by constructing the unique $G$-equivariant gauge
connection on the homogeneous bundles (\ref{indhermbungen}), and then
extend it to equivariant bundles over the product spaces
(\ref{M2ntimesSU3H}).

\bigskip

\noindent
{\bf Symmetric $\mbf{\underline{C}^{1,0}}$ quiver bundles. \ } Let us
describe the $G$-equivariant connection on $\C P^2$. Consider the
principal $\Sp(\utwo\times\uo)$-bundle
\beq
\sut~\xr{\Sp(\utwo\times\uo)}~\C P^2 \ .
\label{Stiefelsut}\eeq
Let $(Y_0~Y_1~Y_2)^\top$ denote homogeneous complex coordinates on $\C
P^2$. The projective plane can be covered by three patches $\C
P^2=\mathcal{U}_0\cup\mathcal{U}_1\cup\mathcal{U}_2$ such that
$Y_i\neq0$ on $\mathcal{U}_i\cong\C^2$. Then
\beq
Y~:=~\begin{pmatrix} y^1\\[4pt] y^2\end{pmatrix} ~\sim~
\begin{pmatrix} 1\\[4pt] Y_1/Y_0\\[4pt] Y_2/Y_0\end{pmatrix} \qquad
\mbox{and} \qquad Y^\dag\=\big(\yb^1~\yb^2\big)
\label{YU0}\eeq
are coordinates on the patch $\mathcal{U}_0$ with
$Y^\dag\,Y=\yb^i\,y^i$ and $i=1,2$. A local section of the fibration 
(\ref{Stiefelsut}) is given by the $3\times3$ matrices
\beq
V\=\frac1\gamma\,\begin{pmatrix}\Lambda & \Yb\\[4pt]
-\Yb^\dag & 1\end{pmatrix} \qquad \mbox{and} \qquad
V^\dag\=\frac1\gamma\,\begin{pmatrix}\Lambda & -\Yb\\[4pt]
\Yb^\dag & 1\end{pmatrix}\ ,
\label{Vlocalsection}\eeq
where
\beq
\Lambda~:=~\gamma~\Idd_2-\frac1{\gamma+1}~Y\,Y^\dag
\qquad \mbox{and} \qquad \gamma~:=~\sqrt{1+Y^\dag\,Y}\=
\sqrt{1+\yb^i\,y^i}
\label{Lambdagammadef}\eeq
obey
\beq
\Lambda\,Y\=Y \ , \quad Y^\dag\,\Lambda\=Y^\dag
\qquad \mbox{and} \qquad \Lambda^2\=\gamma^{2}~\Idd_2-Y\,Y^\dag \ .
\label{Lambdarels}\eeq
Using the identities (\ref{Lambdarels}) it is easy to see that
$V^\dag\,V=V\,V^\dag=\Idd_3$, and hence that $V\in\sut$.

Introduce a flat connection on the trivial bundle $\C
P^2\times\C^3$ by the anti-hermitean one-form
\beq
A_0\=V^\dag~\diff V~=:~\begin{pmatrix} B&\betab\\[4pt]-\beta^\top&-2a
\end{pmatrix}
\label{A0CP2def}\eeq
with $B\in\utwoL$ and $a\in\uoL$, where from (\ref{Vlocalsection}) we
obtain
\bea
B&=&\frac1{\gamma^2}\,\big(-\mbox{$\frac12$}~\diff(Y^\dag\,Y)~
\Idd_2+\Yb~
\diff\Yb^\dag+\Lambda~\diff\Lambda\big) \ , \label{U2instfield}\\[4pt]
a&=&-\frac1{4\gamma^2}\,\big(\Yb^\dag~\diff\Yb-\diff\Yb^\dag~\Yb\big) \
, \label{CP2monfield}\\[4pt]
\betab&=&\frac1{\gamma^2}\,\Lambda~\diff\Yb\=
\frac1\gamma~\diff\Yb-\frac1{\gamma^2\,(\gamma+1)}\,\Yb\,\Yb^\dag~
\diff\Yb \ , \label{betabdef}\\[4pt]
\beta&=&\frac1{\gamma^2}\,\Lambda~\diff Y\=
\frac1\gamma~\diff Y-\frac1{\gamma^2\,(\gamma+1)}\,Y\,Y^\dag~
\diff Y \ . \label{betadef}
\eea
Introducing components of the column one-forms in (\ref{betabdef}) and
(\ref{betadef}), one has
\bea
\betab&:=&\begin{pmatrix}\betab^1\\[4pt]\betab^2\end{pmatrix} \qquad
\mbox{with} \qquad \betab^i\=\frac1{\gamma}~\diff\yb^i-\frac{\yb^i}
{\gamma^2\,(\gamma+1)}\,y^j~\diff\yb^j \ , \label{betabcomps}\\[4pt]
\beta&=&\begin{pmatrix}\beta^1\\[4pt]\beta^2\end{pmatrix} \qquad
\mbox{with} \qquad \beta^i\=\frac1{\gamma}~\diff y^i-\frac{y^i}
{\gamma^2\,(\gamma+1)}\,\yb^j~\diff y^j \ . \label{betacomps}
\eea
The one-forms $B-\frac12\,\tr(B)~\Idd_2$ and $a$ on $\C P^2$ give the 
vertical components of $A_0$ with values in the tangent space 
${\rm su}(2)\oplus\uoL$ to the fibre of
the bundle (\ref{Stiefelsut}), while $\betab$ and $\beta$ are
basis one-forms on $\C P^2$ taking values in the complexified
cotangent bundle of $\C P^2$ and giving the horizontal components of
$A_0$ tangent to the base $\CP^2$. The $(1,0)$-forms $\beta^i$ and the
$(0,1)$-forms $\betab^i$ constitute a $G$-equivariant basis for
the complex vector spaces of forms of type $(1,0)$ and $(0,1)$ on
$\CP^2$, respectively.

Since $V\in\sut$, the canonical one-form (\ref{A0CP2def}) on $\sut$
satisfies the Cartan-Maurer equation
\beq
\diff A_0+A_0\wedge A_0=0 \ .
\label{CartanMaurereq}\eeq
This leads to the component equations
\bea
F_{\utwoL}&:=&\diff B+B\wedge B\=\betab\wedge\beta^\top\=
\begin{pmatrix}\betab^1\wedge\beta^1&\betab^1\wedge\beta^2\\[4pt]
\betab^2\wedge\beta^1&\betab^2\wedge\beta^2\end{pmatrix} \ ,
\label{Fu2inst}\\[4pt]
f_{\uoL}&:=&\diff a\=\mbox{$\frac12$}\,\beta^\dag\wedge\beta\=
\mbox{$\frac12$}\,\big(\betab^1\wedge\beta^1+\bar\beta^2\wedge
\beta^2\big) \label{CP2moncurv}
\eea
along with
\beq
\diff\betab+B\wedge\betab-2\betab\wedge a\=0 \qquad \mbox{and}
\qquad \diff\beta^\top+\beta^\top\wedge B-2a\wedge\beta^\top\=0 \  .
\label{diffbetaBa0}\eeq
The abelian field strength (\ref{CP2moncurv}) can be written
explicitly as
\beq
f_{\uoL}=-\frac1{2\gamma^4}\,\big(\gamma^2~\diff y^i\wedge
\diff\yb^i-\yb^i~\diff y^i\wedge y^j~\diff\yb^j\big)
\label{CP2moncurvexpl}\eeq
while the $\utwoL$-valued curvature (\ref{Fu2inst}) can be reduced to
\beq
F_{\utwoL}=F_{\suL}+f_{\uoL}~\Idd_2 \ ,
\label{Fu2reduce}\eeq
where
\beq
F_{\suL}~:=~\begin{pmatrix}\mbox{$\frac12$}\,\big(
\betab^1\wedge\beta^1-\betab^2\wedge\beta^2\big)&
\betab^1\wedge\beta^2\\[4pt]\betab^2\wedge\beta^1&
-\mbox{$\frac12$}\,\big(
\betab^1\wedge\beta^1-\betab^2\wedge\beta^2\big)\end{pmatrix}\=
\diff B_{(1)}+B_{(1)}\wedge B_{(1)}
\label{Fsu2inst}\eeq
is the curvature of the gauge potential $B_{(1)}=B-a~\Idd_2\in\suL$.

By construction, the fields $a$, $B_{(1)}$, $f_{\uoL}$ and
$F_{\suL}=F_{B_{(1)}}$ are all $G$-equivariant and can be understood
  geometrically as follows. The one-form (\ref{CP2monfield}) is the
  $\uoL$-valued monopole potential on $\C P^2$ which can be described
  as the canonical abelian connection on the Hopf bundle
\beq
S^5=\ut\,\big/\,\utwo~\xr{\uo}~\C P^2 \ .
\label{S5Hopfbundle}\eeq
Choose a linearly embedded projective line $\C P^1\subset\C P^2$
defined by the equation $y^2=\yb^2=0$. Its homology class is the
generator of the group $\HQ_2(\C P^2;\Z)=\Z$ which is Poincar\'e dual
to the generator $[\frac\im{2\pi}\,f_{\uoL}]\in\HQ^2(\C
P^2;\Z)=\Z$. Since the abelian group $\HQ^2(\C P^2;\Z)\cong\HQ^1(\C
P^2;\uo)$ classifies complex line bundles over $\C P^2$, we can pick a
representative $\Lcal\to\C P^2$ of the isomorphism class corresponding
to $f_{\uoL}$. This complex line bundle, associated with the principal
$\uo$-bundle (\ref{S5Hopfbundle}), is the monopole bundle over $\C
P^2$ which we take to be endowed with the same $\uoL$-connection
$a$. Higher degree monopole bundles $\Lcal_m:=\Lcal^{\otimes m}$ are
endowed with the connection $m\,a$ and first Chern number $m\in\Z$,
and are associated to higher irreducible representations
$\underline{(m)}$ of the fibres of (\ref{S5Hopfbundle}). Note that the
monopole field strength $m\,f_{\uoL}$ of charge $m$ is a $(1,1)$-form
proportional to the canonical K\"ahler two-form on $\C P^2$ given
locally on the patch $\mathcal{U}_0$ by
\beq
\omega_{\C P^2}=-\mbox{$\frac\im2$}\,\beta^\top\wedge\betab \ .
\label{KahlerCP2}\eeq

On the other hand, the one-form $B_{(1)}=B-a~\Idd_2$ is the
$\suL$-valued one-instanton field on $\C P^2$ considered as the
canonical connection on the rank two vector bundle $\Ical=\Ical_1$
associated with the Stiefel bundle
\beq
\ut\,\big/\,\uo\times\uo~\xr{\su}~\C P^2 \ .
\label{instbundle}\eeq
Its curvature $F_{\suL}$ is also a $(1,1)$-form on $\C P^2$. Higher
rank instanton bundles $\Ical_n$ are endowed with $G$-equivariant
one-instanton connections $B_{(n)}\in{\rm su}(n+1)$ and fibre spaces
\beq
\mbf{(n+1)}\cong\C^{n+1}
\label{fibrespiso}\eeq
in higher-dimensional  representations of the
$\su$ fibres of the bundle (\ref{instbundle}). For a given
representation $\underline{(n,m)}$ of $H=\Sp(\utwo\times\uo)$, the
corresponding homogeneous vector bundle is given by
$\Hcal_{n,m}=G\times_H\,\underline{(n,m)}$ and can be identified 
with $\Ical_n\otimes\Lcal_m$. It follows that the flat connection
(\ref{A0CP2def}) is a connection on the homogeneous bundle
(\ref{indhermbungen}) induced by the decomposition of the fundamental
$\sut$ representation $\underline{C}^{1,0}$ in
(\ref{fundSU2U1decomp}), with $\Ical\oplus\Lcal_{-2}\cong\C
P^2\times\C^3$. This defines a connection on the elementary quiver
bundle
\beq
\Lcal_{-2}~\xr{\wedge\betab}~\Ical\otimes\Lcal
\label{CP2quivbunfund}\eeq
over $\C P^2$ representing (\ref{fundCP2quiver}). The diagonal
elements $B=B_{(1)}+a~\Idd_2$ and $-2a$ naturally define connections
on $\Hcal_{1,1}\cong\Ical\otimes\Lcal$ and
$\Hcal_{0,-2}\cong\Lcal_{-2}$, respectively. By (\ref{diffbetaBa0}),
the off-diagonal elements $\betab$ and $-\beta^\top$ implement the
$G$-actions which connect the $H$-modules at the vertices of the
quiver.

Let us now extend this quiver bundle with the $H$-equivariant vector
bundle $E^{1,0}\to M_D$ given by
\beq
E^{1,0}=\big(E_{p_{0,-2}}\otimes\,\underline{(0,-2)}\,_{M_D}\big)~
\oplus~\big(E_{p_{1,1}}\otimes\,\underline{(1,1)}\,_{M_D}\big) \ ,
\label{E10decomp}\eeq
where $\underline{(n,m)}\,_{M_D}$ denotes the trivial $H$-equivariant
vector bundle $M_D\times\,\underline{(n,m)}$, and $E_{p_{n,m}}$ are
hermitean vector bundles over $M_D$ of rank $p_{n,m}$ with the trivial
$H$-action and the gauge connections $A^{n,m}=A^{n,m}(x)\in{\rm
  u}(p_{n,m})$ for $x\in M_D$. The action of the $\sut$ operators
$E_{\pm\,\alpha_2}$ and $E_{\pm\,(\alpha_1+\alpha_2)}$ is 
implemented by bundle morphisms given by sections
$\phi_{0,-2}^+=\phi_{0,-2}^+(x)\in\Hom(E_{p_{0,-2}},E_{p_{1,1}})$ and
${\phi_{0,-2}^+}{}^\dag=
{\phi_{0,-2}^+}{}^\dag(x)\in\Hom(E_{p_{1,1}},E_{p_{0,-2}})$, yielding 
a quiver bundle which corresponds to the holomorphic triple
\beq
\xymatrix{ & ~E_{p_{1,1}} \\
E_{p_{0,-2}}~\ar[ur]^{\phi_{0,-2}^+} & }
\label{quiverbunCP2fund}\eeq
over $M_D$ representing (\ref{fundCP2quiver}).

The induced $G$-equivariant bundle over $M_D\times\C P^2$ is
then constructed as the fibred product
\beq
\Ecal^{1,0}~:=~G\times_HE^{1,0}\=\big(E_{p_{0,-2}}\boxtimes
\Lcal_{-2}\big)~\oplus~\big(E_{p_{1,1}}\boxtimes(\Ical\otimes
\Lcal)\big) \ .
\label{CP2Ecalfund}\eeq
A $G$-equivariant connection on this complex vector bundle is
given by naturally extending the flat connection (\ref{A0CP2def}) to
get
\beq
\ca=\begin{pmatrix}A^{1,1}\otimes\Idd_2+\Idd_{p_{1,1}}\otimes
(B_{(1)}+a)&\phi_{0,-2}^+\otimes\betab\\[4pt]-{\phi_{0,-2}^+}^\dag
\otimes\beta^\top&A^{0,-2}\otimes1+\Idd_{p_{0,-2}}\otimes(-2a)
\end{pmatrix} \ .
\label{CP2cafund}\eeq
The curvature of the connection (\ref{CP2cafund}) is given by
\beq
\cf=\diff\ca+\ca\wedge\ca \ .
\label{cfca}\eeq
Using (\ref{Fu2inst})--(\ref{diffbetaBa0}) and abbreviating
$\phi:=\phi_{0,-2}^+$, it is given explicitly by
\beq
\cf=\begin{pmatrix}F^{1,1}\otimes\Idd_2+
\big(\Idd_{p_{1,1}}-\phi\,{\phi}^\dag\big)
\otimes\big(\betab\wedge\beta^\top\,\big)&\big(\diff\phi+
A^{1,1}\,\phi-\phi\,A^{0,-2}\big)\wedge
\betab\\[4pt]-\big(\diff{\phi}^\dag+{\phi}^\dag\,
A^{1,1}-A^{0,-2}\,{\phi}^\dag\big)\wedge\beta^\top&
F^{0,-2}\otimes1+\big(\Idd_{p_{0,-2}}-{\phi}^\dag\,
{\phi}\big)\otimes\big(\beta^\top\wedge\betab\,\big)
\end{pmatrix}
\label{CP2cffund}\eeq
where $F^{n,m}=\diff A^{n,m}+A^{n,m}\wedge A^{n,m}$ is the curvature
of the vector bundle $E_{p_{n,m}}\to M_D$.

\bigskip

\noindent
{\bf Non-symmetric $\mbf{\underline{C}^{1,0}}$ quiver bundles. \ } Now
let us consider quiver bundles associated with the fundamental
representation of $\sut$ and the coset space $Q_3$.  As a complex
manifold, $Q_3$ is a quadric in $\C P^2\times\C P^2_*$. In homogeneous
complex coordinates $p^a$ and $q_a$ with $a=1,2,3$ on $\C P^2$ and $\C
P^2_*$, respectively, it is defined by the equation $p^a\,q_a=0$. This
embedding also identifies $Q_3$ as the twistor space of $\C P^2$
through the sphere fibration
\beq
\pi\,:\,Q_3~\xr{\C P^1}~\C P^2
\label{Q3CP2rel}\eeq
defined by forgetting the dependence on the coordinates $q_a$. This
twistor fibration is a geometric version of the algebraic relationship
derived in the previous section between the quivers associated to the
homogeneous spaces $Q_3$ and $\C P^2$. As we shall see, integration
over the $\C P^1$ fibres of (\ref{Q3CP2rel}) is mimicked by a sum over
monopole charges $q$ for fixed isospin $n$ which maps the
$\underline{(q,m)_n}$ representation onto $\underline{(n,m)}$. This
enables one to map configurations built on the two spaces using the
projection $\pi$ and the homomorphisms induced by it.

Consider the principal torus bundle
\beq
\sut~\xr{\uo\times\uo}~Q_3 \ .
\label{Q3princbun}\eeq
A local section of the bundle (\ref{Q3princbun}) is given 
by $3\times3$ matrices of the form
\beq
\Psi\=\begin{pmatrix}u^1&\epsilon^{1bc}\,\vb_b\,\ub_c&v^1\\[4pt]
u^2&\epsilon^{2bc}\,\vb_b\,\ub_c&v^2\\[4pt]
u^3&\epsilon^{3bc}\,\vb_b\,\ub_c&v^3\end{pmatrix} \qquad \mbox{and}
\qquad \Psi^\dag\=\begin{pmatrix}\ub_1&\ub_2&\ub_3\\[4pt]
\epsilon_{1bc}\,v^b\,u^c&\epsilon_{2bc}\,v^b\,u^c&
\epsilon_{3bc}\,v^b\,u^c\\[4pt]\vb_1&\vb_2&\vb_3\end{pmatrix} \ .
\label{PsisectionQ3}\eeq
To ensure that $\Psi\in\sut$,
i.e. $\Psi^\dag\,\Psi=\Psi\,\Psi^\dag=\Idd_3$, the complex
three-vectors $u=(u^a)$ and $v=(v^a)$ must obey the constraints
\beq
v^\dag\,u\=0 \qquad \mbox{and} \qquad v^\dag\,v\=u^\dag\,u\=1 \ .
\label{uvconstrs}\eeq
The space $Q_3$ can be covered by four $\C^3$ patches, and on one of
these patches we can solve the constraints (\ref{uvconstrs}) as
\bea
u&=&\big(u^a\big)\=\alpha\,\big(p^a\big)\=
\alpha\,\begin{pmatrix}1\\[4pt] y^1\\[4pt] \, y^2
\end{pmatrix} \qquad \mbox{with} \quad \alpha\=\frac1{
\sqrt{1+\yb^i\,y^i}} \ , \label{uvconstrsols} \nonumber\\[4pt]
v^\dag&=&\big(\bar v_a\big)\=\alpha_*\,\big(q_a\big)\\[4pt]
&=&
\alpha_*\,\big(z^1\,,\,z^2\,,\,1\big)~:=~
\alpha_*\,\big(y^1\,y^3-y^2\,,\,-y^3\,,\,1\big)
\qquad \mbox{with} \quad \alpha_*\=\frac1{\sqrt{1+\zb^i\,z^i}} \ .
\nonumber
\eea
Here $y^i$ and $z^i$ with $i=1,2$ are local complex
coordinates on $\CP^2$ and $\CP_*^2$, respectively, such that $z^1$ is
expressed in terms of the other three complex coordinates on the
quadric $Q_3\hookrightarrow\CP^2\times\CP^2_*$.

A flat connection on the trivial bundle $Q_3\times\C^3$ is then given
by the anti-hermitean one-form
\beq
A_0\=\Psi^\dag~\diff\Psi~=:~\begin{pmatrix} a_1&\gammab^3&\gammab^1
\\[4pt] -\gamma^3&-a_1-a_2&\gammab^2\\[4pt] -\gamma^1&-\gamma^2&
a_2\end{pmatrix} \ ,
\label{A0Q3def}\eeq
where from (\ref{PsisectionQ3}) the connection one-forms
$a_1,a_2\in\uoL$ are given by
\beq
a_1\=\ub_a~\diff u^a \qquad \mbox{and} \qquad a_2\=\vb_a~
\diff v^a
\label{a1a2defs}\eeq
while
\beq
\gamma^1\=u^a~\diff\vb_a \ , \qquad \gamma^2\=\epsilon^{abc}\,\vb_a\,
\ub_b~\diff\vb_c \qquad \mbox{and} \qquad \gamma^3\=
\epsilon_{abc}\,u^a\,v^b~\diff u^c \ .
\label{gamma123defs}\eeq
The one-forms $\gamma^a$, $a=1,2,3$, form a $G$-equivariant basis
for the vector space of $(1,0)$-forms on $Q_3$. Since $\Psi\in\sut$,
the canonical connection (\ref{A0Q3def}) again obeys the Cartan-Maurer
equation (\ref{CartanMaurereq}), from which we obtain the abelian
curvature equations
\beq
f_1~:=~\diff a_1\=\gammab^1\wedge\gamma^1+\gammab^3\wedge\gamma^3
\qquad \mbox{and} \qquad f_2~:=~\diff a_2\=-\gammab^1\wedge\gamma^1-
\gammab^2\wedge\gamma^2
\label{f1f2eqs}\eeq
along with the bi-covariant constancy equations
\bea
\diff\gamma^1-(a_1-a_2)\wedge\gamma^1-\gamma^2\wedge\gamma^3
&=&0 \ , \nonumber\\[4pt]
\diff\gamma^2+(a_1+2a_2)\wedge\gamma^2+\gamma^1\wedge\gammab^3
&=&0 \ , \nonumber\\[4pt]
\diff\gamma^3-(2a_1+a_2)\wedge\gamma^3-\gamma^1\wedge\gammab^2
&=&0 \ .
\label{gamma123coveqs}\eea

The geometrical meaning of these $G$-equivariant fields is as
follows. For a weight $\mu=(q,m)_n$ of the maximal torus
$T=\uo\times\uo$, the induced bundle (\ref{indhermbungen}) is the
usual homogeneous line bundle ${}^n\Lcal_{q,m}\to Q_3$ of the
Borel-Weil-Bott theory for the semisimple Lie group $\sut$. Since
topologically $\sut\cong S^5\times S^3$, one has
$\HQ^1(\sut;\Z)=\HQ^2(\sut;\Z)=\{0\}$ and thus the Leray-Serre
spectral sequence for the fibration (\ref{Q3princbun}) gives
\beq
\HQ^2(Q_3;\Z)\=\HQ^1(T;\Z)\=\Z\oplus\Z \ .
\label{H2Q3Z}\eeq
It follows that there are complex line bundles $\Lcal_{(i)}\to Q_3$,
$i=1,2$, whose $\uoL$ fluxes $g_i$ define first Chern classes which are
the generators of the free abelian group (\ref{H2Q3Z}). We can
identify ${}^n\Lcal_{q,m}$ with $(\Lcal_{(1)})^{\otimes
  q}\otimes(\Lcal_{(2)})^{\otimes m}$. By Poincar\'e-Hodge
duality, one also has $\HQ_2(Q_3;\Z)\cong\Z\oplus\Z$. This homology
group is generated by a pair of disjoint projective lines
$\CP^1_{(i)}\subset Q_3$ in the quadric
$Q_3\hookrightarrow\CP^2\times\CP_*^2$. The two-forms $g_i$ intersect
dually with these projective lines as
\beq
\frac\im{2\pi}\,\int_{\CP^1_{(i)}}\,g_j=\delta_{ij} \ ,
\label{fluxP1dual}\eeq
and thus generate the monopole charges of the line bundles 
${}^n\Lcal_{q,m}\to Q_3$. With the change of basis
\beq
f_1~:=~g_1+g_2 \qquad \mbox{and} \qquad f_2~:=~-2g_2
\label{fgQ3rels}\eeq
for the abelian group (\ref{H2Q3Z}), the corresponding monopole gauge
potentials $a_i$ have respective $(H_{\alpha_1},H_{\alpha_2})_n$
charges $(1,1)_1$ and $(0,-2)_0$.
Thus the diagonal entries of the flat connection (\ref{A0Q3def}) are
connections on the respective monopole line bundles ${}^1\Lcal_{1,1}$,
${}^1\Lcal_{-1,1}$ and ${}^0\Lcal_{0,-2}$ over $Q_3$, while from
(\ref{gamma123coveqs}) it follows that the off-diagonal elements
define bundle morphisms
\beq
{}^0\Lcal_{0,-2}~\xr{\wedge\gammab^1}~{}^1\Lcal_{1,1} \ , \qquad
{}^0\Lcal_{0,-2}~\xr{\wedge\gammab^2}~{}^1\Lcal_{-1,1} \qquad \mbox{and}
\qquad {}^1\Lcal_{-1,1}~\xr{\wedge\gammab^3}~{}^1\Lcal_{1,1} \ .
\label{gammabundlemors}\eeq
Hence (\ref{A0Q3def}) naturally defines a connection of the elementary
quiver bundle over $Q_3$ representing (\ref{fundQ3quiver}). Note that
as a K\"ahler form on $Q_3$ one can choose
\beq
\omega_{Q_3}\=-\mbox{$\frac{3\im}4$}\,(f_1-f_2)\=-
\mbox{$\frac{3\im}4$}\,
\big(2\gammab^1\wedge\gamma^1+\gammab^2\wedge\gamma^2+
\gammab^3\wedge\gamma^3\big) \ .
\label{KahlerQ3}\eeq

Analogously to the $\CP^2$ case above, we consider a $T$-equivariant
hermitean bundle $E^{1,0}\to M_D$ given by
\beq
E^{1,0}=\big(E_{{}^1p_{1,1}}\otimes\,\underline{(1,1)_1}\,_{M_D}\big)~
\oplus~\big(E_{{}^1p_{-1,1}}\otimes\,\underline{(-1,1)_1}\,_{M_D}\big)~
\oplus~\big(E_{{}^0p_{0,-2}}\otimes\,\underline{(0,-2)_0}\,_{M_D}\big)
\label{E10Q3}\eeq
along with appropriate bundle morphisms between the $E_{{}^np_{q,m}}$ such
that
\beq
\xymatrix{ E_{{}^1p_{-1,1}}~& \xrightarrow{\phantom{xxxxxx}{}^1
\phi_{-1,1}^0\phantom{xxxxxx}}
 & ~E_{{}^1p_{1,1}} \\ &
E_{{}^0p_{0,-2}}\ar[ul]^{{}^0\phi_{0,-2}^-} 
\ar[ur]_{{}^0\phi_{0,-2}^+} & }
\label{fundQ3quivbun}\eeq
is a quiver bundle over $M_D$ representing (\ref{fundQ3quiver}). Then
a $G$-equivariant connection on the induced bundle
\beq
\Ecal^{1,0}=\big(E_{{}^1p_{1,1}}\boxtimes\,{}^1\Lcal_{1,1}\big)~
\oplus~\big(E_{{}^1p_{-1,1}}\boxtimes\,{}^1\Lcal_{-1,1}\big)~
\oplus~\big(E_{{}^0p_{0,-2}}\boxtimes\,{}^0\Lcal_{0,-2}\big)
\label{Ecal10Q3}\eeq
over $M_D\times Q_3$ is given by
\beq
\ca=\begin{pmatrix} {}_1A^{1,1}\otimes1+\Idd_{{}^1p_{1,1}}\otimes a_1 &
{}^1\phi^0_{-1,1}\otimes\gammab^3 & {}^0\phi^+_{0,-2}\otimes\gammab^1 \\[4pt]
-{{}^1\phi^0_{-1,1}}^\dag\otimes\gamma^3 & {}_1A^{{-1,1}}\otimes1+
\Idd_{{}^1p_{-1,1}}\otimes(-a_1-a_2) & {}^0\phi^-_{0,-2}\otimes\gammab^2
\\[4pt] -{{}^0\phi^+_{0,-2}}^\dag\otimes\gamma^1 &
-{{}^0\phi^-_{0,-2}}^\dag\otimes\gamma^2 & {}_0A^{0,-2}\otimes1+
\Idd_{{}^0p_{0,-2}}\otimes a_2 \end{pmatrix} \ .
\label{fundeqconnQ3}\eeq
For the curvature $\cf=(\cf^{ab})$ given by (\ref{cfca}), from the
identities (\ref{f1f2eqs}) and (\ref{gamma123coveqs}) we obtain
\bea
\cf^{11}&=&{}_1F^{1,1}+\big(\Idd_{{}^1p_{1,1}}-{}^0\phi^+_{0,-2}\,
{{}^0\phi^+_{0,-2}}^\dag\,\big)\,\gammab^1\wedge\gamma^1+
\big(\Idd_{{}^1p_{1,1}}-
{}^1\phi^0_{-1,1}\,{{}^1\phi^0_{-1,1}}^\dag\,\big)\,\gammab^3\wedge\gamma^3 \ ,
\label{fundcurvQ3}\\[4pt]
\cf^{22}&=&{}_1F^{-1,1}-\big(\Idd_{{}^1p_{-1,1}}-{{}^1\phi^0_{-1,1}}^\dag\,
{}^1\phi^0_{-1,1}\big)\,\gammab^3\wedge\gamma^3+\big(\Idd_{{}^1p_{-1,1}}-
{}^0\phi^-_{0,-2}\,{{}^0\phi^-_{0,-2}}^\dag\,\big)\,\gammab^2\wedge\gamma^2 \ ,
\nonumber\\[4pt]
\cf^{33}&=&{}_0F^{0,-2}-\big(\Idd_{{}^0p_{0,-2}}-{{}^0\phi^+_{0,-2}}^\dag\,
{}^0\phi^+_{0,-2}\big)\,\gammab^1\wedge\gamma^1-\big(\Idd_{{}^0p_{0,-2}}-
{{}^0\phi^-_{0,-2}}^\dag\,{}^0\phi^-_{0,-2}\big)\,\gammab^2\wedge\gamma^2 \ ,
\nonumber\\[4pt]
\cf^{12}&=&\big(\diff \,{}^1\phi^0_{-1,1}+{}_1A^{1,1}~{}^1\phi^0_{-1,1}-{}^1
\phi^0_{-1,1}~
{}_1A^{-1,1}\big)\wedge\gammab^3+\big(\,{}^1\phi^0_{-1,1}-{}^0\phi^+_{0,-2}\,
{{}^0\phi^-_{0,-2}}^\dag\,\big)\,\gammab^1\wedge\gamma^3 \ ,
\nonumber\\[4pt]
\cf^{13}&=&\big(\diff\, {}^0\phi^+_{0,-2}+{}_1A^{1,1}~{}^0\phi^+_{0,-2}-
{}^0\phi^+_{0,-2}~{}_0A^{0,-2}\big)\wedge\gammab^1+\big(\,{}^0\phi^+_{0,-2}-
{}^1\phi^0_{-1,1}\,{}^0\phi^-_{0,-2}\big)\,\gammab^2\wedge\gammab^3 \ ,
\nonumber\\[4pt]
\cf^{23}&=&\big(\diff \,{}^0\phi^-_{0,-2}+{}_1A^{-1,1}~{}^0\phi^-_{0,-2}-
{}^0\phi^-_{0,-2}~{}_0A^{0,-2}\big)\wedge\gammab^2+\big(\,{}^0\phi^-_{0,-2}-
{{}^1\phi^0_{-1,1}}^\dag\,{}^0\phi^+_{0,-2}\big)\,\gammab^3\wedge\gamma^1
\nonumber\eea
along with the hermitean conjugates $\cf^{ba}=-(\cf^{ab})^\dag$ for
$a<b$. In (\ref{fundcurvQ3}) we have suppressed tensor products from
the notation in order to simplify the expressions somewhat. Note that
the off-diagonal matrix elements of the field strength carry
information about both holomorphic and non-holomorphic relations for
the quiver (\ref{fundQ3quiver}). In particular, the relation
(\ref{fundQ3quiverrel}) is contained in~$\cf^{13}$.

\subsection{General constructions\label{quivbungen}}

Let $\Ecal^{k,l}\to X$ be a rank~$p$ hermitean vector bundle over the
space (\ref{M2ntimesSU3H}), associated to an irreducible
representation $\underline{C}^{k,l}$ of $\sut$, with the structure group
$\urm(p)$ and gauge connection $\ca\in\urmL(p)$. There is a
one-to-one correspondence between $G$-equivariant hermitean vector
bundles over $X$ and $H$-equivariant hermitean vector bundles over
$M_D$, with $H$ acting trivially on $M_D$~\cite{A-CG-P1}. Given an
$H$-equivariant bundle $E^{k,l}\to M_D$ of rank $p$ associated to the
representation $\underline{C}^{k,l}\big|_H$ of $H$, the corresponding
$G$-equivariant bundle over $X$ is defined by induction as
\beq
\Ecal^{k,l}=G\times_HE^{k,l} \ .
\label{Ecalklind}\eeq
As a holomorphic vector bundle, the action of the Levi subgroup
$L=H\otimes \C_*$ on $E^{k,l}$ is defined by the isotopical
decomposition
\beq
E^{k,l}~\cong~\bigoplus_{v\in\quiver_0(k,l)}\,E_{p_v}\otimes\,
\underline{v}\,_{M_D}\qquad \mbox{with} \quad
E_{p_v}\=\Hom_L\big(\,\underline{v}\,_{M_D}\,,\,E^{k,l}\,\big) \ ,
\label{Eklisotopical}\eeq
where $\underline{v}\,_{M_D}$ denotes the trivial $H$-equivariant bundle 
${M_D}\times\underline{v}$
over $M_D$ corresponding to the irreducible $H$-module $\underline{v}$.
The holomorphic vector bundles $E_{p_v}\to M_D$ of rank $p_v$, at the
vertices $v\in\quiver_0(k,l)$ of the pertinent quiver, have trivial
$L$-actions. The rank $p$ of $E^{k,l}$ is given by
\beq
p\=\sum_{v\in\quiver_0(k,l)}\,d_v\,p_v \qquad \mbox{with} \quad
d_v\=\dim\big(\,\underline{v}\,\big) \ .
\label{rankdimv}\eeq
The extension of the $L$-action to a $P$-action is defined by means of
bi-fundamental Higgs fields
$\phi_{v,\Phi(v)}\in\Hom(E_{p_v},E_{p_{\Phi(v)}})$ which, after
imposing BPS conditions, will satisfy the holomorphic relations
$\rel(k,l)$ of the quiver. These holomorphic bundle morphisms realize
the $G$-action of the coset generators which twists the naive
dimensional reduction by ``off-diagonal'' terms. This construction
breaks the gauge group of the bundle $E^{k,l}$ as
\beq
\urm(p)~\longrightarrow~\prod_{v\in\quiver_0(k,l)}\,\urm(p_v)
\label{gaugegroupbreak}\eeq
via the Higgs effect.

With $\Vcal_v$ the homogeneous bundle (\ref{indhermbungen}) induced by
the irreducible $H$-module $\underline{v}$, the structure group of the
principal bundle associated to
\beq
\Ecal^{k,l}=\bigoplus_{v\in\quiver_0(k,l)}\,E_{p_v}\boxtimes\Vcal_v
\label{Ecalklisotopical}\eeq
is then $H\times\prod_{v\in\quiver_0(k,l)}\,\urm(p_v)$. The space of
  $G$-equivariant sections of a homogeneous bundle
  (\ref{indhermbungen}) induced by a representation $\underline{V}$ of
  $H$ is in a one-to-one correspondence with the set of $H$-invariant
  subspaces of $\underline{V}$. It follows that the vector space of
  $G$-equivariant $\Vcal$-valued $(0,1)$-forms on the homogeneous
  space $G/H$ is given by
\beq
\Omega^{0,1}(\Vcal)^G\cong\big(\mathfrak{u}^\vee\otimes
\underline{V}\,\big)^H \ .
\label{Omega01Hcal}\eeq
To determine the generic form of a $G$-equivariant, holomorphic
connection one-form $\ca$ on the bundle $\Ecal^{k,l}\to X$, we
decompose the space $\Omega^{0,1}(\End(\Ecal^{k,l}))^G$ using the
Whitney sum (\ref{Ecalklisotopical}). Since by Schur's lemma
$\Hom(\,\underline{v}\,,\,\underline{v'}\,)^G\cong\delta_{v,v'}~\C$,
corresponding to each vertex $v\in\quiver_0(k,l)$ there is a
``diagonal'' subspace
\beq
\big(\Omega^{0,1}(\End(E_{p_v}))\otimes\Idd_{d_v}\big)~\oplus~\big(
\Idd_{p_v}\otimes\Omega^{0,1}(\End(\Vcal_{v}))^G\,\big)
\label{diagsubsp}\eeq
in which we can choose a connection $A^v$ on the bundle $E_{p_v}\to
M_D$ twisted by a $G$-equivariant connection on the homogeneous 
vector bundle
$\Vcal_v\to G/H$. To each arrow $\Phi\in\quiver_1(k,l)$ there is an
``off-diagonal'' subspace $\Omega^0(\Hom(E_{p_v},E_{p_{\Phi(v)}}))\otimes
\Omega^{0,1}(\Hom(\Vcal_v,\Vcal_{\Phi(v)}))^G$, with
\beq
\Omega^{0,1}\big(\Hom(\Vcal_v,\Vcal_{\Phi(v)})\big)^G\cong\Big(
\mathfrak{u}^\vee\otimes\Hom\big(\,\underline{v}\,,\,\underline{
\Phi(v)}\,\big)\Big)^H \ ,
\label{Omega01HomHcal}\eeq
in which we twist the Higgs fields $\phi_{v,\Phi(v)}$ by suitable
invariant $d_{\Phi(v)}\times d_v$ matrix-valued $(0,1)$-forms built
from basis $(0,1)$-forms spanning
$\Omega^{0,1}( G/H)^G\cong(\mathfrak{u}^\vee\,)^H$. Thus the
condition of $G$-equivariance dictates the form of the gauge
connection $\ca$ in $p_v\,d_v\times p_{\Phi(v)}\,d_{\Phi(v)}$ blocks.

\bigskip

\noindent
{\bf The Biedenharn basis. \ } The key to making the above construction
explicit is finding a suitable basis for the irreducible representations
$\underline{C}^{k,l}$ of $\sut$ that is tailored to the structure of
the arrows and vertices of the pertinent quiver, which requires
appropriately assembling the invariant $(0,1)$-forms into rectangular
block matrices. A particularly nice basis for this is the Biedenharn
representation~\cite{Biedenharn,MP1} which takes care of both
symmetric and non-symmetric quivers simultaneously. The complete set
of $d^{k,l}$ orthonormal basis vectors in this basis set are denoted
$\big|\noverq,m\big\rangle$ and are labelled by the isospin quantum
numbers $n=2I$, $q=2I_z$ and the hypercharge $m=3Y$. As such, they will
encode the vertices and arrows of the quivers through the actions of
the $\sut$ operators as given in (\ref{CP2arrows}) and
(\ref{Q3arrows}). These states define the spin $\frac n2$
representation of the isospin subgroup $\su\subset\sut$ through
\bea
H_{\alpha_1}\big|\noverq\,,\,m\big\rangle&=&
q\,\big|\noverq\,,\,m\big\rangle \ , \label{Izrep}\\[4pt]
E_{\pm\,\alpha_1}\big|\noverq\,,\,m\big\rangle&=&
\mbox{$\frac12$}\,\sqrt{(n\mp q)\,(n\pm q+2)}~\big|\noverqpmt\,,\,m
\big\rangle \ .
\label{Ipmrep}\eea
They are also hypercharge eigenstates with
\beq
H_{\alpha_2}\big|\noverq\,,\,m\big\rangle=m\,\big|\noverq\,,\,m
\big\rangle \ .
\label{Yrep}\eeq

The remaining matrix elements can be determined by noting that the
generators $E_{\alpha_1+\alpha_2},E_{\alpha_2}$ form the $\pm\,\frac12$
spin components of a spherical tensor operator of rank $\frac12$ with
respect to the isospin subgroup $\su\subset\sut$. All of their matrix
elements can thus be determined by applying the Wigner-Eckart theorem
for $\su$ and relating the resulting reduced matrix elements to an
irreducible $\sut$ tensor~\cite{MP1}. The latter computation
determines numerical coefficient functions
\bea
\lambda_{k,l}^+(n,m)&=&\sqrt{\frac{\big(\frac{k+2l}3+\frac n2+\frac
    m6+2\big)\,\big(\frac{k-l}3+\frac n2+\frac m6+1\big)\,\big(
\frac{2k+l}3-\frac n2-\frac m6\big)}{n+2}} \ , \nonumber\\[4pt]
\lambda_{k,l}^-(n,m)&=&\sqrt{\frac{\big(\frac{k+2l}3-\frac n2+\frac
    m6+1\big)\,\big(\frac{l-k}3+\frac n2-\frac m6\big)\,\big(
\frac{2k+l}3+\frac n2-\frac m6+1\big)}{n}} \ .
\label{lambdaklnm}\eea
The latter constants are defined for $n>0$ and we set
$\lambda_{k,l}^-(0,m):=0$. For fixed $k,l$, many of the constants
(\ref{lambdaklnm}) vanish. Then with the Biedenharn phase convention
one has
\bea
E_{\alpha_2}\big|\noverq\,,\,m\big\rangle&=&
\biggl[\cgstack{\frac n2}{\frac q2}~\cgstack{\frac12}{-\frac12}~
\cgstack{\frac{n+1}2}{\frac{q-1}2}\biggr]\,\lambda_{k,l}^+(n,m)\,
\big|\npoverqmo\,,\,m+3\big\rangle+
\biggl[\cgstack{\frac n2}{\frac q2}~\cgstack{\frac12}{-\frac12}~
\cgstack{\frac{n-1}2}{\frac{q-1}2}\biggr]\,\lambda_{k,l}^-(n,m)\,
\big|\nmoverqmo\,,\,m+3\big\rangle \ , \nonumber\\ &&
\label{Fprep}\\[4pt]
E_{\alpha_1+\alpha_2}\big|\noverq\,,\,m\big\rangle&=&
\biggl[\cgstack{\frac n2}{\frac q2}~\cgstack{\frac12}{\frac12}~
\cgstack{\frac{n+1}2}{\frac{q+1}2}\biggr]\,\lambda_{k,l}^+(n,m)\,
\big|\npoverqpo\,,\,m+3\big\rangle+
\biggl[\cgstack{\frac n2}{\frac q2}~\cgstack{\frac12}{\frac12}~
\cgstack{\frac{n-1}2}{\frac{q+1}2}\biggr]\,\lambda_{k,l}^-(n,m)\,
\big|\nmoverqpo\,,\,m+3\big\rangle \ , \nonumber\\ &&
\label{Fmrep}\eea
where the square brackets denote $\su$ Clebsch-Gordan coefficients
which in the present case can be computed explicitly from the known
values
\beq
\biggl[\cgstack{j}{m}~\cgstack{\frac12}{\alpha}~\cgstack{j+\frac12}
{m+\alpha}\biggr]\=\sqrt{\frac{j+2\alpha\,m+1}{2j+1}}
\qquad \mbox{and} \qquad \biggl[\cgstack{j}{m}~\cgstack{\frac12}
{\alpha}~\cgstack{j-\frac12}
{m+\alpha}\biggr]\=2|\alpha|\,\sqrt{\frac{j-2\alpha\,m}{2j+1}}
\ .
\label{CGcoeffs}\eeq
The analogous relations for $E_{-\alpha_2}$ and
$E_{-\alpha_1-\alpha_2}$ can be derived by hermitean conjugation of
(\ref{Fprep}) and (\ref{Fmrep}), respectively. 

The Biedenharn basis is related to the more conventional basis of
irreducible $\sut$ tensors $\mbf T$ for the representation
$\underline{C}^{k,l}$ as follows. Introducing the $\su$ spins
\beq
j_\pm:=\mbox{$\frac n4\pm\frac m{12}\pm\frac16\,(k-l)$}
\label{jpmSU2spins}\eeq
with $2j_+=0,1,\dots,k$ and $2j_-=0,1,\dots,l$, one has the change of
basis formulas
\bea
\big|\noverq\,,\,m\big\rangle&=&N_3(n,m)\,\sum_{|m_\pm|\leq
j_\pm}\,\biggl[\cgstack{j_+}{m_+}~\cgstack{j_-}{m_-}~
\cgstack{\frac{n}2}{\frac{q}2}\biggr]\,N_1(j_\pm,m_\pm)~
{\mbf T}^{(j_++m_+,j_+-m_+,k-2j_+)}_{(j_--m_-,j_-+m_-,
l-2j_-)} \ , \nonumber\\[4pt]
{\mbf T}^{(j_++m_+,j_+-m_+,k-2j_+)}_{(j_--m_-,j_-+m_-,
l-2j_-)}&=&\frac1{N_1(j_\pm,m_\pm)}\,\sum_{n=0}^{k+l}\,
\biggl[\cgstack{j_+}{m_+}~\cgstack{j_-}{m_-}~
\cgstack{\frac{n}2}{\frac{q}2}\biggr]\,\frac{N_2(j_\pm,n)}
{N_3(n,m)}~\big|\noverq\,,\,m\big\rangle \ ,
\label{BiedenTchange}\eea
where the coefficients $N_1(j_\pm,m_\pm)$, $N_2(j_\pm,n)$ and
$N_3(n,m)$ can be found in~\cite{MP1}, and the brackets $(r,s,t)$
indicate that the tensor $\mbf T$ has $r$ upper or lower indices equal
to~$1$, $s$ equal to~$2$, and $t$ equal to~$3$. From the
decompositions (\ref{fundSU2U1decomp}), (\ref{fundU1U1decomp}) and
their conjugates it follows that the state ${\mbf
  T}^{(j_++m_+,j_+-m_+,k-2j_+)}_{(j_--m_-,j_-+m_-,l-2j_-)}$ possesses
definite values of hypercharge and isospin with the values
\beq
m\=6(j_+-j_-)-2(k-l) \ , \qquad
q\=2(m_++m_-) \qquad \mbox{and} \qquad n\=2(j_++j_-) \ .
\label{mnqjpm}\eeq
Thus the $\su$ spin $j_+$ (resp.~$j_-$) is the value of the isospin
$I$ contributed by the upper (resp.~lower) indices of the $\sut$
tensor $\mbf T$. With this representation it is now straightforward to
write down the $G$-equivariant gauge connections associated to a
generic irreducible representation $\underline{C}^{k,l}$.

\bigskip

\noindent
{\bf Symmetric $\mbf{\underline{C}^{k,l}}$ quiver bundles. \ } We
begin by noting that the flat connection (\ref{A0CP2def}) can be
written in terms of the $3\times3$ matrices of the defining
representation of $\sut$ from Section~\ref{fundquiverreps} as
\beq
A_0=\big[B^{11}\,H_{\alpha_1}+B^{12}\,E_{\alpha_1}-(B^{12}\,
E_{\alpha_1})^\dag\big]+a\,H_{\alpha_2}+
\betab^1\,E_{\alpha_1+\alpha_2}+
\betab^2\,E_{\alpha_2}-\beta^1\,E_{-\alpha_1-\alpha_2}-
\beta^2\,E_{-\alpha_2} \ ,
\label{A0CP2gens}\eeq
where $B^{ij}$ are the matrix elements of the $\suL$-valued instanton
connection $B_{(1)}=B-a~\Idd_2$. As expected, the $(0,1)$-forms
$\betab^i$ on $\C P^2$ are coupled to the generators of the abelian
algebra $\mathfrak{u}$. The corresponding connection of the
quiver bundle over $\C P^2$ associated to the representation
$\underline{C}^{k,l}$ is then obtained by substituting into
(\ref{A0CP2gens}) the $d^{k,l}\times d^{k,l}$ matrices defined by
(\ref{Izrep})--(\ref{CGcoeffs}).

For a fixed vertex $(n,m)\in\quiver_0(k,l)$, one has $d_{(n,m)}=n+1$
and we write
\bea
B_{n,m}&:=&\sum_{q\in\{-n+2j\}_{j=0}^n}\,\Big(q\,B^{11}\,\big|
\noverq\,,\,m\big\rangle\big\langle\noverq\,,\,m\big|+
\mbox{$\frac12$}\,B^{12}\,\sqrt{(n-q)\,(n+q+2)}~\big|
\noverqpt\,,\,m\big\rangle\big\langle\noverq\,,\,m\big|\nonumber\\
&& \hspace{3cm} -\,\mbox{$\frac12$}\,
\overline{B^{12}}\,\sqrt{(n+q)\,(n-q+2)}~\big|
\noverqmt\,,\,m\big\rangle\big\langle\noverq\,,\,m\big|\Big)
\label{Bndef}\eea
for the one-instanton connection $B_{(n)}=B_{n,m}$ in the
$(n+1)$-dimensional irreducible representation of $\su$. Denote by
\beq
\Pi_{n,m}:=\sum_{q\in\{-n+2j\}_{j=0}^n}\,\big|
\noverq\,,\,m\big\rangle\big\langle\noverq\,,\,m\big|
\label{Pinmdef}\eeq
the projection of $\underline{C}^{k,l}\big|_H$ onto the irreducible
representation $\underline{(n,m)}$ of 
$H=\su\times\uo$. We further write
\bea
\betab^{\pm}_{n,m}&:=&\sum_{q\in\{-n+2j\}_{j=0}^n}\,
\lambda_{k,l}^\pm(n,m)\,\bigg(\betab^1\,
\biggl[\cgstack{\frac n2}{\frac q2}~
\cgstack{\frac12}{-\frac12}~\cgstack{\frac{n\pm1}2}{\frac{q-1}2}
\biggr]\,\big|\npmoverqmo\,,\,m+3\big\rangle\big\langle\noverq\,,\,
m\big| \nonumber\\ && \hspace{5cm}
+\,\betab^2\,\biggl[\cgstack{\frac n2}{\frac q2}~
\cgstack{\frac12}{\frac12}~\cgstack{\frac{n\pm1}2}{\frac{q+1}2}
\biggr]\,\big|\npmoverqpo\,,\,m+3\big\rangle\big\langle\noverq\,,\,
m\big|\bigg)
\label{betanmdef}\eea
for the $(n\pm1+1)\times(n+1)$ matrix blocks of $G$-equivariant
elementary bundle morphisms between the nodes of the corresponding
quiver diagram, and their hermitean conjugates
$\beta_{n,m}^\pm:={{\betab}^\pm_{n,m}}{}^\dag$.

Then the elementary flat quiver connection can be written as
\beq
A_0=\sum_{(n,m)\in\quiver_0(k,l)}\,\Big(B_{n,m}+m\,a~\Pi_{n,m}
+\betab^+_{n,m}+\betab^-_{n,m}-\beta^+_{n,m}-\beta^-_{n,m}\Big) \ .
\label{A0CP2Cklrep}\eeq
At each vertex $(n,m)\in\quiver_0(k,l)$ the Cartan-Maurer equation
(\ref{CartanMaurereq}) gives the curvature equations
\beq
F_{B_{(n)}}+m\,f_{\uoL}~\Idd_{n+1}=\beta_{n,m}^+\wedge\betab_{n,m}^++
\beta_{n,m}^-\wedge\betab_{n,m}^-+\betab_{n-1,m-3}^+\wedge
\beta_{n-1,m-3}^++\betab_{n+1,m-3}^-\wedge\beta_{n+1,m-3}^- \ ,
\label{curveqsCklrep}\eeq
along with the bi-covariant constancy conditions
\beq
\diff\betab^\pm_{n,m}+\betab^\pm_{n,m}\wedge B_{(n)}+B_{(n\pm1)}\wedge
\betab^\pm_{n,m}-3\betab^\pm_{n,m}\wedge a=0
\label{betabeqsCklrep}\eeq
and the commutation relations
\beq
\betab_{n,m}^+\wedge\betab_{n+1,m-3}^-\=
\betab_{n+2,m}^-\wedge\betab_{n+1,m-3}^+ \qquad \mbox{and} \qquad
\betab_{n,m}^+\wedge\beta_{n,m}^-\=
\beta_{n+1,m+3}^-\wedge\betab_{n-1,m+3}^+
\label{betarelsCklrep}\eeq
plus their hermitean conjugates. All other operator products vanish
identically. In these equations $B_{(n)}:=B_{n,m}$ while
$B_{(n\pm1)}:=B_{n\pm1,m+3}$.

We can extend (\ref{A0CP2Cklrep}) naturally to a $G$-equivariant
gauge connection $\ca$ on the bundle (\ref{Ecalklisotopical}) over
$M_D\times\C P^2$ by introducing the projection $\pi_{p_{n,m}}$ onto
the sub-bundle $E_{p_{n,m}}\to M_D$ to write
\bea
\ca&=&\sum_{(n,m)\in\quiver_0(k,l)}\,\Big(A^{n,m}\otimes
\Pi_{n,m}+\pi_{p_{n,m}}\otimes\big(B_{n,m}+m\,a~\Pi_{n,m}\big)
\label{calACP2Cklrep}\\ && \hspace{3cm}
+\,\phi_{n,m}^+\otimes\betab^+_{n,m}+\phi_{n,m}^-\otimes
\betab_{n,m}^--
{\phi_{n,m}^+}^\dag\otimes\beta^+_{n,m}-{\phi_{n,m}^-}^\dag
\otimes\beta_{n,m}^-\Big) \ . \nonumber
\eea
This operator acts on the typical fibre space
$\underline{V}^{k,l}\cong\C^{\sum_{(n,m)\in\quiver_0(k,l)}\,p_{n,m}}\otimes
\C^{\sum_{(n,m)\in\quiver_0(k,l)}\,(n+1)}$. The matrix elements of the
curvature two-form (\ref{cfca}) in the Biedenharn basis can be
computed by using the flatness conditions on $A_0$ above to simplify
the diagonal and off-diagonal components. At each vertex
$(n,m)\in\quiver_0(k,l)$ one finds using (\ref{curveqsCklrep}) the
diagonal matrix elements
\bea
\cf^{n,m\,;\,n,m}&=&F^{n,m}\otimes\Idd_{n+1}+
\big(\Idd_{p_{n,m}}-{\phi_{n,m}^+}^\dag\,
\phi_{n,m}^+\big)\otimes\big(\beta_{n,m}^+\wedge\betab_{n,m}^+\big)
\nonumber\\ && +\,\big(
\Idd_{p_{n,m}}-{\phi_{n,m}^-}^\dag\,\phi_{n,m}^-\big)\otimes\big(
\beta_{n,m}^-\wedge\betab_{n,m}^-\big) \nonumber\\ &&
+\,\big(\Idd_{p_{n,m}}-\phi_{n-1,m-3}^+\,{\phi_{n-1,m-3}^+}^\dag
\,\big)\otimes\big(\betab_{n-1,m-3}^+\wedge\beta_{n-1,m-3}^+\big)
\nonumber\\ &&
+\,\big(\Idd_{p_{n,m}}-\phi_{n+1,m-3}^-\,{\phi_{n+1,m-3}^-}^\dag
\,\big)\otimes\big(\betab_{n+1,m-3}^-\wedge\beta_{n+1,m-3}^-\big)
\label{curvdiagCklrep}\eea
with $F^{n,m}=\diff A^{n,m}+A^{n,m}\wedge A^{n,m}$, while from
(\ref{betabeqsCklrep}) and (\ref{betarelsCklrep}) the non-vanishing
off-diagonal matrix elements are given respectively by
\beq
\cf^{n\pm1,m+3\,;\,n,m}=\big(\diff\phi_{n,m}^\pm+A^{n\pm1,m+3}\,
\phi_{n,m}^\pm-
\phi_{n,m}^\pm\,A^{n,m}\big)\wedge\betab_{n,m}^\pm
\label{curvoffdiagCklrep}\eeq
and
\bea
\cf^{n+1,m+3\,;\,n+1,m-3}&=&\big(\phi_{n,m}^+\,\phi_{n+1,m-3}^--
\phi_{n+2,m}^-\,\phi_{n+1,m-3}^+\big)\otimes\big(\betab_{n,m}^+\wedge
\betab^-_{n+1,m-3}\big) \ , \label{curvholrelsCklrep}\\[4pt]
\cf^{n+1,m+3\,;\,n-1,m+3}&=&\big(\phi_{n,m}^+\,{\phi_{n,m}^-}^\dag-
{\phi_{n+1,m+3}^-}^\dag\,\phi_{n-1,m+3}^+\big)\otimes
\big(\betab^+_{n,m}\wedge\beta^-_{n,m}\big)
\label{curvnonholrelsCklrep}\eea
along with their hermitean conjugates
$\cf^{r,s\,;\,n,m}=-(\cf^{n,m\,;\,r,s})^\dag$ for
$(r,s)\neq(n,m)$. Note that the matrix elements
(\ref{curvoffdiagCklrep}) define bi-fundamental covariant derivatives
$D\phi_{n,m}^\pm$ of the Higgs fields, while (\ref{curvholrelsCklrep})
(resp.~(\ref{curvnonholrelsCklrep})) contain holomorphic
(resp.~non-holomorphic) Higgs field relations arising from
commutativity of the quiver diagram.

\bigskip

\noindent
{\bf Non-symmetric $\mbf{\underline{C}^{k,l}}$ quiver bundles. \ } The
flat connection (\ref{A0Q3def}) can be written as
\beq
A_0=\big(a_1+\mbox{$\frac12$}\,a_2\big)\,H_{\alpha_1}-
\mbox{$\frac12$}\,a_2\,H_{\alpha_2}+\gammab^1\,E_{\alpha_1+\alpha_2}+
\gammab^2\,E_{\alpha_2}+\gammab^3\,E_{\alpha_1}-\gamma^1\,
E_{-\alpha_1-\alpha_2}-\gamma^2\,E_{-\alpha_2}-\gamma^3\,
E_{-\alpha_1}
\label{A0Q3gens}\eeq
with the $(0,1)$-forms $\gammab^a$ on $Q_3$ coupled to the generators
of the Heisenberg algebra $\mathfrak{u}$. In this case we label the
vertices of the quiver by $(q,m)_n\in\quiver_0(k,l)$, with 
$d_{(q,m)_n}=1$ for each one. Let
\beq
{}^n\Pi_{q,m}:=\big|
\noverq\,,\,m\big\rangle\big\langle\noverq\,,\,m\big|
\label{Piqmdef}\eeq
be the projection of $\underline{C}^{k,l}\big|_T$ onto the
one-dimensional representation $\underline{(q,m)_n}$ of the maximal
torus $T$. Introduce the corresponding $G$-equivariant
$(0,1)$-forms
\bea
{}^n\gammab^{(\pm)+}_{q,m}&:=&\gammab^1~
\biggl[\cgstack{\frac n2}{\frac q2}~
\cgstack{\frac12}{\frac12}~\cgstack{\frac{n\pm1}2}{\frac{q+1}2}
\biggr]\,\lambda_{k,l}^\pm(n,m)~
\big|\npmoverqpo\,,\,m+3\big\rangle\big\langle\noverq\,,\,
m\big|  \ , \label{gammapqmdef}\\[4pt]
{}^n\gammab^{(\pm)-}_{q,m}&:=&\gammab^2~
\biggl[\cgstack{\frac n2}{\frac q2}~
\cgstack{\frac12}{-\frac12}~\cgstack{\frac{n\pm1}2}{\frac{q-1}2}
\biggr]\,\lambda_{k,l}^\pm(n,m)~
\big|\npmoverqmo\,,\,m+3\big\rangle\big\langle\noverq\,,\,
m\big| \ , \label{gammamqmdef}\\[4pt]
{}^n\gammab^0_{q,m}&:=&\mbox{$\frac12$}\,
\gammab^3~\sqrt{(n-q)\,(n+q+2)}~
\big|\noverqpt\,,\,m\big\rangle\big\langle\noverq\,,\,m\big| \ ,
\label{gamma0qmdef}\eea
and the linear combinations
\beq
{}^n\gammab^{\pm}_{q,m}={}^n\gammab^{(+)\pm}_{q,m}+{}^n
\gammab^{(-)\pm}_{q,m}
\label{gammalincomb}\eeq
along with their hermitean conjugates
${}^n\gamma_{q,m}^{(\kappa)\pm}:={}^n\gammab_{q,m}^{(\kappa)\pm}{\,}^\dag$
for $\kappa=\pm$, plus ${}^n\gamma_{q,m}^\pm:={}^n\gammab_{q,m}^\pm{}^\dag$
and ${}^n\gamma_{q,m}^0:={}^n\gammab_{q,m}^0{}^\dag$. 
The Cartan-Maurer equation (\ref{CartanMaurereq}) for the connection 
(\ref{A0Q3gens}) then yields the curvature constraints
\bea
\big(q\,f_1+\mbox{$\frac12$}\,(q-m)\,f_2\big)~{}^n\Pi_{q,m}
&=&{}^n\gamma_{q,m}^+\wedge\,
{}^n\gammab_{q,m}^++{}^n\gamma_{q,m}^-\wedge\,{}^n\gammab_{q,m}^-+
{}^n\gamma_{q,m}^0\wedge\,{}^n\gammab_{q,m}^0 \nonumber \\ && +\,
{}^{n-1}\gammab_{q-1,m-3}^{(+)+}\wedge\,{}^{n-1}
\gamma_{q-1,m-3}^{(+)+}+{}^{n+1}\gammab_{q-1,m-3}^{(-)+}\wedge\,
{}^{n+1}\gamma_{q-1,m-3}^{(-)+} \nonumber \\ && +\,
{}^{n-1}\gammab^{(+)-}_{q+1,m-3}\wedge\,{}^{n-1}
\gamma^{(+)-}_{q+1,m-3}+{}^{n+1}\gammab^{(-)-}_{q+1,m-3}\wedge\,
{}^{n-1}\gamma^{(-)-}_{q+1,m-3} \nonumber \\ && +\,
{}^n\gammab^0_{q-2,m}\wedge\,{}^n\gamma^0_{q-2,m} \ ,
\label{Q3curveqsCklrep}\eea
along with the bi-covariant constancy conditions
\bea
\diff\,{}^n\gammab_{q,m}^+&=&(a_1-a_2)\wedge\,{}^n\gammab^+_{q,m}+
{}^n\gammab^-_{q+2,m}\wedge\,{}^n\gammab^0_{q,m}+
\big(\,{}^{n+1}\gammab^0_{q-1,m+3}+{}^{n-1}\gammab^0_{q-1,m+3}\big)
\wedge\,{}^n\gammab^-_{q,m} \ , \nonumber\\[4pt]
\diff\,{}^n\gammab_{q,m}^-&=&-(a_1+2a_2)\wedge\,{}^n\gammab^-_{q,m}-
{}^n\gammab^+_{q-2,m}\wedge\,{}^n\gamma^0_{q-2,m}-
\big(\,{}^{n+1}\gamma^0_{q-1,m+3}+{}^{n-1}\gamma^0_{q-1,m+3}\big)
\wedge\,{}^n\gammab^+_{q,m} \ , \nonumber\\[4pt]
\diff\,{}^n\gammab_{q,m}^0&=&(2a_1+a_2)\wedge\,{}^n\gammab^0_{q,m}-
{}^n\gamma^-_{q+2,m}\wedge\,{}^n\gammab^+_{q,m} \nonumber\\ && -\,
{}^{n-1}\gammab^{(+)+}_{q+1,m-3}\wedge\,{}^{n-1}
\gamma^{(+)-}_{q+1,m-3}-{}^{n+1}\gammab^{(-)+}_{q+1,m-3}\wedge\,
{}^{n+1}\gamma^{(-)-}_{q+1,m-3}
\label{Q3bicovCklrep}\eea
and the commutation relations
\bea
{}^n\gammab^+_{q,m}\wedge\big(\,{}^{n-1}\gammab^{(+)-}_{q+1,m-3}
+{}^{n+1}\gammab^{(-)-}_{q+1,m-3}\big)&=&{}^n\gammab^-_{q+2,m}\wedge
\big(\,{}^{n-1}\gammab^{(+)+}_{q+1,m-3}+{}^{n+1}
\gammab^{(-)+}_{q+1,m-3}\big) \ , \nonumber\\[4pt]
{}^n\gammab^+_{q,m}\wedge\,{}^n\gammab^0_{q-2,m}
&=&\big(\,{}^{n-1}\gammab^0_{q-1,m+3}+{}^{n+1}\gammab^0_{q-1,m+3}
\big)\wedge\,{}^{n}\gammab^+_{q-2,m} \ , \nonumber\\[4pt]
{}^n\gammab^-_{q,m}\wedge\,{}^n\gammab^0_{q-2,m}
&=&\big(\,{}^{n-1}\gammab^0_{q-3,m+3}+{}^{n+1}\gammab^0_{q-3,m+3}
\big)\wedge\,{}^n\gammab^-_{q-2,m} \ , \nonumber\\[4pt]
{}^n\gammab^-_{q,m}\wedge\,{}^n\gamma^0_{q,m}
&=&\big(\,{}^{n-1}\gamma^0_{q-1,m+3}+{}^{n+1}\gamma^0_{q-1,m+3}
\big)\wedge\,{}^n\gammab^-_{q+2,m} \ , \nonumber\\[4pt]
{}^{n-1}\gammab^{(+)+}_{q,m}\wedge\,{}^{n-1}\gamma^{(+)-}_{q,m}
+{}^{n+1}\gammab^{(-)+}_{q,m}\wedge\,{}^{n+1}\gamma^{(-)-}_{q,m}
&=&{}^n\gamma^-_{q+1,m+3}\wedge\,
{}^n\gammab^+_{q-1,m+3} \ , \label{gammaCklreprels} \\[4pt]
{}^n\gammab^+_{q,m}\wedge\,{}^n\gamma^0_{q,m}
&=&\big(\,{}^{n-1}\gamma^0_{q+1,m+3}+{}^{n+1}\gamma^0_{q+1,m+3}
\big)\wedge\,{}^n\gammab^+_{q+2,m} \ , \nonumber
\eea
plus their hermitean conjugates.

Then a $G$-equivariant gauge connection $\ca$ on the bundle
(\ref{Ecalklisotopical}) over $M_D\times Q_3$ is given by
\bea
\ca&=&\sum_{(q,m)_n\in\quiver_0(k,l)}\,
\Big({}_nA^{q,m}\otimes\,{}^n\Pi_{q,m}+
\pi_{{}^np_{q,m}}\otimes\big(q\,a_1+\mbox{$\frac12$}\,
(q-m)\,a_2\big)~{}^n\Pi_{q,m} \nonumber \\ && 
\qquad\qquad\qquad+\,
{}^n\phi_{q,m}^+\otimes\,{}^n\gammab^+_{q,m}+
{}^n\phi_{q,m}^-\otimes\,{}^n\gammab^-_{q,m}+{}^n\phi_{q,m}^0\otimes\,
{}^n\gammab^0_{q,m}\label{calAQ3Cklrep}\\ && \qquad\qquad\qquad 
-\,{{}^n\phi_{q,m}^+}^\dag\otimes\,{}^n\gamma^+_{q,m}
-{{}^n\phi_{q,m}^-}^\dag\otimes\,{}^n\gamma^-_{q,m}-
{{}^n\phi_{q,m}^0}^\dag
\otimes\,{}^n\gamma^0_{q,m}\Big) \ . \nonumber
\eea
The respective constraint equations
(\ref{Q3curveqsCklrep})--(\ref{gammaCklreprels}) may then be used to
compute the diagonal field strength matrix elements
\bea
{}_{n}\cf^{q,m\,;\,q,m}&=&{}_nF^{q,m}+\sum_{\kappa=0,\pm}\,
\big(\Idd_{{}^np_{q,m}}-
{{}^n\phi_{q,m}^\kappa}^\dag\,{}^n\phi_{q,m}^\kappa
\big)~{}^n\gamma_{q,m}^\kappa
\wedge\,{}^n\gammab_{q,m}^\kappa \nonumber\\ &&
+\,\big(\Idd_{{}^{n}p_{q,m}}-{}^{n-1}\phi^+_{q-1,m-3}\,
{{}^{n-1}\phi_{q-1,m-3}^+}^\dag\,
\big)~{}^{n-1}\gammab_{q-1,m-3}^{(+)+}\wedge\,{}^{n-1}
\gamma^{(+)+}_{q-1,m-3} \nonumber\\ &&
+\,\big(\Idd_{{}^{n}p_{q,m}}-{}^{n+1}\phi^+_{q-1,m-3}\,
{{}^{n+1}\phi_{q-1,m-3}^+}^\dag\,
\big)~{}^{n+1}\gammab_{q-1,m-3}^{(-)+}\wedge\,{}^{n+1}
\gamma^{(-)+}_{q-1,m-3} \nonumber\\ && 
+\,\big(\Idd_{{}^{n}p_{q,m}}-{}^{n-1}\phi^-_{q+1,m-3}\,
{{}^{n-1}\phi_{q+1,m-3}^-}^\dag\, 
\big)~{}^{n-1}\gammab_{q+1,m-3}^{(+)-}\wedge\,{}^{n-1}
\gamma^{(+)-}_{q+1,m-3} \nonumber\\ && 
+\,\big(\Idd_{{}^{n}p_{q,m}}-{}^{n+1}\phi^-_{q+1,m-3}\,
{{}^{n+1}\phi_{q+1,m-3}^-}^\dag\, 
\big)~{}^{n+1}\gammab_{q+1,m-3}^{(-)-}\wedge\,{}^{n+1}
\gamma^{(-)-}_{q+1,m-3} \nonumber\\ && 
+\,\big(\Idd_{{}^np_{q,m}}-{}^n\phi^0_{q-2,m}\,{{}^n
\phi_{q-2,m}^0}^\dag\,\big)~
{}^n\gammab_{q-2,m}^0\wedge\,{}^n\gamma^0_{q-2,m} \ ,
\label{MdQ3curvdiagCklrep}\eea
together with the non-vanishing off-diagonal elements
\bea
{}_{n\pm1}\cf^{q+1,m+3\,;\,q,m}&=&\big(\diff\,{}^n\phi_{q,m}^++
{}_{n\pm1}A^{q+1,m+3}~{}^n\phi^+_{q,m}-
{}^n\phi_{q,m}^+~{}_nA^{q,m}\big)\wedge\,{}^n\gammab_{q,m}^{(\pm)+}
\nonumber\\ && +\, 
\big({}^n\phi_{q,m}^+-{}^n\phi_{q+2,m}^-\,{}^n\phi^0_{q,m}\big)\,
{}^n\gammab_{q+2,m}^{(\pm)-}\wedge\,{}^n\gammab^0_{q,m} 
\nonumber\\ && +\,
\big({}^n\phi_{q,m}^+-{}^{n\pm1}\phi_{q-1,m+3}^0\,{}^n\phi_{q,m}^-
\big)\,{}^{n\pm1}\gammab^0_{q-1,m+3}\wedge\,{}^n\gammab_{q,m}^- \ ,
\label{Q3cfpCklrep}\\[4pt] 
{}_{n\pm1}\cf^{q-1,m+3\,;\,q,m}&=&\big(\diff\,{}^n\phi_{q,m}^-+
{}_{n\pm1}A^{q-1,m+3}~{}^n\phi^-_{q,m}-
{}^n\phi_{q,m}^-~{}_nA^{q,m}\big)\wedge\,{}^n\gammab_{q,m}^{(\pm)-}
\nonumber\\ && -\, 
\big({}^n\phi_{q,m}^--{}^n\phi_{q-2,m}^+\,{{}^n
\phi^0_{q-2,m}}^\dag\,\big)\,{}^n\gammab_{q-2,m}^{(\pm)+}
\wedge\,{}^n\gamma^0_{q-2,m} \nonumber\\ && -\,
\big({}^n\phi_{q,m}^--{{}^{n\pm1}\phi_{q-1,m+3}^0}^\dag\,{}^n
\phi_{q,m}^+\big)\,
{}^{n\pm1}\gamma^0_{q-1,m+3}\wedge\,{}^n\gammab_{q,m}^+ \ ,
\label{Q3cfmCklrep}\\[4pt] 
{}_n\cf^{q+2,m\,;\,q,m}&=&\big(\diff\,{}^n\phi_{q,m}^0+{}_n
A^{q+2,m}~{}^n\phi^0_{q,m}-
{}^n\phi_{q,m}^0~{}_nA^{q,m}\big)\wedge\,{}^n\gammab_{q,m}^0
\nonumber\\ &&
-\, \big({}^n\phi_{q,m}^0-{{}^n\phi_{q+2,m}^-}^\dag\,{}^n
\phi_{q,m}^+\big)\,
{}^n\gamma^-_{q+2,m}\wedge\,{}^n\gammab_{q,m}^+ \label{Q3cf0Cklrep}
\\ && -\, \big({}^n\phi_{q,m}^0-{}^{n-1}\phi_{q+1,m-3}^+\,{{}^{n-1}
\phi^-_{q+1,m-3}}^\dag\,\big)\,
{}^{n-1}\gammab_{q+1,m-3}^{(+)+}\wedge\,{}^{n-1}
\gamma^{(+)-}_{q+1,m-3} \nonumber\\ && \nonumber -\, 
\big({}^n\phi_{q,m}^0-{}^{n+1}\phi_{q+1,m-3}^+\,{{}^{n+1}
\phi^-_{q+1,m-3}}^\dag\,\big)\,
{}^{n+1}\gammab_{q+1,m-3}^{(-)+}\wedge\,{}^{n+1}
\gamma^{(-)-}_{q+1,m-3}
\eea
and
\bea
{}_{n\pm1}\cf^{q+1,m+3\,;\,q+1,m-3}&=&\big({}^n\phi^+_{q,m}\,
{}^{n\mp1}\phi^-_{q+1,m-3}-
{}^n\phi^-_{q+2,m}\,{}^{n\mp1}\phi^+_{q+1,m-3}\big)\,{}^n 
\gammab_{q,m}^+\wedge\,
{}^{n\mp1}\gammab_{q+1,m-3}^{(\pm)-} \ , \nonumber\\
\label{Q3cfholrel1}\\[4pt]
{}_{n\pm1}\cf^{q+1,m+3\,;\,q-2,m}&=&\big({}^n\phi^+_{q,m}\,{}^n
\phi^0_{q-2,m}-{}^{n\pm1}
\phi^0_{q-1,m+3}\,{}^n\phi^+_{q-2,m}\big)\,{}^n
\gammab_{q,m}^{(\pm)+}\wedge\,
{}^{n}\gammab_{q-2,m}^0 \ , \label{Q3cfholrel2}\\[4pt]
{}_{n\pm1}\cf^{q-1,m+3\,;\,q-2,m}&=&\big({}^n\phi^-_{q,m}\,{}^n
\phi^0_{q-2,m}-{}^{n\pm1}
\phi^0_{q-3,m+3}\,{}^n\phi^-_{q-2,m}\big)\,{}^n
\gammab_{q,m}^{(\pm)-}\wedge\,
{}^{n}\gammab_{q-2,m}^0 \ , \label{Q3cfholrel3}\\[4pt]
{}_{n\pm1}\cf^{q-1,m+3\,;\,q+2,m}&=&\big({}^n\phi^-_{q,m}\,{{}^n
\phi^0_{q,m}}^\dag-
{{}^{n\pm1}\phi^0_{q-1,m+3}}^\dag\,{}^n\phi^-_{q+2,m}\big)\,{}^n
\gammab_{q,m}^{(\pm)-}\wedge\,
{}^{n}\gamma_{q,m}^0 \ , \label{Q3cfnonholrel1}\\[4pt]
{}_{n\pm1}\cf^{q+1,m+3\,;\,q-1,m+3}&=&\big({}^{n\mp1}
\phi^+_{q,m}\,{{}^{n\mp1}\phi^-_{q,m}}^\dag-
{{}^n\phi^-_{q+1,m+3}}^\dag\,{}^n\phi^+_{q-1,m+3}\big)\,{}^{n\mp1}
\gammab_{q,m}^{(\pm)+}\wedge\,
{}^{n\mp1}\gamma_{q,m}^{(\pm)-} \ , \nonumber\\
\label{Q3cfnonholrel2}\\[4pt]
{}_{n\pm1}\cf^{q+1,m+3\,;\,q+2,m}&=&\big({}^n\phi^+_{q,m}\,{{}^n
\phi^0_{q,m}}^\dag-
{{}^{n\pm1}\phi^0_{q+1,m+3}}^\dag\,{}^n\phi^+_{q+2,m}\big)\,{}^n
\gammab_{q,m}^{(\pm)+}\wedge\,{}^{n}\gamma_{q,m}^0
\label{Q3cfnonholrel3}\eea
along with their hermitean conjugates. Note that the matrix elements
(\ref{Q3cfpCklrep})--(\ref{Q3cf0Cklrep}) additionally contain linear
Higgs field relations appropriate to the non-symmetric quiver
diagram.

\subsection{Examples}

In order to illustrate how to unravel the construction of
Section~\ref{quivbungen} above, we now turn to some more explicit
constructions of quiver bundles and the associated equivariant gauge
connections. 

\bigskip

\noindent
{\bf Symmetric $\mbf{\underline{C}^{2,0}}$ quiver bundles. \ } The
generators of ${\rm sl}(3,\C)$ in the six-dimensional representation
$\underline{C}^{2,0}$ are given by
\beq
E_{\alpha_1}\=\sqrt2\,\big(e_{12}+e_{23}\big)+e_{45} \ , \qquad
E_{\alpha_2}\=e_{24}+\sqrt2\,\big(e_{35}+e_{56}\big) \ , \qquad
E_{\alpha_1+\alpha_2}\=\sqrt2\,\big(e_{14}+e_{46}\big)+e_{25}
\label{EC20rep}\eeq
and
\beq
H_{\alpha_1}\=2(e_1-e_3)+e_4-e_5 \qquad \mbox{and} \qquad
H_{\alpha_2}\=2(e_1+e_2+e_3)-e_4-e_5-4e_6
\label{HC20rep}\eeq
along with $E_{-\alpha_i}=E_{\alpha_i}^\top$ and
$E_{-\alpha_1-\alpha_2}=E_{\alpha_1+\alpha_2}^\top$. They satisfy the
same commutation relations as in
Section~\ref{fundquiverreps}. Substituting these $6\times6$ matrices
into (\ref{A0CP2gens}) we find the corresponding flat connection
\beq
A_0=\begin{pmatrix}B_{(2)}+2a~\Idd_3&\betab^+_{1,-1}&0\\[4pt]
-\betab^+_{1,-1}{}^\dag&B_{(1)}-a~\Idd_2&\betab^+_{0,-4}\\[4pt]
0&-\betab^+_{0,-4}{}^\dag&-4a\end{pmatrix} \ ,
\label{A0CP2C20rep}\eeq
where
\beq
B_{(2)}=\begin{pmatrix} 2B^{11}&\sqrt2\,B^{12}&0\\[4pt]
-\sqrt2~\overline{B^{12}}&0&\sqrt2\,B^{12}\\[4pt]
0&-\sqrt2~\overline{B^{12}}&-2B^{11}\end{pmatrix}
\label{instadrep}\eeq
is the one-instanton field on $\C P^2$ in the adjoint representation
of $\suL$ while
\beq
\betab^+_{1,-1}\=\begin{pmatrix}\sqrt2~\betab^1&0\\[4pt]
\betab^2&\betab^1\\[4pt]0&\sqrt2~\betab^1\end{pmatrix}
\qquad \mbox{and} \qquad \betab^+_{0,-4}\=\sqrt2\,
\begin{pmatrix}\betab^1\\[4pt]\betab^2\end{pmatrix} \ .
\label{betaC20rep}\eeq

The corresponding $G$-equivariant connection (\ref{calACP2Cklrep})
is given by
\beq
\ca=\begin{pmatrix} \scriptstyle A^{2,2}\otimes\Idd_3+
\Idd_{p_{ 2,2}}
\otimes(B_{(2)}+2a~\Idd_3)&\scriptstyle\phi^+_{1,-1}\otimes
\betab^+_{1,-1}&\scriptstyle0\\[4pt]\scriptstyle
-{\phi^+_{1,-1}}^\dag\otimes\betab^+_{1,-1}{}^\dag&
\scriptstyle A^{1,-1}\otimes\Idd_2+
\Idd_{p_{ 1,-1}}\otimes(B_{(1)}-a~\Idd_2)&\scriptstyle
\phi^+_{0,-4}\otimes
\betab^+_{0,-4}\\[4pt]\scriptstyle0&\scriptstyle
-{\phi^+_{0,-4}}^\dag\otimes\betab^+_{0,-4}{}^\dag&\scriptstyle
A^{0,-4}\otimes1+\Idd_{p_{ 0,-4}}\otimes(-4a)\end{pmatrix}
\ .
\label{calACP2C20rep}\eeq
After using the flatness condition on (\ref{A0CP2C20rep}) we find
the non-vanishing curvature components
\bea
\cf^{11}&=&F^{2,2}\otimes\Idd_3+\big(\Idd_{p_{2,2}}-\phi_{1,-1}^+\,
{\phi_{1,-1}^+}^\dag\big)\otimes\big(\betab_{1,-1}^+\wedge
\betab^+_{1,-1}{}^\dag\,\big) \ , \label{cf11C20sym}\\[4pt]
\cf^{22}&=&F^{1,-1}\otimes\Idd_2+
\big(\Idd_{p_{1,-1}}-{\phi_{1,-1}^+}^\dag\,
\phi_{1,-1}^+\big)\otimes\big(\betab^+_{1,-1}{}^\dag
\wedge\betab_{1,-1}^+\big)\nonumber\\
&& \qquad\quad +\,\big(
\Idd_{p_{1,-1}}-\phi^+_{0,-4}\,{\phi^+_{0,-4}}^\dag\big)\otimes
\big(\betab^+_{0,-4}\wedge\betab^+_{0,-4}{}^\dag\,\big) \ , 
\label{cf22C20sym}\\[4pt]
\cf^{33}&=&F^{0,-4}\otimes1+
\big(\Idd_{p_{0,-4}}-{\phi_{0,-4}^+}^\dag\,
\phi_{0,-4}^+\big)\otimes\big(\betab_{0,-4}^+{}^\dag
\wedge\betab_{0,-4}^+\big) \ , \label{cf33C20sym}\\[4pt]
\cf^{12}&=&\big(\diff\phi_{1,-1}^++A^{2,2}\,\phi_{1,-1}^+-
\phi^+_{1,-1}\,A^{1,-1}\big)\wedge\betab_{1,-1}^+ \ ,
\label{cf12C20sym}\\[4pt]
\cf^{23}&=&\big(\diff\phi_{0,-4}^++A^{1,-1}\,\phi_{0,-4}^+-
\phi_{0,-4}^+\,A^{0,-4}\big)\wedge\betab_{0,-4}^+
\label{cf23C20sym}\eea
along with $\cf^{ba}=-(\cf^{ab})^\dag$ for $a<b$, where here the
matrix elements refer to the $3\times3$ block decomposition in
(\ref{calACP2C20rep}) acting on
$\underline{V}^{2,0}\cong
(\C^{p_{2,2}}\otimes\C^3)\oplus(\C^{p_{1,-1}}\otimes\C^2)
\oplus(\C^{p_{0,-4}}\otimes\C)$, the typical fibre spaces of the
vector bundle over $M_D\times\CP^2$ induced by the quiver bundle
\beq
\xymatrix{ & & E_{p_{2,2}} \
. \\ & E_{p_{1,-1}}\ar[ur]^{\phi^+_{1,-1}}& \\
E_{p_{0,-4}}~\ar[ur]^{\phi^+_{0,-4}} & & }
\label{C20CP2quivbun}\eeq
From (\ref{betaC20rep}) one has the explicit matrix one-form products
\bea
\betab_{1,-1}^+\wedge\betab_{1,-1}^+{}^\dag&=&\begin{pmatrix}
2\,\betab^1\wedge\beta^1&\sqrt2~\betab^1\wedge\beta^2&0\\[4pt]
\sqrt2~\betab^2\wedge\beta^1&\betab^1\wedge\beta^1+\betab^2\wedge
\beta^2&\sqrt2~\betab^1\wedge\beta^2\\[4pt] 0&\sqrt2~
\betab^2\wedge\beta^1&2\,\betab^2\wedge\beta^2 \end{pmatrix} \ ,
\label{beta11prod1CP2C20}\\[4pt]\betab_{1,-1}^+{}^\dag
\wedge\betab_{1,-1}^+&=&-
\begin{pmatrix}2\,\betab^1\wedge\beta^1+\betab^2\wedge
\beta^2&\betab^1\wedge\beta^2\\[4pt] \betab^2\wedge\beta^1&
\betab^1\wedge\beta^1+2\,\betab^2\wedge\beta^2 \end{pmatrix} \ ,
\label{beta11prod2CP2C20}\\[4pt]
\betab^+_{0,-4}\wedge\betab_{0,-4}^+{}^\dag&=&2\,
\begin{pmatrix}\betab^1\wedge\beta^1&\betab^1\wedge\beta^2\\[4pt]
\betab^2\wedge\beta^1&\betab^2\wedge\beta^2 \end{pmatrix} \ ,
\label{beta04prod1CP2C20}\\[4pt]
\betab^+_{0,-4}{}^\dag\wedge\betab_{0,-4}^+&=&-2\,
\big(\betab^1\wedge\beta^1+\betab^2\wedge\beta^2\big) \ .
\label{beta04prod2CP2C20}\eea

\bigskip

\noindent
{\bf Symmetric $\mbf{\underline{C}^{1,1}}$ quiver bundles. \ } The
$8\times8$ generators of ${\rm sl}(3,\C)$ in the adjoint
representation $\underline{C}^{1,1}$ are given by
\bea
E_{\alpha_1}&=&e_{12}+\sqrt2~(e_{45}+e_{56})+e_{78} \ ,
\nonumber\\[4pt]
E_{\alpha_2}&=&e_{14}+\sqrt{\mbox{$\frac32$}}~(e_{23}+e_{37})+
\sqrt{\mbox{$\frac12$}}~(e_{25}+e_{57})+e_{68} \ , \nonumber\\[4pt]
E_{\alpha_1+\alpha_2}&=&\sqrt{\mbox{$\frac32$}}~(e_{13}-e_{38})-
\sqrt{\mbox{$\frac12$}}~(e_{15}-e_{58})+(e_{47}-e_{26})
\label{Eadrep}\eea
along with
\beq
H_{\alpha_1}\=(e_1-e_2)+2(e_4-e_6)+(e_7-e_8) \qquad \mbox{and} \qquad
H_{\alpha_2}\=3(e_1+e_2-e_7-e_8) \ .
\label{Hadrep}\eeq
The flat connection (\ref{A0CP2gens}) in this basis is given by
\beq
A_0=\begin{pmatrix}B_{(1)}+3a~\Idd_2&\betab^+_{0,0}&\betab^-_{2,0}&
0\\[4pt]-\betab^+_{0,0}{}^\dag&0&0&\betab^-_{1,-3}\\[4pt]-
\betab_{2,0}^-{}^\dag&
0&B_{(2)}&\betab^+_{1,-3}\\[4pt]0&-\betab^-_{1,-3}{}^\dag&-
\betab^+_{1,-3}{}^\dag& B_{(1)}-3a~\Idd_2\end{pmatrix} \ .
\label{A0CP2adrep}\eeq
It acts on the typical fibre space
$\underline{C}^{1,1}\big|_{\su\times\uo}\cong\C^8=
\C^2\oplus\C\oplus\C^3\oplus\C^2$ with
\bea
\betab^+_{0,0}\=\sqrt{\frac32}~\begin{pmatrix}\betab^1\\[4pt]
\betab^2\end{pmatrix} \qquad &\mbox{and}& \qquad
\betab^-_{2,0}\=\begin{pmatrix}\betab^2&-\sqrt{\mbox{$\frac12$}}~
\betab^1&0\\[4pt]0&\sqrt{\mbox{$\frac12$}}~\betab^2&-\betab^1
\end{pmatrix} \ , \nonumber\\[4pt]
\betab^-_{1,-3}\=\sqrt{\mbox{$\frac32$}}~\big(\betab^2\,,\,-
\betab^1\big) \qquad &\mbox{and}& \qquad
\betab^+_{1,-3}\=\begin{pmatrix}\betab^1&0\\[4pt]
\sqrt{\mbox{$\frac12$}}~\betab^2&\sqrt{\mbox{$\frac12$}}~\betab^1
\\[4pt] 0&\betab^2\end{pmatrix} \ .
\label{betabCP2adrep}\eea

The corresponding quiver bundle over $M_D$ is given by
\beq
\xymatrix{ & E_{p_{1,3}} & \\
E_{p_{0,0}}\ar[ur]^{\phi^+_{0,0}}& &
E_{p_{2,0}} \ar[ul]_{\phi_{2,0}^-} \\
& E_{p_{1,-3}}\ar[ur]_{\phi_{1,-3}^+} \ar[ul]^{\phi_{1,-3}^-} & }
\label{quivbunCP2adrep}\eeq
which induces a vector bundle over $M_D\times\CP^2$ with typical fibre
space
\beq
\underline{V}^{1,1}\cong\big(\C^{p_{1,3}}\otimes\C^2\big)~\oplus~
\big(\C^{p_{0,0}}\otimes\C\big)~
\oplus~\big(\C^{p_{2,0}}\otimes\C^3\big)~\oplus~
\big(\C^{p_{1,-3}}\otimes\C^2\big) \ .
\label{C11typfibre}\eeq
The corresponding $G$-equivariant connection is given by
\beq
\ca=\begin{pmatrix}\scriptstyle A^{1,3}\otimes\Idd_2+
\Idd_{p_{ 1,3}}\otimes(B_{(1)}+3a~\Idd_2)&\scriptstyle
\phi^+_{0,0}\otimes\betab^+_{0,0}&\scriptstyle \phi^-_{2,0}
\otimes\betab^-_{2,0}&\scriptstyle 
0\\[4pt]\scriptstyle -{\phi^+_{0,0}}^\dag\otimes\betab^+_{0,0}{}^\dag&
\scriptstyle A^{0,0}\otimes 1&\scriptstyle 0&\scriptstyle
\phi^-_{1,-3}\otimes\betab^-_{1,-3}\\[4pt]
\scriptstyle -{\phi^-_{2,0}}^\dag\otimes\betab_{2,0}^-{}^\dag
&\scriptstyle 
\scriptstyle 0&\scriptstyle A^{2,0}\otimes\Idd_3+
\Idd_{p_{ 2,0}}\otimes B_{(2)}&\scriptstyle \phi^+_{1,-3}
\otimes\betab^+_{1,-3}\\[4pt]\scriptstyle 0&\scriptstyle
-{\phi^-_{1,-3}}^\dag\otimes\betab^-_{1,-3}{}^\dag&\scriptstyle
-{\phi^+_{1,-3}}^\dag\otimes\betab^+_{1,-3}{}^\dag&\scriptstyle 
A^{1,-3}\otimes\Idd_2+\Idd_{p_{ 1,-3}}\otimes
(B_{(1)}-3a~\Idd_2)\end{pmatrix} \ .
\label{caCP2adrep}\eeq
By using the flatness condition on (\ref{A0CP2adrep}) and
(\ref{betabCP2adrep}), the matrix elements of the curvature two-form
(\ref{cfca}) with respect to the $4\times4$ block decomposition in
(\ref{caCP2adrep}) are found to be
\bea
\cf^{11}&=&F^{1,3}\otimes\Idd_2+\big(\Idd_{p_{1,3}}-\phi^+_{0,0}\,
{\phi^+_{0,0}}^\dag\big)\otimes\big(
\betab_{0,0}^+\wedge\betab_{0,0}^+{}^\dag\,\big)
\nonumber\\ && \qquad\quad +\,\big(\Idd_{p_{1,3}}-\phi^-_{2,0}\,
{\phi^-_{2,0}}^\dag\big)\otimes\big(
\betab_{2,0}^-\wedge\betab_{2,0}^-{}^\dag\,\big) \ ,
\label{cf11CP2adrep}\\[4pt]
\cf^{22}&=&F^{0,0}\otimes1+\big({\phi^+_{0,0}}^\dag\,
\phi^+_{0,0}-\phi^-_{1,-3}\,{\phi^-_{1,-3}}^\dag\big)\otimes\big(
\betab^-_{1,-3}\wedge\betab^-_{1,-3}{}^\dag\,\big) \ , 
\label{cf22CP2adrep}\\[4pt]
\cf^{33}&=&F^{2,0}\otimes\Idd_3+
\big(\Idd_{p_{2,0}}-{\phi^-_{2,0}}^\dag\,
\phi^-_{2,0}\big)\otimes\big(
\betab_{2,0}^-{}^\dag\wedge\betab_{2,0}^-\big)
\nonumber\\ && \qquad\quad+\,\big(\Idd_{p_{2,0}}-\phi^+_{1,-3}\,
{\phi^+_{1,-3}}^\dag\big)\otimes\big(
\betab_{1,-3}^+\wedge\betab_{1,-3}^+{}^\dag \,\big)
\ , \label{cf33CP2adrep}\\[4pt]
\cf^{44}&=&F^{1,-3}\otimes\Idd_2+\big(\Idd_{p_{1,-3}}-
{\phi^-_{1,-3}}^\dag\,\phi^-_{1,-3}\big)\otimes\big(
\betab^-_{1,-3}{}^\dag\wedge\betab^-_{1,-3}\big)
\nonumber\\ && \qquad\quad +\,\big(\Idd_{p_{1,-3}}-
{\phi^+_{1,-3}}^\dag\,\phi^+_{1,-3}\big)\otimes\big(
\betab_{1,-3}^+{}^\dag\wedge
\betab_{1,-3}^+\big) \ , \label{cf44CP2adrep}\\[4pt]
\cf^{12}&=&\big(\diff\phi_{0,0}^++
A^{1,3}\,\phi_{0,0}^+-\phi_{0,0}^+\,A^{0,0}\big)\wedge
\betab_{0,0}^+ \ , \label{cf12CP2adrep}\\[4pt]
\cf^{13}&=&\big(\diff\phi_{2,0}^-+
A^{1,3}\,\phi_{2,0}^--\phi_{2,0}^-\,A^{2,0}\big)\wedge
\betab_{2,0}^- \ , \label{cf13CP2adrep}\\[4pt]
\cf^{14}&=&\big(\phi_{0,0}^+\,\phi_{1,-3}^--
\phi_{2,0}^-\,\phi_{1,-3}^+\big)\otimes\big(\betab_{0,0}^+\wedge
\betab_{1,-3}^-\big) \ , \label{cf14CP2adrep}\\[4pt]
\cf^{23}&=&\big({\phi_{0,0}^+}^\dag\,\phi_{2,0}^--\phi_{1,-3}^-\,
{\phi_{1,-3}^+}^\dag\big)\otimes\big(
\betab_{1,-3}^-\wedge\betab_{1,-3}^+{}^\dag \,\big)
\ , \label{cf23CP2adrep}\\[4pt]
\cf^{24}&=&\big(\diff\phi_{1,-3}^-+A^{0,0}\,\phi_{1,-3}^--
\phi_{1,-3}^-\,A^{1,-3}\big)\wedge\betab_{1,-3}^- \ ,
\label{cf24CP2adrep}\\[4pt]
\cf^{34}&=&\big(\diff\phi_{1,-3}^++A^{2,0}\,\phi_{1,-3}^+-
\phi_{1,-3}^+\,A^{1,-3}\big)\wedge\betab_{1,-3}^+
\label{cf34CP2adrep}\eea
plus their hermitean conjugates $\cf^{ba}=-(\cf^{ab})^\dag$ for
$a<b$. The holomorphic relation equation (\ref{adCP2rel}) is contained
in (\ref{cf14CP2adrep}). The explicit matrix one-form products can be
computed from (\ref{betabCP2adrep}) to get
\bea
\betab_{0,0}^+\wedge\betab_{0,0}^+{}^\dag&=&\frac32\,\begin{pmatrix}
\betab^1\wedge\beta^1&\betab^1\wedge\beta^2\\\betab^2\wedge\beta^1&
\betab^2\wedge\beta^2\end{pmatrix} \ , \label{adbetap00}\\[4pt]
\betab^-_{2,0}\wedge\betab^-_{2,0}{}^\dag&=&\begin{pmatrix}
\frac12\,\betab^1\wedge\beta^1+\betab^2\wedge\beta^2&
-\frac12\,\betab^1\wedge\beta^2\\-\frac12\,\betab^2\wedge\beta^1&
\betab^1\wedge\beta^1+\frac12\,\betab^2\wedge\beta^2\end{pmatrix} \ ,
\label{adbetam20}\\[4pt] 
\betab^-_{2,0}{}^\dag\wedge\betab^-_{2,0}&=&\begin{pmatrix}
\beta^2\wedge\betab^2&-\sqrt{\frac12}~\beta^2\wedge\betab^1&0\\
-\sqrt{\frac12}~\beta^1\wedge\betab^2&\frac12\,\big(\beta^1\wedge
\betab^1+\beta^2\wedge
\betab^2\big)&-\sqrt{\frac12}~\beta^2\wedge\betab^1\\
0&-\sqrt{\frac12}~\beta^1\wedge\betab^2&\beta^1\wedge\betab^1
\end{pmatrix} \ , \label{adbetam20dag}\\[4pt]
\betab^-_{1,-3}\wedge\betab^-_{1,-3}{}^\dag&=&\mbox{$\frac32$}\,\big(
\betab^1\wedge\beta^1+\betab^2\wedge\beta^2\big) \ , 
\label{adbetam13}\\[4pt]
\betab^-_{1,-3}{}^\dag\wedge\betab^-_{1,-3}&=&\frac32\,\begin{pmatrix}
\beta^2\wedge\betab^2&-\beta^2\wedge\betab^1\\-\beta^1\wedge\betab^2&
\beta^1\wedge\betab^1\end{pmatrix} \ , \label{adbetam13dag}\\[4pt]
\betab^+_{1,-3}\wedge\betab^+_{1,-3}{}^\dag&=&\begin{pmatrix}
\betab^1\wedge\beta^1&\sqrt{\frac12}~\betab^1\wedge\beta^2&0\\
\sqrt{\frac12}~\betab^2\wedge\beta^1&\frac12\,\big(
\betab^1\wedge\beta^1+\betab^2\wedge\beta^2\big)&\sqrt{\frac12}~
\betab^1\wedge\beta^2\\0&\sqrt{\frac12}~\betab^2\wedge\beta^1&
\betab^2\wedge\beta^2\end{pmatrix} \ , \label{adbetap13}\\[4pt]
\betab^+_{1,-3}{}^\dag\wedge\betab^+_{1,-3}&=&\begin{pmatrix}
\beta^1\wedge\betab^1+\frac12\,\beta^2\wedge\betab^2&\frac12\,
\beta^2\wedge\betab^1\\\frac12\,\beta^1\wedge\betab^2&
\frac12\,\beta^1\wedge\betab^1+\beta^2\wedge\betab^2\end{pmatrix} \ ,
\label{adbetap13dag}\\[4pt] \betab_{0,0}^+\wedge\betab^-_{1,-3}&=&
\mbox{$\frac32$}\,\betab^1\wedge\betab^2~\Idd_2 \ , 
\label{adbeta0013}\\[4pt]
\betab_{1,-3}^-\wedge\betab_{1,-3}^+{}^\dag&=&\sqrt{\mbox{$\frac32$}}~
\Big(\betab^2\wedge\beta^1 \ , \ -\sqrt{\mbox{$\frac12$}}~\big(
\betab^1\wedge\beta^1+\betab^2\wedge\beta^2\big) \ , \ 
-\betab^1\wedge\beta^2\Big) \ .
\label{adbetapm13}\eea

\bigskip

\noindent
{\bf Non-symmetric $\mbf{\underline{C}^{2,0}}$ quiver bundles. \ } For
the space $Q_3$ and the $\underline{C}^{2,0}$ representation of
$\sut$, the flat connection (\ref{A0Q3gens}) is obtained by
substituting in (\ref{EC20rep}) and (\ref{HC20rep}). The quiver bundle
over $M_D$ in this case is given by
\beq
\xymatrix{E_{{}^2p_{-2,2}}&\xrightarrow{\phantom{xxxx}{}^2
\phi^0_{-2,2}
\phantom{xxxx}}&E_{{}^2p_{0,2}}&\xrightarrow{\phantom{xxxx}
{}^2\phi^0_{0,2}\phantom{xxxx}}&E_{{}^2p_{2,2}}\\
 & E_{{}^1p_{-1,-1}}\ar[ul]^{{}^1\phi^-_{-1,-1}}
\ar[ur]^{{}^1\phi^+_{-1,-1}}&\xrightarrow{\phantom{xxxx}{}^1
\phi^0_{-1,-1}\phantom{xxxx}}&
E_{{}^1p_{1,-1}}\ar[ul]_{{}^1\phi^-_{1,-1}}\ar[ur]_{{}^1
\phi^+_{1,-1}}& \\ & &E_{{}^0p_{0,-4}}\ar[ul]^{{}^0\phi^-_{0,-4}}
\ar[ur]_{{}^0\phi^+_{0,-4}}& & }
\label{Q3C20quivbun}\eeq
and the corresponding $G$-equivariant connection one-form on
$M_D\times Q_3$ is given by
\beq
\ca=\begin{pmatrix}\scriptstyle {}_2A^{2,2}+
2a_1&\scriptstyle \sqrt2~{}^2\phi^0_{0,2}\,\gammab^3&\scriptstyle 0&
\scriptstyle\sqrt2~{}^1\phi^+_{1,-1}\,\gammab^1&\scriptstyle 0&
\scriptstyle 0\\[4pt] \scriptstyle -\sqrt2~{}^2{\phi^0_{0,2}}^\dag\,
\gamma^3&\scriptstyle {}_2A^{0,2}-a_2&
\scriptstyle \sqrt2~{}^2\phi^0_{-2,2}\,\gammab^3&\scriptstyle
{}^1\phi^-_{1,-1}\,\gammab^2&\scriptstyle{}^1\phi^+_{-1,-1}\,
\gammab^1&\scriptstyle 0\\[4pt] \scriptstyle 0&\scriptstyle-
\sqrt2~{{}^2\phi^0_{-2,2}}^\dag\,\gamma^3&\scriptstyle {}_2A^{-2,2}
-2(a_1+a_2)&\scriptstyle 0 &\scriptstyle
\sqrt2~{}^1\phi^-_{-1,-1}\,\gammab^2&\scriptstyle 0\\[4pt]
\scriptstyle -\sqrt2~{{}^1\phi^+_{1,-1}}^\dag\,\gamma^1&
\scriptstyle -{{}^1\phi^-_{1,-1}}^\dag\,\gamma^2&\scriptstyle 0&
\scriptstyle {}_1A^{1,-1}+(a_1+a_2)&
\scriptstyle {}^1\phi^0_{-1,-1}\,\gammab^3&\scriptstyle
\sqrt2~{}^0\phi_{0,-4}^+\,\gammab^1\\[4pt]\scriptstyle 0 &
\scriptstyle -{{}^1\phi^+_{-1,-1}}^\dag\,\gamma^1&\scriptstyle -
\sqrt2~{{}^1\phi^-_{-1,-1}}^\dag\,\gamma^2&\scriptstyle -
{{}^1\phi^0_{-1,-1}}^\dag\,\gamma^3 &\scriptstyle {}_1A^{-1,-1}-
a_1&\scriptstyle \sqrt2~{}^0\phi^-_{0,-4}\,\gammab^2\\[4pt]
\scriptstyle 0&\scriptstyle 0&\scriptstyle 0&\scriptstyle -
\sqrt2~{{}^0\phi_{0,-4}^+}^\dag\,\gamma^1&\scriptstyle -
\sqrt2~{{}^0\phi^-_{0,-4}}^\dag\,\gamma^2&\scriptstyle
{}_0A^{0,-4}+2a_2\end{pmatrix} \ .
\label{caQ3C20rep}\eeq
Its curvature (\ref{cfca}) with respect to this $6\times6$ block
decomposition has the non-vanishing matrix elements
\bea
\cf^{11}&=&{}_2F^{2,2}+2\big(\Idd_{{}^2p_{2,2}}-{}^1\phi^+_{1,-1}\,
{{}^1\phi^+_{1,-1}}^\dag\,\big)\,\gammab^1\wedge\gamma^1+2\big(
\Idd_{{}^2p_{2,2}}-{}^2\phi^0_{0,2}\,{{}^2\phi^0_{0,2}}^\dag\,\big)\,
\gammab^3\wedge\gamma^3 \ , \label{cf11Q3C20rep}\\[4pt]
\cf^{12}&=&\sqrt2\,\big(\diff\,{}^2\phi^0_{0,2}+{}_2A^{2,2}~{}^2\phi^0_{0,2}-
{}^2\phi^0_{0,2}~{}_2A^{0,2}\big)\wedge\gammab^3\nonumber\\ && \qquad\quad +\,
\sqrt2\,\big(\,{}^2\phi^0_{0,2}-{}^1\phi^+_{1,-1}\,{{}^1
\phi^-_{1,-1}}^\dag\,\big)\,
\gammab^1\wedge\gamma^2 \ , \label{cf12Q3C20rep}\\[4pt]
\cf^{14}&=&\sqrt2\,\big(\diff\,{}^1\phi^+_{1,-1}+{}_2A^{2,2}~{}^1\phi^+_{1,-1}-
{}^1\phi^+_{1,-1}~{}_1A^{1,-1}\big)\wedge\gammab^1\nonumber\\ && \qquad\quad+\,
\sqrt2\,\big(\,{}^1\phi^+_{1,-1}-{}^2\phi^0_{0,2}\,{}^1\phi^-_{1,-1}\big)\,
\gammab^2\wedge\gammab^3 \ , \label{cf14Q3C20rep}\\[4pt]
\cf^{15}&=&\sqrt2\,\big(\,{}^2\phi^0_{0,2}\,{}^1\phi^+_{-1,-1}-
{}^1\phi^+_{1,-1}\,{}^1\phi^0_{-1,-1}\big)\,\gammab^3\wedge\gammab^1 \ ,
\label{cf15Q3C20rep}\\[4pt] \cf^{22}&=&{}_2F^{0,2}+\big(
\Idd_{{}^2p_{0,2}}-{}^1\phi^+_{-1,-1}\,{{}^1\phi^+_{-1,-1}}^\dag\,\big)\,
\gammab^1\wedge\gamma^1+\big(\Idd_{{}^2p_{0,2}}-{}^1\phi^-_{1,-1}\,
{{}^1\phi^-_{1,-1}}^\dag\,\big)\,\gammab^2\wedge\gamma^2 \nonumber\\
&&\qquad\quad +\,2\,\big(\,{}^2{\phi^0_{0,2}}^\dag\,{}^2\phi^0_{0,2}-
{}^2\phi^0_{-2,2}\,{{}^2\phi^0_{-2,2}}^\dag\,\big)\,\gammab^3\wedge\gamma^3 \ ,
\label{cf22Q3C20rep}\\[4pt] \cf^{23}&=&\sqrt2\,\big(\diff\,{}^2\phi^0_{-2,2}+
{}_2A^{0,2}~{}^2\phi^0_{-2,2}-{}^2\phi^0_{-2,2}~{}_2A^{-2,2}\big)
\wedge\gammab^3
\nonumber\\ && \qquad\quad+\,\sqrt2\,\big(\,{}^2\phi^0_{-2,2}-
{}^1\phi^+_{-1,-1}\,{{}^1\phi_{-1,-1}^-}^\dag\,\big)\,\gammab^1\wedge
\gamma^2 \ , \label{cf23Q3C20rep}\\[4pt] \cf^{24}&=&\big(\diff\,{}^1
\phi^-_{1,-1}+{}_2A^{0,2}~{}^1\phi^-_{1,-1}-{}^1\phi^-_{1,-1}~{}_1A^{1,-1}
\big)\wedge\gammab^2 \nonumber\\ &&\qquad\quad -\,\big(\,{}^1\phi^-_{1,-1}-2
~{{}^2\phi^0_{0,2}}^\dag\,{}^1\phi_{1,-1}^++{}^1\phi^+_{-1,-1}\,{
{}^1\phi^0_{-1,-1}}^\dag\,\big)\gammab^1\wedge\gamma^3 \ ,
\label{cf24Q3C20rep}\\[4pt] \cf^{25}&=&\big(\diff\,{}^1\phi_{-1,-1}^++
{}_2A^{0,2}~{}^1\phi_{-1,-1}^+-{}^1\phi_{-1,-1}^+~{}_1A^{-1,-1}\big)\wedge
\gammab^1 \nonumber\\ &&\qquad\quad+\,\big(\,{}^1\phi^+_{-1,-1}+
{}^1\phi^-_{1,-1}\,{}^1\phi^0_{-1,-1}-2~
{}^2\phi^0_{-2,2}\,{}^1\phi^-_{-1,-1}
\big)\,\gammab^2\wedge\gammab^3 \ , \label{cf25Q3C20rep}
\\[4pt] \cf^{26}&=&\sqrt2\,\big(\,{}^1\phi^+_{-1,-1}\,
{}^0\phi^-_{0,-4}-{}^1\phi^-_{1,-1}\,{}^0\phi^+_{0,-4}\big)\,\gammab^1
\wedge\gammab^2 \ ,
\label{cf26Q3C20rep}\\[4pt] \cf^{33}&=&{}_2F^{-2,2}+2\big(
\Idd_{{}^2p_{-2,2}}-{}^1\phi^-_{-1,-1}\,{{}^1\phi^-_{-1,-1}}^\dag\,\big)\,
\gammab^2\wedge\gamma^2\nonumber\\ && \qquad\quad -\,
2\big(\Idd_{{}^2p_{-2,2}}-{{}^2\phi^0_{-2,2}}^\dag\,
{}^2\phi^0_{-2,2}\big)\,\gammab^3\wedge\gamma^3 \ ,
\label{cf33Q3C20rep}\\[4pt] \cf^{34}&=&
\sqrt2\,\big(\,{}^2{\phi^0_{-2,2}}^\dag\,{}^1\phi^-_{1,-1}-
{}^1\phi^-_{-1,-1}\,{{}^1\phi^0_{-1,-1}}{}^\dag\,\big)\,\gammab^2
\wedge\gamma^3 \ ,
\label{cf34Q3C20rep}\\[4pt] \cf^{35}&=&\sqrt2\,\big(\diff\,{}^1
\phi^-_{-1,-1}+{}_2A^{-2,2}~{}^1\phi^-_{-1,-1}-{}^1\phi^-_{-1,-1}~
{}_1A^{-1,-1}\big)\wedge\gammab^2\nonumber\\ &&\qquad\quad-\,
\sqrt2\,\big(\,{}^1\phi^-_{-1,-1}-{{}^2\phi^-_{-2,2}}^\dag\,
{}^1\phi^+_{-1,-1}\big)\,\gammab^1\wedge\gamma^3 \ ,
\label{cf35Q3C20rep}\\[4pt] \cf^{44}&=&{}_1F^{1,-1}-\big(
\Idd_{{}^1p_{1,-1}}-{{}^1\phi^-_{1,-1}}^\dag\,{}^1\phi^-_{1,-1}\big)\,
\gammab^2\wedge\gamma^2+\big(\Idd_{{}^1p_{1,-1}}-{}^1\phi^0_{-1,-1}\,
{{}^1\phi^0_{-1,-1}}^\dag\big)\,\gammab^3\wedge\gamma^3 \nonumber\\
&&\qquad\quad +\,2\big(\,{}^1{\phi^+_{1,-1}}^\dag\,{}^1\phi^+_{1,-1}-
{}^0\phi^+_{0,-4}\,{{}^0\phi^+_{0,-4}}^\dag\,\big)\,\gammab^1\wedge\gamma^1 \ ,
\label{cf44Q3C20rep}\\[4pt] \cf^{45}&=&\big(\diff\,{}^1\phi^0_{-1,-1}+
{}_1A^{1,-1}~{}^1\phi^0_{-1,-1}-{}^1\phi^0_{-1,-1}~{}_1A^{-1,-1}\big)\wedge
\gammab^3 \nonumber\\ &&\qquad\quad +\,\big(\,{}^1\phi^0_{-1,-1}+
{{}^1\phi^-_{1,-1}}^\dag\,{}^1\phi^+_{-1,-1}-2~{}^0\phi^+_{0,-4}\,
{{}^0\phi^-_{0,-4}}^\dag\,\big)\,\gammab^1\wedge\gamma^2 \ ,
\label{cf45Q3C20rep}\\[4pt] \cf^{46}&=&\sqrt2\,\big(\diff\,{}^0
\phi^+_{0,-4}+{}_1A^{1,-1}~{}^0\phi^+_{0,-4}-{}^0\phi^+_{0,-4}~{}_0A^{0,-4}
\big)\wedge\gammab^1 \nonumber\\ &&\qquad\quad +\,
\sqrt2\,\big(\,{}^0\phi^+_{0,-4}-
{}^1\phi^0_{-1,-1}\,{}^0\phi^-_{0,-4}\big)\,\gammab^2\wedge\gammab^3 \ ,
\label{cf46Q3C20rep}\\[4pt] \cf^{55}&=&{}_1F^{-1,-1}-\big(
\Idd_{{}^1p_{-1,-1}}-{{}^1\phi^+_{-1,-1}}^\dag\,{}^1\phi^+_{-1,-1}\big)\,
\gammab^1\wedge\gamma^1-\big(\Idd_{{}^1p_{-1,-1}}-
{{}^1\phi^0_{-1,-1}}^\dag\,{}^1\phi^0_{-1,-1}\big)\,\gammab^3\wedge\gamma^3
\nonumber\\ &&\qquad\quad +\,2\big(\,{{}^1\phi^-_{-1,-1}}^\dag\,
{}^1\phi^-_{-1,-1}-{}^0\phi^-_{0,-4}\,{{}^0\phi^-_{0,-4}}^\dag\,\big)\,
\gammab^2\wedge\gamma^2 \ , \label{cf55Q3C20rep}\\[4pt] \cf^{56}&=&
\sqrt2\,\big(\diff\,{}^0\phi^-_{0,-4}+{}_1A^{-1,-1}~{}^0\phi^-_{0,-4}-
{}^0\phi^-_{0,-4}~{}_0A^{0,-4}\big)\wedge\gammab^2
\nonumber\\ &&\qquad\quad -\,\sqrt2\,\big(
{}^0\phi^-_{0,-4}-{{}^1\phi^0_{-1,-1}}^\dag\,{}^0\phi^+_{0,-4}\big)\,
\gammab^1\wedge\gamma^3 \ , \label{cf56Q3C20rep}\\[4pt] \cf^{66}&=&
{}_0F^{0,-4}-2\big(\Idd_{{}^0p_{0,-4}}-{{}^0\phi^+_{0,-4}}^\dag\,{}^0
\phi^+_{0,-4}\big)\,\gammab^1\wedge\gamma^1-2\big(
\Idd_{{}^0p_{0,-4}}-{{}^0\phi^-_{0,-4}}^\dag\,{}^0\phi^-_{0,-4}\big)\,
\gammab^2\wedge\gamma^2
\label{cf66Q3C20rep}\eea
plus their hermitean conjugates $\cf^{ba}=-(\cf^{ab})^\dag$ for
$a<b$.

\bigskip

\noindent
{\bf Non-symmetric $\mbf{\underline{C}^{1,1}}$ quiver bundles. \ } Our
final example of this section is the $G$-equivariant connection on
a $G$-bundle over $M_D\times Q_3$ related to the adjoint
representation $\underline{C}^{1,1}$. It has the form
\bea
&\ca=\begin{pmatrix}\scriptstyle 0&\scriptstyle{}^1\phi^0_{-1,3}\,
\gammab^3&\scriptstyle\sqrt{\mbox{$\scriptstyle\frac32$}}~{}^0
\phi^+_{0,0}\,\gammab^1&
\scriptstyle {}^2\phi^-_{2,0}\,\gammab^2&\scriptstyle -
\sqrt{\mbox{$\scriptstyle\frac12$}}~{}^2\phi^+_{0,0}\,\gammab^1&
\scriptstyle 0&
\scriptstyle 0&\scriptstyle 0\\[4pt] &\scriptstyle 0&
\scriptstyle \sqrt{\mbox{$\scriptstyle\frac32$}}~
{}^0\phi_{0,0}^-\,\gammab^2&
\scriptstyle 0&\scriptstyle \sqrt{\mbox{$\scriptstyle\frac12$}}~
{}^2\phi_{0,0}^-\,
\gammab^2&\scriptstyle -{}^2\phi^+_{-2,0}\,\gammab^1&\scriptstyle 0&
\scriptstyle 0\\[4pt] & & \scriptstyle 0& \scriptstyle 0&
\scriptstyle 0&\scriptstyle 0&\scriptstyle
\sqrt{\mbox{$\scriptstyle\frac32$}}~
{}^0\phi_{1,-3}^-\,\gammab^2&\scriptstyle -
\sqrt{\mbox{$\scriptstyle\frac32$}}~
{}^0\phi_{-1,-3}^+\,\gammab^1\\[4pt] & & & \scriptstyle 0&\scriptstyle
\sqrt2~{}^2\phi^0_{0,0}\,\gammab^3&\scriptstyle 0&\scriptstyle
{}^1\phi_{1,-3}^+\,\gammab^1&\scriptstyle 0\\[4pt] & &\scriptstyle & &
\scriptstyle 0&\scriptstyle \sqrt2~{}^2\phi_{-2,0}^0\,
\gammab^3&\scriptstyle \sqrt{\mbox{$\scriptstyle\frac12$}}~
{}^1\phi_{1,-3}^-\,
\gammab^2&\scriptstyle \sqrt{\mbox{$\scriptstyle\frac12$}}~
{}^1\phi_{-1,-3}^+\,\gammab^1\\[4pt]
 & &-\,\mbox{h.c.}  & & &\scriptstyle 0&\scriptstyle 0&\scriptstyle
{}^1\phi_{-1,-3}^-\,
\gammab^2\\[4pt] & & & & & &\scriptstyle 0&\scriptstyle 
{}^1\phi_{-1,-3}^0\,\gammab^3\\[4pt] & & & & & & &\scriptstyle 0
\end{pmatrix} \nonumber\\[4pt] & +\,{\rm diag}(\scriptstyle{
{}_1A^{1,3}+a_1-a_2,{}_1A^{-1,3}-a_1-2a_2,{}_0A^{0,0},
{}_2A^{2,0}+2a_1+a_2,{}_2A^{0,0},{}_2A^{-2,0}-2a_1-a_2,
{}_1A^{1,-3}+a_1+2a_2,{}_1A^{-1,-3}-a_1+a_2}\big) \nonumber \\
\label{caQ3adrep}\eea
where $\mbox{h.c.}$ indicates the hermitean conjugate of the upper
triangular matrix. We omit the rather complicated list of curvature
matrix elements.

\bigskip

\section{Nonabelian quiver vortex equations\label{NCinst}}

\noindent
In this section we will describe explicitly the dimensional reduction
of gauge theory equations on manifolds of the form
(\ref{M2ntimesSU3H}), in the case that $M_D$ is a K\"ahler manifold of
(real) dimension $D=2d$. Using the quiver bundle constructions of the
previous section, we will find that the Yang-Mills equations on $X$
dimensionally reduce to equations on $M_{2d}$ due to the $ G/H$
dependence of fields prescribed by $G$-equivariance. In particular,
we will see that the BPS equations which give the natural analog of
instantons on $X$ reduce to nonabelian coupled vortex equations on
$M_{2d}$.

\subsection{BPS equations\label{BPSeqs}}

Let us pick a complex structure on the K\"ahler manifold $M_{2d}$ and
local complex coordinates $(z^1,\dots,z^d)\in\C^{d}$. In these
coordinates the riemannian metric on $X$ takes the form
\beq
\diff s^2 =2\,g_{a\bb}\ \diff z^a~\diff \zb^{\bb} + 2\,G_{i\bar\jmath}
~\diff y^i~\diff\yb^{\jb} \ ,
\label{metricX}\eeq
where $g_{a\bb}$, $1\leq a,b\leq d$, (resp. $G_{i\jb}$, $1\leq
i,j\leq d_H/2$, $d_H:=\dim( G/H)$) is the K\"ahler metric on $M_{2d}$
(resp. $ G/H$) and $y^i$ denote local complex coordinates on the
homogeneous space $ G/H$. The Hodge duality operator on $X$
constructed from (\ref{metricX}) is denoted $*$. The K\"ahler two-form
$\Omega $ on $X$ is given by
\beq\label{kahlerX}
\Omega=- 2\im g_{a\bb}\ \diff z^a\wedge\diff \zb^{\bb}
- 2\im G_{i\jb}\ \diff y^i\wedge\diff \yb^\jb \ .
\eeq

Let $\Ecal\to X$ be a hermitean vector bundle of rank $p$ with the
structure group ${\rm U}(p)$ and gauge connection $\ca$. The
corresponding curvature (\ref{cfca}) obeys the Bianchi identity
\beq
D_\ca\cf=0 \ ,
\label{Bianchiid}\eeq
where $D_\ca$ is the gauge covariant derivative constructed from
$\ca$. The vacuum Yang-Mills equations are
\beq
D_\ca(*\cf)=0 \ .
\label{YMeqs}\eeq
One can write the curvature two-form via its K\"ahler decomposition as
$\cf=\cf^{2,0}+\cf^{1,1}+\cf^{0,2}$ with
$\cf^{r,s}\in\Omega^{r,s}(X,{\rm u}(p))$. Stable gauge bundles $\Ecal$
are then those which solve the Donaldson-Uhlenbeck-Yau (DUY) 
equations~\cite{DUY1}
\begin{equation}\label{DUY}
*\Omega\wedge {\cf}\ =\ 0 \qquad\textrm{and}\qquad
{\cf}^{2,0}\=0\=\cf^{0,2}\ .
\end{equation}
These equations generalize the usual self-duality equations in four
dimensions, and they imply the Yang-Mills equations given by
(\ref{YMeqs}).

In the local complex coordinates $(z^a,y^i)$ the DUY equations take
the form
\begin{eqnarray}\label{DUY1}
g^{a\bb}\,{\cf}_{z^a\zb^{\bb}}+ G^{i\jb}\,{\cf}_{y^i\yb^\jb}&=&0 \ ,
\\[4pt]\label{DUY2}
{\cf}_{\zb^{\ab}\zb^{\bb}}&=&0\=\cf_{z^az^b} \ , \\[4pt] 
 \label{DUY3} {\cf}_{\zb^{\ab}\yb^\ib}&=&0\=\cf_{z^ay^i} \ , \\[4pt]
\cf_{\yb^\ib\yb^\jb}&=&0\=\cf_{y^iy^j} \label{DUY4}
\end{eqnarray}
for $a,b=1,\dots,d$ and $i,j=1,\dots,d_H/2$. For a given quiver
bundle $\Ecal^{k,l}\to X$, both (\ref{YMeqs}) and (\ref{DUY})
dimensionally reduce to equations on $M_{2d}$ alone for the
anti-hermitean $\urmL(p_v)$ gauge connections
\beq
A^v=A_a^v~\diff z^a+A^v_\ab~\diff\zb^\ab
\label{gaugeconnred}\eeq
with curvatures
\beq
F^v=F^v_{ab}~\diff z^a\wedge\diff z^b+2F_{a\bb}^v~\diff z^a
\wedge\diff\zb^\bb+F^v_{\ab\bb}~\diff\zb^\ab\wedge\diff\zb^\bb
\label{curvred}\eeq
coupled to the Higgs fields $\phi_{v,\Phi(v)}$, for each vertex
$v\in\quiver_0(k,l)$ and arrow $\Phi\in\quiver_1(k,l)$. With the form
of the gauge potentials in the previous section, the equation
(\ref{DUY2}) implies
\beq
F^v_{\ab\bb}\=0\=F^v_{ab}
\label{Fvholgen}\eeq
which expresses holomorphicity of the vector bundle $E_{p_v}\to
M_{2d}$. The remaining equations (\ref{DUY1}), (\ref{DUY3}) and
(\ref{DUY4}) lead to nonabelian quiver vortex equations on $M_{2d}$.

\bigskip

\noindent
{\bf Symmetric $\mbf{\underline{C}^{k,l}}$ quiver vortices. \ } Let us
fix a vertex $v=(n,m)\in\quiver_0(k,l)$ of the symmetric quiver
associated to the $\sut$ representation
$\underline{C}^{k,l}$ and substitute the field strength matrix
elements (\ref{curvdiagCklrep})--(\ref{curvnonholrelsCklrep}) in the
Biedenharn basis into the DUY equations
(\ref{DUY1})--(\ref{DUY4}). The only contribution to (\ref{DUY3})
comes from (\ref{curvoffdiagCklrep}) which yields the bi-covariant
Higgs field derivatives
\beq
\partial_\ab\phi_{n,m}^\pm+A_\ab^{n\pm1,m+3}\,
\phi^\pm_{n,m}-\phi^\pm_{n,m}\,A_\ab^{n,m}=0
\label{bicovHiggsderiv}\eeq
plus their hermitean conjugates, where
$\partial_\ab:=\frac\partial{\partial\zb^\ab}$. This expresses the
fact that the BPS bundle morphisms $\phi_{n,m}^\pm$ are holomorphic
maps. Likewise, only the matrix element (\ref{curvholrelsCklrep})
contributes to (\ref{DUY4}) which implies the holomorphic relations
\beq
\phi_{n,m}^+\,\phi^-_{n+1,m-3}-\phi^-_{n+2,m}\,\phi^+_{n+1,m-3}=0 \ .
\label{holrelCkl}\eeq
Note that non-holomorphic relations (such as those implied by the
vanishing of (\ref{curvnonholrelsCklrep})) do not arise as BPS
conditions.

The remaining instanton equation (\ref{DUY1}) must be satisfied by the
matrix elements (\ref{curvdiagCklrep}). A K\"ahler two-form on $\CP^2$
can be identified locally as in (\ref{KahlerCP2}). Using
(\ref{betanmdef}) the pertinent matrix one-form products are given by
\bea
\betab^\pm_{n,m}{}^\dag\wedge\betab_{n,m}^\pm&=&
\sum_{q\in\{-n+2j\}_{j=0}^n}\,
\lambda_{k,l}^\pm(n,m)^2\nonumber\\ &&
\times\,\bigg\{\bigg(
\biggl[\cgstack{\frac n2}{\frac q2}~
\cgstack{\frac12}{-\frac12}~\cgstack{\frac{n\pm1}2}{\frac{q-1}2}
\biggr]^2~\beta^1\wedge\betab^1
+\biggl[\cgstack{\frac n2}{\frac q2}~
\cgstack{\frac12}{\frac12}~\cgstack{\frac{n\pm1}2}{\frac{q+1}2}
\biggr]^2~\beta^2\wedge\betab^2\bigg)~
\big|\noverq\,,\,m\big\rangle\big\langle\noverq\,,\,
m\big| \nonumber\\ &&  +\,
\biggl[\cgstack{\frac n2}{\frac q2}~
\cgstack{\frac12}{-\frac12}~\cgstack{\frac{n\pm1}2}{\frac{q-1}2}
\biggr]\,\biggl[\cgstack{\frac n2}{\frac{q-2}2}~
\cgstack{\frac12}{\frac12}~\cgstack{\frac{n\pm1}2}{\frac{q-1}2}
\biggr]~\beta^1\wedge\betab^2~\big|\noverq\,,\,m\big\rangle
\big\langle{\stackrel{\scriptstyle n}{\scriptstyle q-2}}\,,\,m
\big| \nonumber\\ &&  +\,
\biggl[\cgstack{\frac n2}{\frac q2}~
\cgstack{\frac12}{\frac12}~\cgstack{\frac{n\pm1}2}{\frac{q+1}2}
\biggr]\,\biggl[\cgstack{\frac n2}{\frac {q+2}2}~
\cgstack{\frac12}{-\frac12}~\cgstack{\frac{n\pm1}2}{\frac{q+1}2}
\biggr]~\beta^2\wedge\betab^1~\big|\noverq\,,\,m\big\rangle
\big\langle{\stackrel{\scriptstyle n}{\scriptstyle q+2}}\,,\,m
\big|\bigg\} \ , \label{betaprodCkl1} \\[4pt]
\betab^\pm_{n\mp1,m-3}\wedge\betab^\pm_{n\mp1,m-3}{}^\dag&=&
\sum_{q\in\{-n+2j\}_{j=0}^n}\,
\lambda_{k,l}^\pm(n\mp1,m-3)^2\nonumber\\ && \times\,\bigg\{\bigg(
\biggl[\cgstack{\frac {n\mp1}2}{\frac {q+1}2}~
\cgstack{\frac12}{-\frac12}~\cgstack{\frac{n}2}{\frac{q}2}
\biggr]^2~\betab^1\wedge\beta^1
+\,\biggl[\cgstack{\frac {n\mp1}2}{\frac {q-1}2}~
\cgstack{\frac12}{\frac12}~\cgstack{\frac{n}2}{\frac{q}2}
\biggr]^2~\betab^2\wedge\beta^2\bigg)~
\big|\noverq\,,\,m\big\rangle\big\langle\noverq\,,\,
m\big| \nonumber\\ && +\, 
\biggl[\cgstack{\frac {n\mp1}2}{\frac q2}~
\cgstack{\frac12}{-\frac12}~\cgstack{\frac{n}2}{\frac{q-1}2}
\biggr]\,\biggl[\cgstack{\frac {n\mp1}2}{\frac{q}2}~
\cgstack{\frac12}{\frac12}~\cgstack{\frac{n}2}{\frac{q+1}2}
\biggr] \nonumber\\ && \times\,\Big(\betab^1\wedge\beta^2~
\big|{\stackrel{\scriptstyle n}{\scriptstyle q-1}}\,,\,m\big\rangle
\big\langle{\stackrel{\scriptstyle n}{\scriptstyle q+1}}\,,\,m
\big|+\betab^2\wedge\beta^1~\big|
{\stackrel{\scriptstyle n}{\scriptstyle q+1}}\,,\,m\big\rangle
\big\langle{\stackrel{\scriptstyle n}{\scriptstyle q-1}}\,,\,m
\big|\Big)\bigg\}
\label{betaprodCkl2}\eea
with $\lambda_{k,l}^\pm(n,m):=0$ for $n<0$.

Upon contracting with (\ref{KahlerCP2}), only the diagonal matrix
elements survive in (\ref{betaprodCkl1}) and (\ref{betaprodCkl2}). It
is straightforward to see from (\ref{CGcoeffs}) that the sum of the
squares of the Clebsch-Gordan coefficients in each case is independent
of $q\in\{-n+2j\}_{j=0}^n$. Using the explicit coefficient functions in
(\ref{lambdaklnm}) one establishes the identity
\beq
\lambda_{k,l}^+(n-1,m-3)^2+\lambda_{k,l}^-(n+1,m-3)^2-
\mbox{$\frac{n+2}{n+1}$}\,\lambda_{k,l}^+(n,m)^2-
\mbox{$\frac{n}{n+1}$}\,\lambda_{k,l}^-(n,m)^2=m
\label{lambdasymid}\eeq
and we arrive finally at the curvature equations
\bea
g^{a\bb}\,F_{a\bb}^{n,m}&=&m+\mbox{$\frac{n+2}{n+1}$}\,
\lambda_{k,l}^+(n,m)^2~\phi_{n,m}^+{}^\dag\,\phi_{n,m}^++
\mbox{$\frac{n}{n+1}$}\,\lambda_{k,l}^-(n,m)^2~
\phi^-_{n,m}{}^\dag\,\phi_{n,m}^-  \nonumber\\ && -\,
\lambda_{k,l}^+(n-1,m-3)^2~\phi_{n-1,m-3}^+\,
\phi_{n-1,m-3}^+{}^\dag \nonumber\\ && -\,\lambda_{k,l}^-(n+1,m-3)^2~
\phi_{n+1,m-3}^-\,\phi_{n+1,m-3}^-{}^\dag \ .
\label{symcurveqs}\eea
The constant perturbation in (\ref{symcurveqs}) is just the magnetic
charge $m\in\Z$ at a given symmetric vertex
$v=(n,m)\in\quiver_0(k,l)$. This is a typical feature of quiver vortex
equations which will play an important role in the following.

\bigskip

\noindent
{\bf Non-symmetric $\mbf{\underline{C}^{k,l}}$ quiver vortices. \ } Let
us now fix a vertex $v=(q,m)_n\in\quiver_0(k,l)$ of the non-symmetric
$\underline{C}^{k,l}$ quiver. Then (\ref{DUY3}) together with the
off-diagonal field strength matrix elements
(\ref{Q3cfpCklrep})--(\ref{Q3cf0Cklrep}) yield the bi-covariant Higgs
field derivatives
\bea
\partial_\ab{}^n\phi_{q,m}^++
{}_{n\pm1}A_\ab^{q+1,m+3}~{}^n\phi^+_{q,m}-
{}^n\phi_{q,m}^+~{}_nA_\ab^{q,m}&=&0 \ , \nonumber\\[4pt]
\partial_\ab{}^n\phi_{q,m}^-+
{}_{n\pm1}A_\ab^{q-1,m+3}~{}^n\phi^-_{q,m}-
{}^n\phi_{q,m}^-~{}_nA_\ab^{q,m}&=&0 \ , \nonumber\\[4pt]
\partial_\ab{}^n\phi_{q,m}^0+{}_n
A_\ab^{q+2,m}~{}^n\phi^0_{q,m}-
{}^n\phi_{q,m}^0~{}_nA_\ab^{q,m}&=&0
\label{Cklcovderivs}\eea
plus their hermitean conjugates. From (\ref{DUY4}) and
(\ref{Q3cfpCklrep}), along with the explicit forms (\ref{CGcoeffs})
and (\ref{gammapqmdef})--(\ref{gammalincomb}), we obtain the linear
holomorphic relations
\bea
{}^n\phi_{q,m}^+-\mbox{$\frac12$}\,(q-n)~
{}^n\phi_{q+2,m}^-\,{}^n\phi^0_{q,m}+\mbox{$\frac12$}\,(q-n-2)~
{}^{n+1}\phi_{q-1,m+3}^0\,{}^n\phi_{q,m}^-&=&0 \ ,
\nonumber\\[4pt]
{}^n\phi_{q,m}^+-\mbox{$\frac12$}\,(q+n+2)~
{}^n\phi_{q+2,m}^-\,{}^n\phi^0_{q,m}+\mbox{$\frac12$}\,(q+n)~
{}^{n-1}\phi_{q-1,m+3}^0\,{}^n\phi_{q,m}^-&=&0 \ .
\label{Ckllinrels}\eea
These equations hold whenever $\lambda_{k,l}^\pm(n,m)\neq0$, and
for $q=n$ the second equation is absent. Likewise, from
(\ref{DUY4}) and the curvature matrix elements
(\ref{Q3cfholrel1})--(\ref{Q3cfholrel3}) we find the quadratic
holomorphic relations
\bea
{}^n\phi^+_{q,m}\,{}^{n\pm1}\phi^-_{q+1,m-3}-
{}^n\phi^-_{q+2,m}\,{}^{n\pm1}\phi^+_{q+1,m-3}&=&0 \ ,
\nonumber\\[4pt]
{}^n\phi^+_{q,m}\,{}^n\phi^0_{q-2,m}-{}^{n\pm1}
\phi^0_{q-1,m+3}\,{}^n\phi^+_{q-2,m}&=&0 \ , \nonumber\\[4pt]
{}^n\phi^-_{q,m}\,{}^n\phi^0_{q-2,m}-{}^{n\pm1}
\phi^0_{q-3,m+3}\,{}^n\phi^-_{q-2,m}&=&0 \ .
\label{Cklquadrels}\eea

Finally, we substitute the diagonal matrix elements
(\ref{MdQ3curvdiagCklrep}) into (\ref{DUY1}) and use the K\"ahler
form (\ref{KahlerQ3}) on $Q_3$. Using the explicit coefficient
functions (\ref{lambdaklnm}), one finds the identity
\beq
\mbox{$\frac{n-q+2}{n+1}$}\,\lambda_{k,l}^+(n,m)^2+
\mbox{$\frac{n+q}{n+1}$}\,\lambda_{k,l}^-(n,m)^2-
\mbox{$\frac{n-q}{n}$}\,\lambda_{k,l}^+(n-1,m-3)^2-
\mbox{$\frac{n+q+2}{n+2}$}\,\lambda_{k,l}^-(n+1,m-3)^2=q-m
\label{nonsymlambdaid}\eeq
which gives the curvature equations
\bea
g^{a\bb}\,{}_nF^{q,m}_{a\bb}&=&(q+m)+
\frac{\lambda_{k,l}^+(n,m)^2}{3(n+1)}\,\Big((n+q+2)~{}^n
\phi_{q,m}^+{}^\dag\,{}^n\phi_{q,m}^++2(n-q+2)~{}^n
\phi_{q,m}^-{}^\dag\,{}^n\phi_{q,m}^-\Big) \nonumber\\ && +\,
\frac{\lambda_{k,l}^-(n,m)^2}{3(n+1)}\,\Big((n-q)~{}^n
\phi_{q,m}^+{}^\dag\,{}^n\phi_{q,m}^++2(n+q)~{}^n
\phi_{q,m}^-{}^\dag\,{}^n\phi_{q,m}^-\Big) \nonumber\\ && -\,
\frac{\lambda_{k,l}^+(n-1,m-3)^2}{3n}\,\Big((n+q)~
{}^{n-1}\phi^+_{q-1,m-3}\,{{}^{n-1}\phi_{q-1,m-3}^+}^\dag
\nonumber\\ && \hspace{4cm} +\,
2(n-q)~{}^{n-1}\phi^-_{q+1,m-3}\,
{{}^{n-1}\phi_{q+1,m-3}^-}^\dag\,\Big) \label{Cklcurveqs}\\ && -\,
\frac{\lambda_{k,l}^-(n+1,m-3)^2}{3(n+2)}\,\Big((n-q+2)~
{}^{n+1}\phi^+_{q-1,m-3}\,{{}^{n+1}\phi_{q-1,m-3}^+}^\dag
\nonumber\\ && \hspace{4cm} +\,
2(n+q+2)~{}^{n+1}\phi^-_{q+1,m-3}\,
{{}^{n+1}\phi_{q+1,m-3}^-}^\dag\,\Big) \nonumber\\ && +\,
\mbox{$\frac13$}\,(n-q)\,(n+q+2)~{}^n
\phi_{q,m}^0{}^\dag\,{}^n\phi_{q,m}^0-\mbox{$\frac13$}\,
(n+q)\,(n-q+2)~{}^n\phi^0_{q-2,m}\,{{}^n
\phi_{q-2,m}^0}^\dag \ . \nonumber
\eea
Again the constant perturbation is just the total magnetic
charge $(q+m)\in2\,\Z$ at the non-symmetric vertex 
$v=(q,m)_n\in\quiver_0(k,l)$. The expressions 
(\ref{Cklcovderivs})--(\ref{Cklquadrels}) and (\ref{Cklcurveqs})
all naturally incorporate the
contributions from multiple arrows at a given vertex whenever
degenerate weight vectors of the $\sut$ representation
$\underline{C}^{k,l}$ exist.

\subsection{Examples}

Let us now look at some explicit instances of the BPS equations
from Section~\ref{BPSeqs} above.

\bigskip

\noindent
{\bf Symmetric $\mbf{\underline{C}^{1,0}}$ quiver vortices. \ } Using
(\ref{CP2cffund}) and abbreviating $\phi:=\phi_{0,-2}^+$ again, the
DUY equations in this case become vortex equations on $M_{2d}$ given
by
\bea
g^{a\bb}\,F_{a\bb}^{1,1}&=&\Idd_{p_{1,1}}-\phi\,\phi^\dag \ ,
\nonumber\\[4pt] g^{a\bb}\,F_{a\bb}^{0,-2}&=&-2\big(\Idd_{p_{0,-2}}-
\phi^\dag\,\phi\big) \ , \nonumber\\[4pt]\pa_{\ab}\phi+A^{1,1}\,
\phi-\phi\,A^{0,-2}&=& 0 \ .
\label{vortexeqCP2C10}\eea
This system is a generalization of the standard holomorphic triple
$(E_1,E_2,\phi)$~\cite{LPS,Garcia1}. For $d=2$, it is related to the perturbed
Seiberg-Witten monopole equations on the K\"ahler four-manifold
$M_4$~\cite{Witten,group1}.

\bigskip

\noindent
{\bf Symmetric $\mbf{\underline{C}^{2,0}}$ quiver vortices. \ } Using
the K\"ahler two-form (\ref{KahlerCP2}) along with
(\ref{cf11C20sym})--(\ref{cf23C20sym}) and
(\ref{beta11prod1CP2C20})--(\ref{beta04prod2CP2C20}), we obtain the
DUY equations
\bea
g^{a\bb}\,F_{a\bb}^{2,2}&=&2\big(\Idd_{p_{2,2}}-\phi_{1,-1}^+\,
{\phi_{1,-1}^+}^\dag\big) \ , \nonumber\\[4pt]
g^{a\bb}\,F_{a\bb}^{1,-1}&=&-\big(\Idd_{p_{1,-1}}-
3{\phi_{1,-1}^+}^\dag\,
\phi_{1,-1}^++2\phi_{0,-4}^+\,{\phi_{0,-4}^+}^\dag\big) \ ,
\nonumber\\[4pt]
g^{a\bb}\,F_{a\bb}^{0,-4}&=&-4\big(\Idd_{p_{0,-4}}-
{\phi_{0,-4}^+}^\dag\,\phi_{0,-4}^+\big) \ , \nonumber\\[4pt]
\pa_{\ab}\phi_{1,-1}^++A_{\ab}^{2,2}\,\phi_{1,-1}^+-\phi_{1,-1}^+\,
A_{\ab}^{1,-1}&=&0 \ , \nonumber\\[4pt]
\pa_{\ab}\phi_{0,-4}^++A_{\ab}^{1,-1}\,\phi_{0,-4}^+-\phi_{0,-4}^+\,
A_{\ab}^{0,-4}&=&0 \ .
\label{vortexeqCP2C20}\eea
This system is an extension of the basic holomorphic chain vortex
equations~\cite{PS1,Garcia1}.

\bigskip

\noindent
{\bf Symmetric $\mbf{\underline{C}^{1,1}}$ quiver vortices. \ } Using
(\ref{cf11CP2adrep})--(\ref{adbetapm13}) we find the curvature
equations 
\bea
g^{a\bb}\,F_{a\bb}^{1,3}&=&3\big(\Idd_{p_{1,3}}-\mbox{$\frac12$}\,
\phi_{0,0}^+\,\phi_{0,0}^+{}^\dag-\mbox{$\frac12$}\,
\phi_{2,0}^-\,\phi_{2,0}^-{}^\dag\big) \ , \nonumber\\[4pt]
g^{a\bb}\,F_{a\bb}^{0,0}&=&3\big(\phi_{0,0}^+{}^\dag\,\phi_{0,0}^+-
\phi_{1,-3}^-\,\phi_{1,-3}^-{}^\dag\big) \ , \nonumber\\[4pt]
g^{a\bb}\,F_{a\bb}^{2,0}&=&\phi_{2,0}^-{}^\dag\,\phi_{2,0}^--
\phi_{1,-3}^+\,\phi_{1,-3}^+{}^\dag \ , \nonumber\\[4pt]
g^{a\bb}\,F_{a\bb}^{1,-3}&=&-3\big(\Idd_{p_{1,-3}}-\mbox{$\frac12$}\,
\phi_{1,-3}^-{}^\dag\,\phi_{1,-3}^--\mbox{$\frac12$}\,
\phi_{1,-3}^+{}^\dag\,\phi_{1,-3}^+\big) \ ,
\label{intcondCP2C11}\eea
the bi-covariant Higgs field derivatives
\bea
\partial_\ab\phi_{0,0}^++A_\ab^{1,3}\,\phi_{0,0}^+-\phi_{0,0}^+\,
A_\ab^{0,0}&=&0 \ , \nonumber\\[4pt]
\partial_\ab\phi_{2,0}^-+A_\ab^{1,3}\,\phi_{2,0}^--\phi_{2,0}^-\,
A_\ab^{2,0}&=&0 \ , \nonumber\\[4pt]
\partial_\ab\phi_{1,-3}^-+A_\ab^{0,0}\,\phi_{1,-3}^--\phi_{1,-3}^-\,
A_\ab^{1,-3}&=&0 \ , \nonumber\\[4pt]
\partial_\ab\phi_{1,-3}^++A_\ab^{2,0}\,\phi_{1,-3}^+-\phi_{1,-3}^+\,
A_\ab^{1,-3}&=&0 \ ,
\label{HiggsCP2C11}\eea
and the holomorphic relation
\beq
\phi_{0,0}^+\,\phi_{1,-3}^--\phi_{2,0}^-\,\phi_{1,-3}^+=0 \ .
\label{holrelCP2C11}\eeq
We refer to this system as a ``holomorphic square''. It is the basic
building block for higher representation symmetric quiver vortices.

\bigskip

\noindent
{\bf Non-symmetric $\mbf{\underline{C}^{1,0}}$ quiver vortices. \ }
Using the K\"ahler form (\ref{KahlerQ3}) on $Q_3$ and
(\ref{fundcurvQ3}), the DUY equations reduce on $M_{2d}$ in this case
to
\bea
g^{a\bb}\,{}_1F_{a\bb}^{1,1}&=&2\big(\Idd_{{}^1p_{1,1}}-
\mbox{$\frac13$}~{}^0\phi_{0,-2}^+\,
{{}^0\phi_{0,-2}^+}^\dag-\mbox{$\frac23$}~{}^1
\phi^0_{-1,1}\,{{}^1\phi^0_{-1,1}}^\dag\big) \ ,
\nonumber\\[4pt]
g^{a\bb}\,{}_1F_{a\bb}^{-1,1}&=&\mbox{$\frac43$}\,
\big(\,{{}^1\phi^0_{-1,1}}^\dag\,{}^1\phi^0_{-1,1}-
{}^0\phi_{0,-2}^-\,{{}^0\phi_{0,-2}^-}^\dag\,\big) \ , \nonumber\\[4pt]
g^{a\bb}\,{}_0F_{a\bb}^{0,-2}&=&-2\big(\Idd_{{}^0p_{0,-2}}-
\mbox{$\frac13$}~
{{}^0\phi_{0,-2}^+}^\dag\,{}^0\phi_{0,-2}^+-
\mbox{$\frac23$}~{{}^0\phi_{0,-2}^-}^\dag\,
{}^0\phi_{0,-2}^-\big) \ , \nonumber\\[4pt]
\pa_{\ab}{}^0\phi_{0,-2}^++{}_1A_{\ab}^{1,1}~{}^0\phi_{0,-2}^+-
{}^0\phi_{0,-2}^+~{}_0A_{\ab}^{0,-2}&=&0 \ , \nonumber\\[4pt]
\pa_{\ab}{}^0\phi_{0,-2}^-+{}_1A_{\ab}^{-1,1}~{}^0\phi_{0,-2}^--
{}^0\phi_{0,-2}^-~{}_0A_{\ab}^{0,-2}&=&0 \ , \nonumber\\[4pt]
\pa_{\ab}{}^1\phi^0_{-1,1}+{}_1A_{\ab}^{1,1}~{}^1\phi^0_{-1,1}-
{}^1\phi^0_{-1,1}~{}_1A_{\ab}^{-1,1}&=&0 \ ,
\label{vortexeqQ3C10}\eea
together with the holomorphic relation
\beq
{}^0\phi^+_{0,-2}-
{}^1\phi^0_{-1,1}\,{}^0\phi^-_{0,-2}=0 \ .
\label{nonsymC10holrel}\eeq
We call this system a ``holomorphic triangle''. It is the basic
building block for higher representation non-symmetric quiver
vortices.

\bigskip

\noindent
{\bf Non-symmetric $\mbf{\underline{C}^{2,0}}$ quiver vortices. \ }
From (\ref{cf11Q3C20rep})--(\ref{cf66Q3C20rep}) the quiver vortex
equations on $M_{2d}$ in this instance are found to comprise the
curvature equations
\bea
g^{a\bb}\,{}_2F_{a\bb}^{2,2}&=&
4\big(\Idd_{{}^2p_{2,2}}-\mbox{$\frac13$}~{}^1\phi^+_{1,-1}\,
{{}^1\phi^+_{1,-1}}^\dag-\mbox{$\frac23$}~
{}^2\phi^0_{0,2}\,{{}^2\phi^0_{0,2}}^\dag\,\big) \ , \nonumber\\[4pt]
g^{a\bb}\,{}_2F_{a\bb}^{0,2}&=&2\big(
\Idd_{{}^2p_{0,2}}-\mbox{$\frac13$}~
{}^1\phi^+_{-1,-1}\,{{}^1\phi^+_{-1,-1}}^\dag\nonumber\\ &&
-\,\mbox{$\frac23$}~{}^1\phi^-_{1,-1}\,
{{}^1\phi^-_{1,-1}}^\dag+\mbox{$\frac43$}~
{}^2{\phi^0_{0,2}}^\dag\,{}^2\phi^0_{0,2}+\mbox{$\frac43$}~
{}^2\phi^0_{-2,2}\,{{}^2\phi^0_{-2,2}}^\dag\,\big) \ ,
\nonumber\\[4pt]
g^{a\bb}\,{}_2F_{a\bb}^{-2,2}&=&-\mbox{$\frac83$}\,\big(\,
{}^1\phi^-_{-1,-1}\,{{}^1\phi^-_{-1,-1}}^\dag
-{{}^2\phi^0_{-2,2}}^\dag\,{}^2\phi^0_{-2,2}\big) \ ,
\nonumber\\[4pt]
g^{a\bb}\,{}_1F_{a\bb}^{1,-1}&=&\mbox{$\frac43$}\,\big(\,
{{}^1\phi^-_{1,-1}}^\dag\,{}^1\phi^-_{1,-1}-{}^1\phi^0_{-1,-1}\,
{{}^1\phi^0_{-1,-1}}^\dag+
{}^1{\phi^+_{1,-1}}^\dag\,{}^1\phi^+_{1,-1}-
{}^0\phi^+_{0,-4}\,{{}^0\phi^+_{0,-4}}^\dag\,\big) \ ,
\nonumber\\[4pt]
g^{a\bb}\,{}_1F_{a\bb}^{-1,-1}&=&-2\big(
\Idd_{{}^1p_{-1,-1}}-\mbox{$\frac13$}~
{{}^1\phi^+_{-1,-1}}^\dag\,{}^1\phi^+_{-1,-1}\nonumber\\ &&
-\,\mbox{$\frac23$}~{{}^1\phi^0_{-1,-1}}^\dag\,{}^1\phi^0_{-1,-1}
-\mbox{$\frac43$}~
{{}^1\phi^-_{-1,-1}}^\dag\,{}^1\phi^-_{-1,-1}+\mbox{$\frac43$}~
{}^0\phi^-_{0,-4}\,{{}^0\phi^-_{0,-4}}^\dag\,\big) \ ,
\nonumber\\[4pt]
g^{a\bb}\,{}_0F_{a\bb}^{0,-4}&=&-4\big(\Idd_{{}^0p_{0,-4}}-
\mbox{$\frac13$}~{{}^0\phi^+_{0,-4}}^\dag\,{}^0
\phi^+_{0,-4}-\mbox{$\frac23$}
~{{}^0\phi^-_{0,-4}}^\dag\,{}^0\phi^-_{0,-4}\big) \ ,
\label{C20nonsymcurv}\eea
the bi-covariant Higgs field derivatives
\bea
\partial_\ab{}^2\phi^0_{0,2}+{}_2A_\ab^{2,2}~{}^2\phi^0_{0,2}-
{}^2\phi^0_{0,2}~{}_2A_\ab^{0,2}&=&0 \ , \nonumber\\[4pt]
\partial_\ab{}^1\phi^+_{1,-1}+{}_2A_\ab^{2,2}~{}^1\phi^+_{1,-1}-
{}^1\phi^+_{1,-1}~{}_1A_\ab^{1,-1}&=&0 \ , \nonumber\\[4pt]
\partial_\ab{}^2\phi^0_{-2,2}+
{}_2A_\ab^{0,2}~{}^2\phi^0_{-2,2}-{}^2\phi^0_{-2,2}~
{}_2A_\ab^{-2,2}&=&0 \ , \nonumber\\[4pt]
\partial_\ab{}^1\phi^-_{1,-1}+{}_2A_\ab^{0,2}~{}^1\phi^-_{1,-1}-{}^1
\phi^-_{1,-1}~{}_1A_\ab^{1,-1}&=&0 \ , \nonumber\\[4pt]
\partial_\ab{}^1\phi_{-1,-1}^++
{}_2A_\ab^{0,2}~{}^1\phi_{-1,-1}^+-{}^1\phi_{-1,-1}^+~
{}_1A_\ab^{-1,-1}&=&0 \ , \nonumber\\[4pt]
\partial_\ab{}^1\phi^-_{-1,-1}+{}_2A_\ab^{-2,2}~{}^1
\phi^-_{-1,-1}-{}^1\phi^-_{-1,-1}~
{}_1A_\ab^{-1,-1}&=&0 \ , \nonumber\\[4pt]
\partial_\ab{}^1\phi^0_{-1,-1}+
{}_1A_\ab^{1,-1}~{}^1\phi^0_{-1,-1}-{}^1\phi^0_{-1,-1}~{}_1
A_\ab^{-1,-1}&=&0 \ , \nonumber\\[4pt]
\partial_\ab{}^0\phi^+_{0,-4}+{}_1A_\ab^{1,-1}~{}^0
\phi^+_{0,-4}-{}^0\phi^+_{0,-4}~{}_0A_\ab^{0,-4}&=&0 \ ,
\nonumber\\[4pt]
\partial_\ab{}^0\phi^-_{0,-4}+{}_1A_\ab^{-1,-1}~{}^0\phi^-_{0,-4}-
{}^0\phi^-_{0,-4}~{}_0A_\ab^{0,-4}&=&0 \ ,
\label{C20nonsymHiggs}\eea
the linear holomorphic relations
\bea
{}^1\phi^+_{1,-1}-{}^2\phi^0_{0,2}\,{}^1\phi^-_{1,-1}&=&0 \ ,
\nonumber\\[4pt]
{}^1\phi^+_{-1,-1}+{}^1\phi^-_{1,-1}\,{}^1\phi^0_{-1,-1}-2~
{}^2\phi^0_{-2,2}\,{}^1\phi^-_{-1,-1}&=&0 \ , \nonumber\\[4pt]
{}^0\phi^+_{0,-4}-
{}^1\phi^0_{-1,-1}\,{}^0\phi^-_{0,-4}&=&0 \ ,
\label{C20linhol}\eea
and the quadratic holomorphic relations
\bea
{}^2\phi^0_{0,2}\,{}^1\phi^+_{-1,-1}-
{}^1\phi^+_{1,-1}\,{}^1\phi^0_{-1,-1}&=&0 \ , \nonumber\\[4pt]
{}^1\phi^+_{-1,-1}\,
{}^0\phi^-_{0,-4}-{}^1\phi^-_{1,-1}\,{}^0\phi^+_{0,-4}&=&0 \ .
\label{C20quadhol}\eea

\bigskip

\section{Noncommutative quiver vortices\label{NCquivvort}}

\noindent
In this section we shall construct explicit solutions to the
nonabelian quiver vortex equations of the previous section. For this,
we specialize to the manifold $M_{2d}=\C^d$ with the standard flat
K\"ahler metric $g_{a\bb}=\delta_{ab}$. Although it is known that 
solutions to the DUY equations in $2d>4$ dimensions exist on~$\C^d$ 
\cite{DUY1,Garcia1,A-CG-P1,A-CG-P2},
they cannot be constructed explicitly as far as we know. To get non-trivial 
solutions on this space, we will need to apply a noncommutative deformation.

\subsection{Noncommutative quiver gauge theory\label{NCGTquiver}}

The Moyal deformation of $\C^d$ is realized by mapping Schwartz
functions $f$ on $\C^d$ to compact Weyl operators $\hat f$ acting on a
separable Hilbert space $\Hcal$. The coordinates $z^a$ and $\zb^\bb$
of $\C^d$ are mapped to operators $\zh^a$ and $\zbh^\bb$ subject to
the commutation relations
\beq
\big[\zh^a\,,\,\zbh^{\bb}\,\big] =\th^{a\bb}
\label{zcommrels}\eeq
with a constant real-valued antisymmetric matrix $(\theta^{a\bb})$ of
maximal rank. All other commutators vanish. We may rotate the
coordinates so that $(\theta^{a\bb})$ assumes its canonical form with
\beq
\th^{a\bb}=2\,\de^{ab}\,\th^a
\label{thetacanform}\eeq
for $\theta^a\in(0,\infty)$, $a=1,\dots,d$. This defines the
noncommutative space $\C_\theta^d$ with isometry group ${\rm USp}(d)$
and carrying $d$ commuting copies of the Heisenberg algebra
\beq
\Bigl[\,\frac{\zh^a}{\sqrt{2\,\th^a}}\ ,\ 
\frac{\zbh^{\bb}}{\sqrt{2\,\th^b}}\,
\Bigr] = \de^{ab} \ .
\label{Heisenalg}\eeq
We will represent this algebra on the standard irreducible Fock module
$\Hcal$. Recall~\cite{Harvey} that derivatives and integrals of fields
on $\C^d$ are mapped according to
\beq
\widehat{\pa_{{\ab}} f}\=\th_{\ab b}\,\big[\zh^b \,,\, \fh\,\big]
\qquad \mbox{and} \qquad \int_{\C^{d}}\, \diff V~f\=
(2\pi)^d~{\rm Pf}(\th)~\textrm{Tr}^{~}_\Hcal\big(\fh\,\big) \ ,
\label{derivintmap}\eeq
where $\th^{a\bb}\,\th_{\bb c}=\delta^a{}_c$. For simplicity, we will
drop the hats in the notation from now on.

To rewrite the nonabelian quiver vortex equations of the previous
section using the Moyal deformation, we define the covariant
coordinates
\beq
X_{a}^{v}\ :=\ A_{a}^{v} + \th_{a\bb}\,\zb^{\bb}
\qquad\textrm{and}\qquad
X_{{\ab}}^{v}\ :=\ A_{{\ab}}^{v} + \th^{~}_{\ab b}\,z^b
\label{covcoords}\eeq
at each vertex $v\in\quiver_0(k,l)$ of the pertinent quiver. Then the
components of the field strength tensor can be expressed as
\beq
F_{{a}{b}}^{v}\ =\ \big[X_{{a}}^{v}\,,\,
X_{b}^{v}\,\big]\qquad\textrm{and}\qquad
F_{{a}{\bb}}^{v}\ =\ \big[X_{a}^{v}\,,\,X_{{\bb}}^{v}\,\big] +
\th_{a\bb} \ .
\label{FabXexpress}\eeq
Likewise, the Higgs field gradients $D_\ab\phi_{v,\Phi(v)}$ can be
expressed through the $X_\ab^v$ for each arrow
$\Phi\in\quiver_1(k,l)$. Then the vortex equations reduce to algebraic
equations for the collection of operators
$\{X^v,\phi_{v,\Phi(v)}\}$ acting on the projective module
$\underline{V}^{k,l}\otimes\Hcal$ of rank $p$ over $\C_\theta^d$, where
\beq
\underline{V}^{k,l}=\bigoplus_{v\in\quiver_0(k,l)}\,\underline{V}\,_v
\otimes\,\underline{v}
\label{Vklmodule}\eeq
with $\underline{V}\,_v=\C^{p_v}$ the fibre space of the
vector bundle $E_{p_v}\to\C^d$. For example, the holomorphicity
equations (\ref{Fvholgen}) become the commutativity equations
\beq
\big[X_{a}^{v}\,,\,X_{b}^{v}\,\big] \= 0 \=
\big[X_{\ab}^{v}\,,\,X_{\bb}^{v}\,\big] \ .
\label{commeqsgen}\eeq

\bigskip

\noindent
{\bf Symmetric $\mbf{\underline{C}^{k,l}}$ quiver modules. \ } In the
symmetric case, the system of additional algebraic equations at each
vertex $v=(n,m)\in\quiver_0(k,l)$ reads
\bea
X_\ab^{n\pm1,m+3}\,\phi^\pm_{n,m}&=&\phi^\pm_{n,m}\,X_\ab^{n,m} \ ,
\label{symCklXphi1}\\[4pt]
\phi_{n,m}^+\,\phi^-_{n+1,m-3}&=&\phi^-_{n+2,m}\,\phi^+_{n+1,m-3} \ ,
\label{symCklhol}\\[4pt]
\delta^{ab}\,\big(\big[X_{a}^{n,m}\,,\,X_{{\bb}}^{n,m}\,\big] +
\th_{a\bb}\big)&=&m+\mbox{$\frac{n+2}{n+1}$}\,
\lambda_{k,l}^+(n,m)^2~\phi_{n,m}^+{}^\dag\,\phi_{n,m}^++
\mbox{$\frac{n}{n+1}$}\,\lambda_{k,l}^-(n,m)^2~
\phi^-_{n,m}{}^\dag\,\phi_{n,m}^- \nonumber\\ && -\,
\lambda_{k,l}^+(n-1,m-3)^2~\phi_{n-1,m-3}^+\,
\phi_{n-1,m-3}^+{}^\dag \nonumber 
\\ && -\,\lambda_{k,l}^-(n+1,m-3)^2~
\phi_{n+1,m-3}^-\,\phi_{n+1,m-3}^-{}^\dag \ .
\label{symCklmods}\eea

\bigskip

\noindent
{\bf Non-symmetric $\mbf{\underline{C}^{k,l}}$ quiver modules. \ } In
the non-symmetric case, the system of additional algebraic equations
at each vertex $v=(q,m)_n\in\quiver_0(k,l)$ reads
\bea
{}_{n\pm1}X_\ab^{q+1,m+3}~{}^n\phi^+_{q,m}&=&
{}^n\phi_{q,m}^+~{}_nX_\ab^{q,m} \ , \label{nonsymCklXphi1}\\[4pt]
{}_{n\pm1}X_\ab^{q-1,m+3}~{}^n\phi^-_{q,m}&=&
{}^n\phi_{q,m}^-~{}_nX_\ab^{q,m} \ , \label{nonsymCklXphi2}\\[4pt]
{}_nX_\ab^{q+2,m}~{}^n\phi^0_{q,m}&=&
{}^n\phi_{q,m}^0~{}_nX_\ab^{q,m} \ , \label{nonsymCklXphi3}\\[4pt]
{}^n\phi_{q,m}^+&=&\mbox{$\frac12$}\,(q-n)~
{}^n\phi_{q+2,m}^-\,{}^n\phi^0_{q,m}\nonumber\\ &&-\,
\mbox{$\frac12$}\,(q-n-2)~
{}^{n+1}\phi_{q-1,m+3}^0\,{}^n\phi_{q,m}^- \ , 
\label{nonsymCkllin1}\\[4pt]
{}^n\phi_{q,m}^+&=&\mbox{$\frac12$}\,(q+n+2)~
{}^n\phi_{q+2,m}^-\,{}^n\phi^0_{q,m} \nonumber\\ &&-\,
\mbox{$\frac12$}\,(q+n)~
{}^{n-1}\phi_{q-1,m+3}^0\,{}^n\phi_{q,m}^- \ ,
\label{nonsymCkllin2}\\[4pt]
{}^n\phi^+_{q,m}\,{}^{n\pm1}\phi^-_{q+1,m-3}&=&
{}^n\phi^-_{q+2,m}\,{}^{n\pm1}\phi^+_{q+1,m-3} \ , 
\label{nonsymCklquad1}\\[4pt]
{}^n\phi^+_{q,m}\,{}^n\phi^0_{q-2,m}&=&{}^{n\pm1}
\phi^0_{q-1,m+3}\,{}^n\phi^+_{q-2,m} \ , \label{nonsymCklquad2}\\[4pt]
{}^n\phi^-_{q,m}\,{}^n\phi^0_{q-2,m}&=&{}^{n\pm1}
\phi^0_{q-3,m+3}\,{}^n\phi^-_{q-2,m} \ , \label{nonsymCklquad3}\\[4pt]
\delta^{ab}\,\big(\big[{}_nX_{a}^{q,m}\,,\,
{}_nX_{{\bb}}^{q,m}\,\big] +\th_{a\bb}\big)&=& \nonumber \\
&& \hspace{-2cm} (q+m) +
\frac{\lambda_{k,l}^+(n,m)^2}{3(n+1)}\,\Big((n+q+2)~{}^n
\phi_{q,m}^+{}^\dag\,{}^n\phi_{q,m}^++2(n-q+2)~{}^n
\phi_{q,m}^-{}^\dag\,{}^n\phi_{q,m}^-\Big) \nonumber\\ &&
\hspace{-2cm}+\, 
\frac{\lambda_{k,l}^-(n,m)^2}{3(n+1)}\,\Big((n-q)~{}^n
\phi_{q,m}^+{}^\dag\,{}^n\phi_{q,m}^++2(n+q)~{}^n
\phi_{q,m}^-{}^\dag\,{}^n\phi_{q,m}^-\Big) \nonumber\\ &&
\hspace{-2cm}-\, 
\frac{\lambda_{k,l}^+(n-1,m-3)^2}{3n}\,\Big((n+q)~
{}^{n-1}\phi^+_{q-1,m-3}\,{{}^{n-1}\phi_{q-1,m-3}^+}^\dag
\nonumber\\ &&  \hspace{2cm}+\,
2(n-q)~{}^{n-1}\phi^-_{q+1,m-3}\,
{{}^{n-1}\phi_{q+1,m-3}^-}^\dag\,\Big) \nonumber\\ && \hspace{-2cm}-\,
\frac{\lambda_{k,l}^-(n+1,m-3)^2}{3(n+2)}\,\Big((n-q+2)~
{}^{n+1}\phi^+_{q-1,m-3}\,{{}^{n+1}\phi_{q-1,m-3}^+}^\dag
\nonumber\\ &&  \hspace{2cm}+\,
2(n+q+2)~{}^{n+1}\phi^-_{q+1,m-3}\,
{{}^{n+1}\phi_{q+1,m-3}^-}^\dag\,\Big) \nonumber\\ && \hspace{-2cm}+\,
\mbox{$\frac13$}\,(n-q)\,(n+q+2)~{}^n
\phi_{q,m}^0{}^\dag\,{}^n\phi_{q,m}^0 \nonumber\\ && \hspace{-2cm}-\,
\mbox{$\frac13$}\,(n+q)\,(n-q+2)~{}^n\phi^0_{q-2,m}\,{{}^n
\phi_{q-2,m}^0}^\dag \ .
\label{nonsymCklmods}\eea

\subsection{Finite energy solutions\label{Finitesols}}

We will begin by constructing finite energy solutions of Yang-Mills
theory on the noncommutative space $X=\C_\theta^d\times G/H$. In the
generic (non-BPS) case, the gauge group of the quiver gauge theory of
Section~\ref{NCGTquiver} above is
\beq
{\cal G}^{k,l}=\prod_{v\in\quiver_0(k,l)}\,\urm(p_v) \ .
\label{gaugegpnonBPS}\eeq
The corresponding reduction of the Yang-Mills action on $X$, regarded
as an energy functional for static quiver gauge fields on
$\R^{0,1}\times\C_\theta^d$ in the temporal gauge, is given by computing
\beq
E_{\rm YM}:=\mbox{$\frac14$}~\Pf(2\pi\,\theta)\,
\int_{ G/H}\,\Tr_{\underline{V}^{k,l}\otimes\Hcal}\big(\cf\wedge*
\cf\big)  \ .
\label{EYMgen}\eeq

Fix an integer $r$ with $0<r\leq p$ and introduce a collection of
partial isometries $T_v$, $v\in\quiver_0(k,l)$ on $\Hcal$ realized by
$p_v\times r$ Toeplitz operators obeying
\beq
T_v^\dag\,T_v\=\Idd_r \qquad \mbox{and} \qquad
T_v\,T_v^\dag\=\Idd_{p_v}-P_v \ ,
\label{partialisom}\eeq
where $P_v=P_v^\dag=P_v^2$ is a hermitean projector of finite rank
\beq
N_v:=\Tr^{~}_{\underline{V}\,_v\otimes\Hcal}(P_v) \ .
\label{Nvrank}\eeq
Such operators $T_v$ can be constructed explicitly from an
$\sut$-equivariant version of the noncommutative ABS construction,
analogously to the $\su$ case of~\cite{PS1}. We will return to this
point in the next section. We make the ansatz for
the gauge connections given by
\beq
A_a^v=\th_{a\bb}~\big(T_v\,\zb^\bb\,T_v^\dag-\zb^\bb\,\big) \ ,
\label{Xavansatz}\eeq
which yields the field strength components
\beq
F_{ab}^v\=0\=F_{\ab\bb}^v \qquad \mbox{and} \qquad 
F_{a\bb}^v\=\th_{a\bb}~P_v\=\frac1{2\theta^a}\,\delta_{ab}~P_v
\label{Fabvcomps}\eeq
at each vertex $v\in\quiver_0(k,l)$. The details of the ansatz for the
module morphisms $\phi_{v,\Phi(v)}$ depend on the particular
quiver. In the following we will use the projector
\beq
\Pcal=\sum_{v\in\quiver_0(k,l)}\,P_v\otimes\Pi_v \ ,
\label{Pcaldef}\eeq
where $\Pi_v$ is the projection of $\underline{C}^{k,l}\big|_H$ onto
the irreducible $H$-module $\underline{v}$.

\bigskip

\noindent
{\bf Symmetric $\mbf{\underline{C}^{k,l}}$ quiver modules. \ } At a
given symmetric vertex $v=(n,m)\in\quiver_0(k,l)$ we use the partial
isometries $T_{n,m}$ above to construct the operators
\beq
\phi_{n,m}^\pm=T_{n\pm1,m+3}\,T_{n,m}^\dag \ .
\label{phisymansatz}\eeq
With the ansatz (\ref{phisymansatz}), one sees that both the holomorphic 
relations (\ref{symCklhol}) and the non-holomorphic relations
\beq
\phi_{n,m}^+\,\phi_{n,m}^-{}^\dag=
\phi_{n+1,m+3}^-{}^\dag\,\phi_{n-1,m+3}^+
\label{symCklnonhol}\eeq
are satisfied. These conditions are necessary to yield a finite
Yang-Mills action below. Moreover, with this ansatz one easily checks
that the covariant constancy equations (\ref{symCklXphi1}) are
satisfied, along with
\beq
\phi_{n,m}^\pm{}^\dag\,
\phi_{n,m}^\pm\=\Idd_{p_{n,m}}-P_{n,m}\=\phi_{n\mp1,m-3}^\pm\,
\phi_{n\mp1,m-3}^\pm{}^\dag \ .
\label{symphiprodrels}\eeq

Thus for our ansatz the off-diagonal field strength components
(\ref{curvoffdiagCklrep})--(\ref{curvnonholrelsCklrep}) all
vanish. The reduction of the energy functional (\ref{EYMgen}) is
therefore given by substituting (\ref{curvdiagCklrep}) in the basis
$\{\beta^i\wedge\betab^j\}$ of $(1,1)$-forms on $\C P^2$ using
the K\"ahler metric given by (\ref{KahlerCP2}) and (\ref{metricX})
with $g_{a\bb}=\de_{ab}$ to get
\bea
E_{\rm YM}&=&2~\Pf(2\pi\,\theta)~\vol\big(\C P^2\big)\,
\sum_{(n,m)\in\quiver_0(k,l)}\,\bigg[\,(n+1)\,\sum_{a,b=1}^d\,
\Tr^{~}_{\underline{V}\,_{n,m}\otimes\Hcal}
\big|F_{a\bb}^{n,m}\big|^2 \label{EYMsymdef}
\\ && \hspace{3cm}
+\,\Tr^{~}_{\underline{V}\,_{n,m}\otimes
\underline{(n,m)}\otimes\Hcal}\Big(\big|
\cf_{1\bar1}^{n,m\,;\,n,m}\big|^2+\big|
\cf_{2\bar2}^{n,m\,;\,n,m}\big|^2+2\big|
\cf_{1\bar2}^{n,m\,;\,n,m}\big|^2\Big)\bigg] \ , \nonumber
\eea
where we use the matrix notation
$|\cf|^2:=\frac12\,(\cf^\dag\,\cf+\cf\,\cf^\dag)$. We substitute
(\ref{Fabvcomps}), (\ref{symphiprodrels}) along with
(\ref{betaprodCkl1}), (\ref{betaprodCkl2}) and the explicit
Clebsch-Gordan coefficients (\ref{CGcoeffs}). Using the identity
(\ref{nonsymlambdaid}) along with the actions of the $\sut$ generators
in (\ref{Izrep}) and (\ref{Yrep}), one finds the curvature components
\beq
\cf_{1\bar1}\=-\mbox{$\frac12$}\,\Pcal\,(H_{\alpha_1}+H_{\alpha_2})
\qquad \mbox{and} \qquad
\cf_{2\bar2}\=\mbox{$\frac12$}\,\Pcal\,(H_{\alpha_1}-H_{\alpha_2}) \
.
\label{CP2F11F22}\eeq
Using the identity
\beq
\mbox{$\frac1n$}\,\lambda_{k,l}^+(n-1,m-3)^2+
\mbox{$\frac1{n+2}$}\,\lambda_{k,l}^-(n+1,m-3)^2-
\mbox{$\frac1{n+1}$}\,\lambda_{k,l}^+(n,m)^2-
\mbox{$\frac1n$}\,\lambda_{k,l}^-(n,m)^2=1
\label{1newid}\eeq
along with (\ref{Ipmrep}), one also finds
\beq
\cf_{1\bar2}=-\Pcal\,E_{-\alpha_1} \ .
\label{CP2F12}\eeq
It follows that
\beq
\big|\cf_{1\bar1}\big|^2+\big|
\cf_{2\bar2}\big|^2+2\big|
\cf_{1\bar2}\big|^2=\Pcal\,\mbf C_2(H) \ ,
\label{CP2FsumC2}\eeq
where
\beq
\mbf C_2(H)=\big(E_{\alpha_1}\,E_{-\alpha_1}+E_{-\alpha_1}\,
E_{\alpha_1}+\mbox{$\frac12$}\,H_{\alpha_1}^2\big)+\mbox{$\frac12$}\,
H_{\alpha_2}^2
\label{quadCasH}\eeq
is the quadratic Casimir operator of the holonomy group
$H=\su\times\uo$. In the irreducible representation
$\underline{(n,m)}$, it acts as the multiplication operator by the
eigenvalue $\frac12\,n\,(n+2)+\frac12\,m^2$, where we have used the
fact that the summation range over isospin is symmetric under
reflection $q\to-q$.

After tracing over the representation spaces
$\underline{(n,m)}\cong\C^{n+1}$ in the last line of
(\ref{EYMsymdef}), we arrive finally at the Yang-Mills energy
\beq
E_{\rm YM}=2~\Pf(2\pi\,\theta)~\vol\big(\C P^2\big)\,
\sum_{(n,m)\in\quiver_0(k,l)}\,\Tr^{~}_{\underline{V}\,_{n,m}\otimes
\Hcal}\big(\varepsilon_{n,m}~P_{n,m}\big) \ ,
\label{EYMsymfinal}\eeq
where
\beq
\varepsilon_{n,m}=(n+1)\,\Big(\,\frac14\,
\sum_{a=1}^d\,\frac1{(\theta^a)^2}+\mbox{$\frac12$}\,n\,(n+2)+
\mbox{$\frac12$}\,m^2\,\Big) \ .
\label{symenergynm}\eeq
The quantity (\ref{symenergynm}) is the finite energy density at the
symmetric vertex $v=(n,m)$. The $\theta$-dependent term is the
Yang-Mills energy of vortices on $\C_\theta^d$~\cite{DMR1}, and in the
total energy (\ref{EYMsymfinal}) it can be interpreted as the tension
of $N$ D0-branes bound inside a collection of D$(2d)$-branes in the
Seiberg-Witten decoupling limit~\cite{Harvey}, where
\beq
N:=\sum_{(n,m)\in\quiver_0(k,l)}\,(n+1)\,N_{n,m} \ .
\label{NsymD0}\eeq

The second term in (\ref{symenergynm}) is the angular momentum
contribution from the isospin $I=\frac n2$ of the instanton gauge
potential $B_{n,m}$. The third term is recognized as the Yang-Mills
energy of a monopole on the projective plane $\CP^2$ of magnetic
charge $m\in\Z$ (in suitable area units for the embedded two-cycle
$\CP^1\subset\CP^2$). The charge one
$\su$ instanton on $\CP^2$ also contributes overall multiplicity
factors $(n+1)$ corresponding to the dimension of the irreducible
$\su$ representation it lives in at the vertex $(n,m)$. The
correlation between monopole and instanton quantum numbers is a
consequence of the fact that the instanton bundle
can be realized as the $\su$-bundle
\beq
\Ical=S^5\times_\rho\,\su
\label{ISU2bundle}\eeq
associated to the Hopf bundle (\ref{S5Hopfbundle}) by the diagonal
representation $\rho:\uo\to\su$.

Note that in contrast to the $\su$-equivariant quiver gauge theories
based on the symmetric space $\CP^1$~\cite{PS1,LPS2}, the ansatz
(\ref{phisymansatz}) automatically yields a finite Yang-Mills energy,
without the need of multiplying them by suitable coefficient functions
on $\quiver_0(k,l)$. We are free to multiply the field configurations
(\ref{phisymansatz}) by arbitrary complex numbers of modulus one,
which are functions on $\quiver_0(k,l)$. By working in the basis
generated by the canonical one-forms $\beta^i$ and $\betab^i$ on
$\CP^2$, one can straightforwardly adapt the proof of~\cite{PS1} to
show that our ansatz solves the full Yang-Mills equations
(\ref{YMeqs}) on $X=\C_\theta^d\times\CP^2$.

\bigskip

\noindent
{\bf Non-symmetric $\mbf{\underline{C}^{k,l}}$ quiver modules. \ } At
a non-symmetric vertex $v=(q,m)_n\in\quiver_0(k,l)$, we use the
partial isometries ${}^nT_{q,m}$ above to construct the operators
\beq
{}^n\phi^+_{q,m}\={}^{n\pm1}T_{q+1,m+3}\,{}^nT_{q,m}{}^\dag \ , \quad
{}^n\phi^-_{q,m}\={}^{n\pm1}T_{q-1,m+3}\,{}^nT_{q,m}{}^\dag \ , \quad
{}^n\phi^0_{q,m}\=
{}^{n}T_{q+2,m}\,{}^nT_{q,m}{}^\dag \ ,
\label{nonsymphiansatz}\eeq
where the values $n\pm1$ depend on the particular weight vectors
$(q,m)_n$ as in (\ref{Q3arrows}). With this ansatz one has the linear
holomorphic relations
\beq
{}^n\phi^+_{q,m}\={}^n\phi^-_{q+2,m}\,{}^n\phi^0_{q,m} \qquad
\mbox{and} \qquad {}^n\phi^+_{q,m}\={}^{n\pm1}\phi^0_{q-1,m+3}\,
{}^n\phi^-_{q,m}
\label{linholweak}\eeq
which together imply (\ref{nonsymCkllin1}) and (\ref{nonsymCkllin2}),
and one also has the quadratic holomorphic relations
(\ref{nonsymCklquad1})--(\ref{nonsymCklquad3}). Moreover, one
easily checks the linear non-holomorphic relations
\bea
{}^n\phi^-_{q,m}\={}^n\phi^+_{q-2,m}\,{}^n\phi^0_{q-2,m}{}^\dag \qquad
&\mbox{and}& \qquad {}^n\phi^-_{q,m}\={}^{n\pm1}\phi^0_{q-1,m+3}{}^\dag\,
{}^n\phi^+_{q,m} \ , \nonumber\\[4pt]
{}^n\phi^0_{q,m}\={}^n\phi^-_{q+2,m}{}^\dag\,{}^n\phi^+_{q,m} \qquad
&\mbox{and}& \qquad {}^n\phi^0_{q,m}\={}^{n\pm1}\phi^+_{q+1,m-3}\,
{}^{n\pm1}\phi^-_{q+1,m-3}{}^\dag
\label{linnonholrels}\eea
as well as the quadratic non-holomorphic relations
\bea
{}^n\phi^-_{q,m}\,{}^n\phi^0_{q,m}{}^\dag&=&{}^{n\pm1}
\phi_{q-1,m+3}^0{}^\dag\,{}^n\phi^-_{q+2,m} \ , \nonumber\\[4pt]
{}^{n\pm1}\phi^+_{q,m}\,{}^{n\pm1}\phi^-_{q,m}{}^\dag&=&
{}^n\phi^-_{q+1,m+3}{}^\dag\,{}^n\phi^+_{q-1,m+3} \ , \nonumber\\[4pt]
{}^n\phi_{q,m}^+\,{}^n\phi^0_{q,m}{}^\dag&=&
{}^{n\pm1}\phi^0_{q+1,m+3}{}^\dag\,{}^n\phi^+_{q+2,m} \ .
\label{quadnonholrels}\eea
As the covariant constancy equations
(\ref{nonsymCklXphi1})--(\ref{nonsymCklXphi3}) are also easily
verified, it follows again that for our ansatz all off-diagonal field
strength matrix elements (\ref{Q3cfpCklrep})--(\ref{Q3cfnonholrel3})
vanish.

The reduction of the Yang-Mills energy functional
(\ref{EYMgen}) is given by substituting (\ref{MdQ3curvdiagCklrep}) in
the basis $\{\gamma^a\wedge\gammab^b\}$ of $(1,1)$-forms on $Q_3$ and
using the K\"ahler metric given by (\ref{KahlerQ3}) to get
\bea
E_{\rm YM}&=&2~\Pf(2\pi\,\theta)~\vol(Q_3)\,
\sum_{(q,m)_n\in\quiver_0(k,l)}\,\Tr^{~}_{{}^n\underline{V}\,_{q,m}
\otimes\underline{(q,m)_n}\otimes\Hcal}\bigg[\,\sum_{a,b=1}^d\,
\big|{}_nF_{a\bb}^{q,m}\big|^2 \label{EYMnonsymdef}
\\ && \hspace{5cm}
+\,\mbox{$\frac43$}\,\Big(\mbox{$\frac14$}\,\big|
{}_n\cf_{1\bar1}^{q,m\,;\,q,m}\big|^2+\big|
{}_n\cf_{2\bar2}^{q,m\,;\,q,m}\big|^2+\big|
{}_n\cf_{3\bar3}^{q,m\,;\,q,m}\big|^2\Big)\bigg] \ . \nonumber
\eea
We use the explicit expressions (\ref{Fabvcomps}),
(\ref{gammapqmdef})--(\ref{gammalincomb}) and (\ref{CGcoeffs}), along
with the projector identities
\bea
\Idd_{{}^np_{q,m}}-{}^n\phi_{q,m}^\pm{}^\dag\,
{}^n\phi_{q,m}^\pm&=&{}^nP_{q,m}\=\Idd_{{}^np_{q,m}}-{}^n
\phi_{q,m}^0{}^\dag\,{}^n\phi_{q,m}^0 \ , \nonumber\\[4pt]
\Idd_{{}^np_{q,m}}-{}^{n\pm1}\phi_{q-1,m-3}^+\,
{}^{n\pm1}\phi_{q-1,m-3}^+{}^\dag&=&{}^nP_{q,m}\=
\Idd_{{}^np_{q,m}}-{}^{n\pm1}\phi_{q+1,m-3}^-\,
{}^{n\pm1}\phi_{q+1,m-3}^-{}^\dag \ , \nonumber\\[4pt]
\Idd_{{}^np_{q,m}}-{}^n
\phi_{q-2,m}^0\,{}^n\phi_{q-2,m}^0{}^\dag&=&{}^nP_{q,m} \ .
\label{nonsymprojids}\eea
Using (\ref{Izrep}), (\ref{Yrep}) and the identity
(\ref{nonsymlambdaid}), we find the curvature components
\beq
\cf_{1\bar1}\=-\mbox{$\frac12$}\,\Pcal\,(H_{\alpha_1}+H_{\alpha_2}) \
, \qquad
\cf_{2\bar2}\=\mbox{$\frac12$}\,\Pcal\,(H_{\alpha_1}-H_{\alpha_2})
\qquad \mbox{and} \qquad \cf_{3\bar3}\=-\Pcal\,H_{\alpha_1} \ .
\label{Q3F11F22F33}\eeq
In this way we arrive finally at the Yang-Mills energy
\beq
E_{\rm YM}=\mbox{$\frac12$}~\Pf(2\pi\,\theta)~\vol(Q_3)\,
\sum_{(q,m)_n\in\quiver_0(k,l)}\,\Tr^{~}_{{}^n\underline{V}\,_{q,m}
\otimes\Hcal}\big(\varepsilon_{q,m}~
{}^nP_{q,m}\big) \ ,
\label{EYMnonsymfinal}\eeq
where
\beq
\varepsilon_{q,m}=\sum_{a=1}^d\,\frac1{(\theta^a)^2}+\mbox{$\frac43$}\,
\big(q^2+q_U^2+\mbox{$\frac14$}\,q_V^2\big)
\label{nonsymenergynm}\eeq
is the finite energy density at the non-symmetric vertex $v=(q,m)_n$,
and we have introduced the additional charges
\beq
q_U~:=~-\mbox{$\frac12$}\,(q-m) \qquad \mbox{and} \qquad
q_V~:=~\mbox{$\frac12$}\,(q+m) \ .
\label{qUdefs}\eeq

The meaning of the new charges (\ref{qUdefs}) can be understood from
the concept of ``$U$-spin'', an electromagnetic analog of
isospin. Recall that the quantum number $q\in\Z$ is twice the third
component of isospin in the subgroup $\su\subset\sut$ generated by
$E_{\pm\,\alpha_1},H_{\alpha_1}$, with $E_{\alpha_1}$ acting by shifts
$q\mapsto q+2$ and leaving the hypercharge quantum number $m$
invariant. The $U$-spin subgroup of $\sut$ is the $\su$ subgroup
generated by the operators
$E_{\pm\,\alpha_2},\frac14\,(H_{\alpha_2}-H_{\alpha_1})$. All states in
a $U$-spin multiplet have the same electric charge eigenvalue
$Y_U:=-\frac12\,(q+\frac m3)$, the $U$-spin analog of
hypercharge. Thus the magnetic charge $q_U\in\Z$ is twice the third
component of $U$-spin. The Weyl subgroup $S_3\subset\sut$ takes
isospin into $U$-spin, and the explicit transformation of states in
the Biedenharn basis for $\underline{C}^{k,l}$ can be found
in~\cite{MP1}. The operator $E_{\alpha_2}$ shifts $q_U\mapsto q_U+2$
and leaves $m_U:=3Y_U$ invariant. Likewise, the ``$V$-spin''
subgroup $\su\subset\sut$ is generated by
$E_{\pm\,(\alpha_1+\alpha_2)},\frac14\,(H_{\alpha_1}+H_{\alpha_2})$,
with associated quantum numbers $q_V\in\Z$ shifted by
$q_V\mapsto q_V+2$ and $m_V:=\frac32\,(q-\frac m3)$
invariant under the action of $E_{\alpha_1+\alpha_2}$.

Thus the vertex energy (\ref{nonsymenergynm}) contains that of
\emph{three} non-interacting charges on $Q_3$, one for each arrow of
the non-symmetric quiver. The sum of these charges is the total
magnetic charge
\beq
q+q_U+q_V=q+m
\label{totalqmcharge}\eeq
at the vertex $v=(q,m)_n$. Note that the energy density is independent
of the degeneracy label~$n$. The energy of the $V$-spin charge is down
by $\frac14$ due to the area of the embedded two-cycle $\CP^1$ dual to
the $(1,1)$-form $\gamma^1\wedge\gammab^1$ on $Q_3$ (see
(\ref{KahlerQ3})). All of this is qualitatively similar to the quiver
energies associated to the symmetric space
$\CP^1\times\CP^1$~\cite{LPS2}, which also carry abelian vertex
charges but only two arrows per vertex. For the present solutions the
noncommutative vortex number is now
\beq
N=\sum_{(q,m)_n\in\quiver_0(k,l)}\,{}^nN_{q,m} \ ,
\label{nonsymvortexnum}\eeq
and the verification of the Yang-Mills equations on
$X=\C_\theta^d\times Q_3$ proceeds exactly as outlined before in the
basis generated by the canonical one-forms $\gamma^a$ and $\gammab^a$
on $Q_3$. As expected from the construction of
Section~\ref{quivbungen}, the sum over monopole charges $q$ at fixed
isospin $n$, mimicking the integration over the $\C P^1$ fibres of the
twistor bundle (\ref{Q3CP2rel}), maps the non-symmetric energy density
(\ref{nonsymenergynm}) onto the symmetric one (\ref{symenergynm}) via 
the summation identity
\beq
\sum_{q\in\{-n+2j\}_{j=0}^n}\,q^2=\mbox{$\frac13$}\,{n\,(n+1)\,(n+2)}
\ .
\label{sumqid}\eeq

\subsection{BPS solutions\label{BPSsols}}

The solutions obtained thus far are generically unstable non-BPS
solutions of the Yang-Mills equations on $X=\C_\theta^d\times G/H$,
due to the presence of non-trivial vacua at more than one quiver
vertex in $\quiver_0(k,l)$. Using the identities (\ref{lambdasymid})
and (\ref{nonsymlambdaid}) it is easy to see from the projector
equations (\ref{Fabvcomps}), (\ref{symphiprodrels}) and
(\ref{nonsymprojids}) that the ans\"atze of Section~\ref{Finitesols}
above are incompatible with the BPS equations (\ref{symCklmods}) and
(\ref{nonsymCklmods}) if more than one projector $P_v$ is non-zero. To
obtain non-trivial BPS solutions we need a more specialized ansatz
inside our previous ones. Let us set $p_v=r$ for all
$v\neq v_0$, where $v_0\in\quiver_0(k,l)$ is a distinguished vertex of
the pertinent quiver. A natural choice for $v_0$ comes from the choice
of trivial $\su$ spins $j_\pm=0$ in (\ref{jpmSU2spins}) and
(\ref{mnqjpm}), for which $n=q=0$ and $m=-2(k-l)$. The gauge group is
now 
\beq
{\cal G}^{k,l}_{\rm BPS}=\urm(p_{v_0})\times
\urm(r)^{\#\quiver_0(k,l)-1} \ ,
\label{gaugegpBPS}\eeq
where the number of quiver vertices $\#\quiver_0(k,l)=(k+1)\,(l+1)$ in
the symmetric case while $\#\quiver_0(k,l)=d^{k,l}$ in the
non-symmetric case is the dimension (\ref{dimkl}) of the irreducible
$\sut$ representation $\underline{C}^{k,l}$.

At each vertex $v\neq v_0$, we take the vacuum solution with $A_a^v=0$
and trivial module morphisms $\phi_{v,\Phi(v)}=\Idd_r$. At $v=v_0$ we
take the partial isometry ans\"atze of Section~\ref{Finitesols}
above. In both symmetric and non-symmetric cases, the vortex number of
this solution is $N=N_{v_0}$, and the noncommutative vortex equations
require a non-trivial relation
\beq
\sum_{a=1}^d\,\frac1{\theta^a}=-4(k-l)
\label{thetacosetrel}\eeq
between the noncommutative geometry and the geometry of the coset
space $ G/H$. By putting $P_v=0$ for all $v\neq v_0$ in
(\ref{EYMsymfinal}) and (\ref{EYMnonsymfinal}), from
(\ref{thetacosetrel}) one finds that the BPS energies are proportional
to the vortex number $N$ with proportionality constant dependent only
on the geometry of the noncommutative space
$X=\C_\theta^d\times G/H$. Note that the constraint
(\ref{thetacosetrel}) requires $k<l$. For $k\geq l$ one can
straightforwardly modify the constructions of this section with some
$\theta^a<0$.

\bigskip

\section{Topological charges\label{Dcharges}}

\noindent
The class of solutions constructed in the previous section only exist
when the system is subjected to a noncommutative deformation. In this
final section we will compute their topological charges in order to
further unravel their physical meaning. We will do this in three
independent but equivalent ways. Firstly, we shall calculate the
instanton charges directly in the original Yang-Mills gauge theory on
$X=\C_\theta^d\times G/H$. Secondly, we will demonstrate how the
topological charges can be elegantly understood in terms of the
K-theory class determined by the partial isometries which parametrize
the noncommutative instantons. Finally, we will show that these
charges also coincide with the Euler characteristic of a certain
complex which is canonically associated to the vortex solutions of the
quiver gauge theory on $M_{2d}=\C_\theta^d$. These latter two
constructions illustrate a novel interpretation of our classical field
configurations in terms of D-branes.

Although our treatment of the DUY equations in general dimensions
primarily relates to the stability of holomorphic vector bundles over
K\"ahler manifolds, for appropriate values of~$d$ one has contact with
the D-brane picture in bosonic or supersymmetric string theory.
The standard interpretation of the noncommutative configurations at
each quiver vertex is that of an unstable system of D0-branes inside
D$(2d)$-branes (this is done in detail in~\cite{DMR1}), in the
presence of a constant $B$-field background and in the Seiberg-Witten
scaling limit. This is done by calculating both the tension (first term
in (\ref{symenergynm}) and~(\ref{nonsymenergynm})) and the spectrum of
small fluctuations about each instanton configuration, in a (boundary)
conformal field theory analysis. The difference in our case is that
a system of D-branes excited at only a single vertex is stable on its own
(consistently with what is found in Section~5.3). This is due to the
monopole/instanton fields carried by the branes via the dimensional
reduction, whose topological charges prevent a decay to the closed-string
vacuum. It is a generalization of the flux stabilization mechanism of
Bachas-Douglas-Schweigert~\cite{BDS}.

The D-brane interpretation works in principle both in bosonic string theory 
and in type IIA superstring theory. The latter is probably more desirable, 
as only in that case is there a notion of Ramond-Ramond charge, as is
implicit in our interpretation. As for the fermionic sector, we are
following the standard prescription in obtaining BPS solutions -- we set
all fermionic and auxiliary fields to zero. There is no problem in adding
in fermions, supersymmetric or not. The D-branes are thought to wrap
topologically trivial worldvolumes~$\R^{2d}$ (only for those and not
for general K\"ahler manifolds we put forward the D-brane interpretation)
in a \emph{constant} $B$-field background, hence the Freed-Witten anomaly
automatically vanishes.

\subsection{Chern-Weil invariants\label{Chernnums}}

We will first present a field theoretic derivation of the topological
charges of the multi-instanton solutions in the original Yang-Mills
gauge theory on $X=\C_\theta^d\times G/H$. Given the
$G$-equivariant vector bundle (\ref{Ecalklisotopical}) over the
K\"ahler manifold $X$, we may construct various topological charges
classifying our gauge field configurations by taking products in
complimentary degrees of the K\"ahler two-form (\ref{kahlerX}) and of
the curvature two-form (\ref{cfca}) representing the usual Chern
characteristic classes of the bundle $\Ecal^{k,l}$. For each
$j=1,\dots,d+\frac{d_H}2$, we define Chern-Weil topological invariants
of $\Ecal^{k,l}$ by
\beq
\chern_j\big(\Ecal^{k,l}\big)= \frac{1}{j!}\ \Bigl(\frac{\im}{2\pi}
\Bigr)^j~{\rm Pf}(2\pi\,\th)~\int_{ G/H}\,
\Tr^{~}_{\underline{V}^{k,l}\otimes{\cal H}}\big(
{\cf}^j\big) \wedge
\frac{\Omega^{d+d_H/2-j}}{(d+d_H/2-j)!} \ .
\label{ChjEdef}\eeq
The quantity $\chern_1(\Ecal^{k,l})$ is proportional to the degree of
the bundle $\Ecal^{k,l}$, while $\chern_2(\Ecal^{k,l})$ is
proportional to the Yang-Mills action (\ref{EYMgen}) in the BPS
limit~\cite{A-CG-P2,LPS2}. Many of these charges will in fact vanish
for topological reasons.

In this section we will consider only the top degree $j=d+\frac{d_H}2$
Chern number $\chern_{d+d_H/2}(\Ecal^{k,l})$, and refer to this as the
instanton charge $Q$ for brevity. Explicitly, one has
\beq
Q:=\frac{1}{\big(d+{d_H}/2\big)!}\ \Bigl(\frac{\im}{2\pi}
\Bigr)^{d+d_H/2}~
{\rm Pf}(2\pi\,\th)~\int_{ G/H}\,
\Tr^{~}_{\underline{V}^{k,l}\otimes{\cal H}}\big(\,
\underbrace{{\cf}\wedge\cdots\wedge {\cf}}_{d+d_H/2}\,\big) \ .
\label{topchargeChdef}\eeq
To compute the integral (\ref{topchargeChdef}), we note that generally
for the gauge field configurations (\ref{Xavansatz}),
(\ref{Fabvcomps}), the non-vanishing components of the field strength
tensor (\ref{cfca}) on $X$ along $\C_\theta^d$ are given by
\beq
{\cf}^{~}_{2a-1~ 2a} \= 2\im\, {\cf}^{~}_{a\ab} \=
-\frac{\im}{\th^a}~\Pcal \ ,
\label{cFalongC}\eeq
where $\Pcal$ is the projector (\ref{Pcaldef}). The remaining details
of the computation depend explicitly on the particular quiver.

\bigskip

\noindent
{\bf Symmetric $\mbf{\underline{C}^{k,l}}$ quiver charges. \ } By
working as before in the basis $\{\beta^i\wedge\betab^j\}$ of
$(1,1)$-forms on $\C P^2$, the instanton density in
(\ref{topchargeChdef}) may be calculated as
\bea
&& \mbox{$\frac1{(d+2)!}$}\,\epsilon^{\mu_1\cdots\mu_{2d+4}}\,
\cf_{\mu_1\mu_2}\cdots\cf_{\mu_{2d+3}\mu_{2d+4}}\= (-\im)^2\,\big(\cf_{12}\,
\cf_{34}\cdots\cf_{2d-1~2d}\big)\,\big(\cf_{1\bar1}\,\cf_{2\bar2}-
|\cf_{1\bar2}|^2\big) \nonumber \\[4pt] && \hspace{2cm} 
\= \Big(\,\frac{(-\im)^{d+2}}{{\rm Pf}(\theta)}\,\Big)~\Pcal~\Big(
-\mbox{$\frac14$}\,\big(H_{\alpha_1}^2-H_{\alpha_2}^2\big)-
\mbox{$\frac12$}\,\big(E_{\alpha_1}\,E_{-\alpha_1}+E_{-\alpha_1}\,
E_{\alpha_1}\big)\Big) \label{CP2instdensity}
\eea
after substituting the field strength components (\ref{cFalongC}),
(\ref{CP2F11F22}) and (\ref{CP2F12}). After tracing over the
representation spaces $\underline{(n,m)}\cong\C^{n+1}$ using
(\ref{Izrep})--(\ref{Yrep}) and symmetry of the isospin summation over
$q$, one finds
\beq
Q=\Big(\frac{1}{2\pi}\Big)^2~{\rm vol}\big(\C P^2\big)~
\sum_{(n,m)\in\quiver_0(k,l)}\,\mbox{$\frac14$}\,(n+1)\,\big(m^2-
n\,(n+2)\big)~\Tr^{~}_{\underline{V}\,_{n,m}\otimes
\Hcal}\big(P_{n,m}\big) \ ,
\label{CP2Qtrace}\eeq
where $\vol(\C P^2):=\int_{\C
  P^2}\,\beta^1\wedge\betab^1\wedge \beta^2\wedge\betab^2$. The
normalization can be fixed by recalling from
Section~\ref{Quivbunfund} that the K\"ahler two-form (\ref{KahlerCP2})
on $\C P^2$ determines the generator $[\eta]$ of the integer
cohomology ring $\HQ^{2\bullet}(\C P^2;\Z)$ through
\beq
\eta=\frac{\omega_{\C P^2}}{2\pi} \ ,
\label{CP2etagen}\eeq
with the intersection numbers
\beq
\int_{\C P^1}\,\eta\=1\=\int_{\C P^2}\,\eta\wedge\eta
\label{CP2intnums}\eeq
for any linearly embedded projective line $\C P^1\subset\C P^2$. This
fixes $\vol(\C P^2)=2(2\pi)^2$, and the topological charge is
finally given by
\beq
Q=\sum_{(n,m)\in\quiver_0(k,l)}\,\mbox{$\frac12$}\,(n+1)\,\big(m^2-
n\,(n+2)\big)\,N_{n,m} \ .
\label{CP2Qfinal}\eeq
Note that (\ref{CP2Qfinal}) is indeed integer-valued as $(n,m)$ have
the same parity.

\bigskip

\noindent
{\bf Non-symmetric $\mbf{\underline{C}^{k,l}}$ quiver charges. \ } In
the $G$-equivariant basis of $(1,1)$-forms on $Q_3$, the instanton
density in (\ref{topchargeChdef}) is given by
\bea
\mbox{$\frac1{(d+3)!}$}\,\epsilon^{\mu_1\cdots\mu_{2d+6}}\,
\cf_{\mu_1\mu_2}\cdots\cf_{\mu_{2d+5}\mu_{2d+6}}&=& (-\im)^3\,\big(\cf_{12}\,
\cf_{34}\cdots\cf_{2d-1~2d}\big)\,\big(\cf_{1\bar1}\,\cf_{2\bar2}\,
\cf_{3\bar3}\big) \nonumber \\[4pt] 
&=& \Big(\,\frac{(-\im)^{d+3}}{{\rm Pf}(\theta)}\,\Big)~\Pcal~\Big(
\mbox{$\frac14$}\,\big(H_{\alpha_1}^2-H_{\alpha_2}^2\big)\,
H_{\alpha_1}\Big) 
\label{Q3instdensity}\eea
after substituting the field strength components (\ref{cFalongC}) and
(\ref{Q3F11F22F33}). After tracing using (\ref{Izrep}), (\ref{Yrep})
and (\ref{qUdefs}), one finds 
\beq
Q=\Big(\frac{1}{2\pi}\Big)^3~{\rm vol}(Q_3)~
\sum_{(q,m)_n\in\quiver_0(k,l)}\,q\,q_U\,q_V~
\Tr^{~}_{{}^n\underline{V}\,_{q,m}\otimes
\Hcal}\big({}^nP_{q,m}\big) \ ,
\label{Q3Qtrace}\eeq
where
\beq
\vol(Q_3):=\int_{Q_3}~\bigwedge_{a=1}^3\,\gamma^a\wedge\gammab^a \ . 
\label{volQ3def}\eeq
The normalization can again be determined by examining the generators
of the integer cohomology ring of the space $Q_3$.

For this, it is convenient to use the description of $Q_3$ as the
twistor fibration (\ref{Q3CP2rel}). Since this is a sphere bundle, we
may write down the corresponding Gysin long exact sequence relating
the cohomology groups of $Q_3$ and $\C P^2$. Since the cohomology of
the projective plane is concentrated in even degree, it implies that
all odd degree cohomology groups of $Q_3$ vanish and the even degree
ones are determined by the short exact sequences
\beq
0~\longrightarrow~\HQ^{2j}\big(\C P^2\,;\,\Z\big)~\xrightarrow{\pi^*}~
\HQ^{2j}(Q_3;\Z)~\xrightarrow{\pi_*}~\HQ^{2j-2}\big(\C
P^2\,;\,\Z\big)~\longrightarrow~0
\label{Q3CP2exseqs}\eeq
for $j=0,1,2,3$, where $\pi^*$ is the usual pullback and $\pi_*$ is
integration along the $\C P^1$ fibre. Setting $j=3$ shows that the
generator $[\tau]$ of $\HQ^6(Q_3;\Z)\cong\HQ^4(\C P^2;\Z)$ is
given in terms of the generating element $[\eta]$ of the cohomology
ring of $\C P^2$ through
\beq
\pi_*(\tau)=\eta\wedge\eta \ ,
\label{tauetarel}\eeq
with
\beq
\int_{Q_3}\,\tau=1 \ .
\label{tauintnum}\eeq
Recall from Section~\ref{Quivbunfund} that
$\HQ^2(Q_3;\Z)=\Z[\sigma_1]\oplus\Z[\sigma_2]$ where
\beq
\sigma_i=\mbox{$\frac\im{2\pi}$}\,g_i \ .
\label{sigmaigidef}\eeq
Setting $j=1$ in (\ref{Q3CP2exseqs}) shows that a consistent choice of
basis is given by setting $\pi^*(\eta)=\sigma_1$, $\pi_*(\sigma_1)=0$
and $\pi_*(\sigma_2)=1$, so that $\pi_*$ can be represented by
integration over the two-cycle $\C P_{(2)}^1\subset Q_3$. Then from
(\ref{Q3CP2exseqs}) with $j=2$ it follows that
$\pi^*(\eta\wedge\eta)=\sigma_1\wedge\sigma_1$ and
$\sigma_1\wedge\sigma_2$ generate $\HQ^4(Q_3;\Z)\cong\Z\oplus\Z$, with
$\pi_*(\sigma_1\wedge\sigma_1)=0$ and
$\pi_*(\sigma_1\wedge\sigma_2)=\eta$. Hence by (\ref{tauetarel}) we
have
\beq
\tau=\sigma_1\wedge\sigma_1\wedge\sigma_2 \ .
\label{tausigmarel}\eeq

On substituting (\ref{sigmaigidef}), (\ref{fgQ3rels}), and
(\ref{f1f2eqs}) into (\ref{tausigmarel}) and (\ref{tauintnum}), we fix
the normalization (\ref{volQ3def}) as $\vol(Q_3)=2(2\pi)^3$. For
the topological charge (\ref{Q3Qtrace}) we thus find
\beq
Q\=-2\sum_{(q,m)_n\in\quiver_0(k,l)}\,q\,q_U\,q_V~{}^nN_{q,m} \=
\sum_{(q,m)_n\in\quiver_0(k,l)}\,\mbox{$\frac12$}\,q\,
\big(q^2-m^2\big)~{}^nN_{q,m} \ .
\label{Q3Qfinal}\eeq
Now $Q\in2\,\Z$ since the integers $(q,m)_n$ have the same parity.

\subsection{K-theory charges\label{Kcharge}}

We will now describe a natural interpretation of the instanton charges
(\ref{CP2Qfinal}) and (\ref{Q3Qfinal}) in terms of equivariant
K-theory and brane-antibrane annihilation on the manifold
$M_{2d}$, using the explicit solutions we have constructed in
Section~\ref{Finitesols}. The main physical idea is that the
instantons on $X=\C_\theta^d\times G/H$ can be interpreted as a
system of $p$ coincident D$(2d+d_H)$-branes wrapping the coset space
$ G/H$, with the $G$-equivariance condition splitting the rank
$p$ as in (\ref{rankdimv}) and wrapping $ G/H$ with instanton and
monopole fields. After dimensional reduction, we are left with an
equivalent system of D$(2d)$ branes and antibranes carrying the
appropriate topological quantum numbers, which stabilize the
marginally bound space-filling brane configurations. On the subset of
branes at vertex $v\in\quiver_0(k,l)$ there lives a Chan-Paton gauge
potential $A^v\in\End(E_{p_v})$, and neighbouring subsets are
connected by Higgs fields
$\phi_{v,\Phi(v)}\in\Hom(E_{p_v},E_{p_{\Phi(v)}})$
corresponding to massless open string excitations. The BPS vortex
configurations of Section~\ref{BPSsols} are stable bound states of
$M$ D0-branes inside the system of D$(2d)$-branes, where $M=(\#
\quiver_0(k,l)-1)\,N$. We will now explain how a particular K-theory
construction naturally leads to this physical picture.

The noncommutative instantons are classified by the $G$-equivariant
K-theory group of $X$ which may be computed via the equivariant
excision theorem to get
\beq
\K_G(X)\=\K_G\big(G\times_HM_{2d}\big)~\cong~
\K_H(M_{2d}) \ .
\label{SU3eqKXexcision}\eeq
Since $H$ acts trivially on $M_{2d}$, this group reduces to the
product
\beq
\K_G(X)\cong\K(M_{2d})\otimes \rep(H)
\label{SU3eqKXprod}\eeq
of the ordinary K-theory of $M_{2d}$ with the representation ring of
the subgroup $H$. The K-theory charge is thus computed by finding
appropriate representatives for each of these factors in the quiver
gauge theory.

We start with the second factor. Collapsing $M_{2d}$ to a point in
(\ref{SU3eqKXprod}) shows that 
\beq
\rep(H)\cong\K_G\big( G/H\big) \ .
\label{repHcoset}\eeq
It follows that the classes in $\rep(H)$ may be constructed from
appropriate representatives of the homogeneous vector bundles
(\ref{indhermbungen}), and (\ref{SU3eqKXprod}) is equivalent to the
isotopical bundle decompositions given by (\ref{Eklisotopical}) and
(\ref{Ecalklisotopical}). Under the isomorphism
(\ref{Omega01Hcal}), we can use the index class of the equivariant
Dirac operator $\Dirac$ in the background of quiver gauge fields on
the homogeneous space $ G/H$ which live in the second component of
the subspaces (\ref{diagsubsp}). Denote by $\Dirac_v$,
$v\in\quiver_0(k,l)$ the spin$^c$ Dirac operator in the background
$G$-equivariant gauge fields corresponding to the irreducible
$H$-module $\underline{v}$ (see e.g.~\cite{DolanNash}--\cite{Dolan}
for explicit constructions). In a suitable basis, there are chiral
decompositions
\beq
\Dirac\=\sum_{v\in\quiver_0(k,l)}\,\Dirac_v\otimes\Pi_v\=
\sum_{v\in\quiver_0(k,l)}\,\begin{pmatrix}
0 & \Dirac_v^+ \\ \Dirac_v^- & 0 \end{pmatrix}\otimes\Pi_v
\label{Diracchiraldecomp}\eeq
into twisted Dolbeault-Dirac operators $\Dirac_v^\pm$.

For each vertex $v$, there is an index class
\beq
\underline{\ind}(\Dirac_v):=\ker\big(\Dirac_v^+\big)\ominus\ker
\big(\Dirac_v^-\big)
\label{indexclassdef}\eeq
in the $G$-equivariant K-theory of $ G/H$, whose virtual
dimension is the index of the Dirac operator $\Dirac_v$. In analogy to
the quiver gauge theories based on the symmetric space $\C
P^1$~\cite{PS1,LPS2}, we will call the gauge field excitations on
$M_{2d}$ associated to vertex $v$ a \emph{brane} if
$\ker(\Dirac_v^-)=\{0\}$ and an \emph{antibrane} if
$\ker(\Dirac_v^+)=\{0\}$. This associates K-theory charges on $ G/H$
to D-brane charges on $M_{2d}$. Note that these are \emph{not} the
same as the topological charges associated to the gauge field
configurations of the homogeneous vector bundle
(\ref{indhermbungen}). By the Atiyah-Singer index theorem
\beq
\ind(\Dirac_v)=\int_{ G/H}\,\ch(\Vcal_v\otimes\Lcal_c)
\wedge\widehat{A}\big( G/H\big) \ ,
\label{indexthm}\eeq
where the complex line bundle $\Lcal_c\to G/H$ determines the spin$^c$
structure. Hence the K-theory charge generally couples topological
charges with both the $\uo$ charge $c_1(\Lcal_c)$ of the spin$^c$
fermion and the Pontrjagin numbers of the tangent bundle to the coset
space $ G/H$. Such curvature couplings are a standard feature of
D-brane charge. (In the $\C P^1$ case, the two types of charges agree
as the index of the Dirac operator coincides with the first Chern
number of the $m$-monopole line bundle over $\C P^1$.) The class
(\ref{indexclassdef}) represents the symmetric spinors which survive
the $G$-invariant dimensional reduction from $X$ to $M_{2d}$ in the
$G$-equivariant ABS construction of K-theory classes on
$\R^{2d}$~\cite{PS1,LPS2}. We will denote the respective disjoint
vertex subsets corresponding to branes and antibranes by
$\quiver_0^\pm\subset\quiver_0(k,l)$.

Let us now describe the first factor in (\ref{SU3eqKXprod}). The
Toeplitz operators $T_v$ obeying (\ref{partialisom}) determine an
index class
\beq
\underline{\ind}\big(T_v^\dag\big)~:=~
\ker\big(T_v^\dag\big)\ominus\ker(T_v)\=
\ker\big(T_v^\dag\big)~\cong~\C^{N_v}
\label{IndexTvclass}\eeq
in the K-theory group $\K(\C_\theta^d)$. The rank of the corresponding
projector $P_v$ is the index of $T_v^\dag$,
$N_v=\ind\big(T_v^\dag\big)$, and $P_v$ projects the noncommutative
quiver bundle $\underline{V}^{k,l}\otimes\Hcal$ onto the
finite-dimensional quiver module
\beq
\underline{T}:=\bigoplus_{v\in\quiver_0(k,l)}\,\underline{\ind}
\big(T_v^\dag\big) \ .
\label{Tvquivmodule}\eeq
Using the chirality grading introduced above, we define a
$\Z_2$-grading of the fibre space (\ref{Vklmodule}) by
\beq
\underline{V}^{k,l}\=\underline{V}_{\,+}\,\oplus\,
\underline{V}_{\,-} \qquad \mbox{with} \quad
\underline{V}_{\,\pm}~:=~\bigoplus_{v\in\quiver_0^\pm}\,
\underline{V}\,_v\otimes\,\underline{v} \ .
\label{Vklmodulegrading}\eeq
Using the explicit instanton solutions of Section~\ref{Finitesols}, we
will demonstrate how to construct odd operators
\beq
\mbf T\,:\,\underline{V}_{\,+}\otimes\Hcal~\longrightarrow~
\underline{V}_{\,-}\otimes\Hcal \qquad \mbox{with} \quad
\mbf T^2\=0
\label{tachyoncomplex}\eeq
from the Toeplitz operators $T_v$.

This defines a two-term complex corresponding to the basic
brane-antibrane system with tachyon field $\mbf T$. Its cohomology is
a representative of the K-theory Euler class generating
$\K(\C_\theta^d)$. The single brane-antibrane system is obtained by a
``folding'' of the component branes and antibranes at the vertices of
the quiver, as dictated by $G$-equivariance. A description of the
moduli involved in this folding process will be given in
Section~\ref{ERform} below. The tachyon field (\ref{tachyoncomplex})
has an isotopical decomposition $\mbf T=\sum_{v\in\quiver_0(k,l)}\,\mbf
T_v\otimes\Pi_v$, and putting everything together the virtual module
\beq
\underline{\mbf\Tcal}:=
\bigoplus_{v\in\quiver_0(k,l)}\,\underline{\ind}(\mbf T_v)\,
\otimes\,\ker(\Dirac_v)
\label{virmodfinal}\eeq
is the K-theory class in (\ref{SU3eqKXprod}) we are looking for. We
will see that the associated \emph{K-theory} charge on $M_{2d}$,
i.e. the virtual dimension $\Qcal$ of this module, is canonically
related to the \emph{topological} charge $Q$ on $X$ computed in
Section~\ref{Chernnums} above, i.e. the topological charges of the
instanton gauge fields before the dimensional reduction.

The crux of the construction, in addition to an explicit determination
of the Dirac index, is thus an explicit model for the tachyon operator
(\ref{tachyoncomplex}). It naturally appears in a graded connection
formalism~\cite{PS1,LPS2} which is a rewriting of the equivariant
gauge theory on $X$ as an ordinary Yang-Mills gauge theory on the
corresponding quiver bundle over $M_{2d}$, appropriate to its
interpretation in terms of brane-antibrane systems on $M_{2d}$. With
respect to the $\Z_2$-grading (\ref{Vklmodulegrading}), the even
``diagonal'' parts of the graded connections are built from the gauge
connection one-forms as
\beq
\mbf A=\sum_{v\in\quiver_0(k,l)}\,A^v\otimes\Pi_v
\label{evengradedconn}\eeq
in $\bigoplus_{v\in\quiver_0(k,l)}\,\Omega^{0,1}(\End(E_{p_v}))$,
along with similar formulas for the equivariant gauge potentials in
$\bigoplus_{v\in\quiver_0(k,l)}\,\Omega^{0,1}(\End(\Vcal_v))^G$. The
odd zero-form components of the graded connections determine the
tachyon fields (\ref{tachyoncomplex}) and are associated with the
``off-diagonal'' subspace of $\Omega^{0,1}(\End(E^{k,l}))^H$ given by
\beq
\bigoplus_{v\in\quiver_0(k,l)}~\bigoplus_{\Phi\in\quiver_1(k,l)}\,
\Omega^0\big(\Hom(E_{p_v},E_{p_{\Phi(v)}})\big)\otimes
\Hom_H\big(\,\underline{v}\,,\,\underline{\Phi(v)}\,\big) \ ,
\label{oddgradedconnsubsp}\eeq
where we have used $H$-equivariance. The details of the construction
again depend on the particular quiver.

\bigskip

\noindent
{\bf Symmetric $\mbf{\underline{C}^{k,l}}$ quiver charges. \ } By
K\"unneth's theorem, the representation ring of the holonomy group
$H=\su\times\uo$ is the product
$\rep(H)=\rep(\su)\otimes\rep(\uo)$. As in the $\C P^1$ cases, using
the isomorphism (\ref{repHcoset}) we can identify the representation
ring of $\uo$ with the formal Laurent polynomial ring
$\Z[\Lcal,\Lcal^\vee\,]$ generated by classes of the monopole line
bundle over $\CP^2$. The class of $\Lcal$ is also tied to the
(reduced) K-theory of the projective plane itself, which is generated
by $\Lcal$ and $\Lcal\otimes\Lcal$. By constructing irreducible
representations of $\su$ in the standard way through symmetrizations
of the fundamental representation $\mbf2$, there is an isomorphism
$\rep(\su)\cong\Z[\mbf2]$ which under (\ref{repHcoset}) can be
identified with the formal polynomial ring in classes of the $\su$
instanton bundle $\Ical\to\CP^2$. The representation ring may thus be
presented as
\beq
\rep(H)\cong\Z[\Ical]\otimes\Z\big[\Lcal\,,\,\Lcal^\vee\,\big] \ .
\label{repringCP2}\eeq

We will represent classes in (\ref{repringCP2}) by using the Dirac
index on $\CP^2$. As we now demonstrate, the index of the spin$^c$
Dirac operator on $\CP^2$ in the background instanton and monopole
fields associated to the representation $\underline{(n,m)}$ of $H$ is
given by
\beq
\ind(\Dirac_{n,m})\=\mbox{$\frac18$}\,(n+1)\,(m+n+1)\,(m-n-1)\=
\mbox{$\frac18$}\,(n+1)\,\big(m^2-n\,(n+2)-1\big) \ .
\label{indexCP2}\eeq
The Dirac spectrum was determined in~\cite{Dolan} in the following
way. By identifying the Lie algebra $\mathfrak{g}^\C=\sltcL$ with the
Lie algebra of complex left-invariant vector fields on the group $G$,
we get a natural action of the universal enveloping algebra
$\mathcal{U}(\mathfrak{g}^\C)$ on ${\rm C}^\infty(G)$. The quadratic
Casimir element
\beq
\mbf C_2=\sum_{\alpha\in\Delta^+}\,\big(E_\alpha\,E_{-\alpha}+
E_{-\alpha}\,E_\alpha\big)+\mbox{$\frac12$}\,\big(H_{\alpha_1}^2+
H_{\alpha_2}^2\big)
\label{quadCasop}\eeq
thereby induces an invariant second order differential operator, which
coincides with the metric Laplace-Beltrami operator on $\C P^2$. Up to
curvature terms, this operator coincides with the square of the Dirac
operator. Its spectrum is given by the quadratic Casimir eigenvalues
\beq
C_2(k,l)=\mbox{$\frac13$}\,\big(k\,(k+3)+l\,(l+3)+k\,l\big)
\label{C2kl}\eeq
in the irreducible $\sut$ representation $\underline{C}^{k,l}$. After
coupling to the background gauge fields, the twisted Laplace-Beltrami
operator can be expressed as a difference of the quadratic Casimir
operators (\ref{quadCasop}) and (\ref{quadCasH}).

Using the construction of Section~\ref{AlgQuiverGen}, the Dirac
kernels may thus be determined by the $G$-module
$\underline{C}^{k_{n,m},l_{n,m}}$ which minimizes the Casimir
invariant (\ref{C2kl}), subject to the constraints (\ref{mnqjpm}) that
$\underline{C}^{k_{n,m},l_{n,m}}$ contain the representation
$\underline{(n,m)}$ of $H=\su\times\uo$. In~\cite{Dolan} it was shown
that the chiral case $\ker(\Dirac_{n,m}^-)=\{0\}$ corresponds to the
configurations with $m^2>n^2$, for which $\ker(\Dirac_{n,m}^+)$ is
isomorphic to the $\sut$ representation
$\underline{C}^{k,l}$ having $|m|=k+2l$, $n=k$ in
(\ref{Q3genvertices}). The dimension (\ref{dimkl}) of this module
coincides with the index (\ref{indexCP2}) after the shift $m\to
m-c_1(T\CP^2)$, where $c_1(T\CP^2)=3$ is the first Chern number of the
tangent bundle over $\CP^2$. This shift arises from the fact that the
chiral symmetric spinors couple only to the $\uo$ part of the spin$^c$
connection from which the Dirac operator is
constructed~\cite{DolanNash,Bal}. The antichiral case
$\ker(\Dirac_{n,m}^+)=\{0\}$ corresponds to $m^2\leq n^2$, for which
$\ker(\Dirac_{n,m}^-)$ is isomorphic to the module
$\underline{C}^{k,l}$ with $|m|=k-l$, $n=k+l$ in
(\ref{Q3genvertices}). The corresponding dimension (\ref{dimkl})
coincides with minus the index (\ref{indexCP2}) after the shift $n\to
n-1$. This shift accounts for the fact that the antichiral symmetric
spinors couple only to the $\su$ part of the spin$^c$ connection. Note
that because of these shifts, the index (\ref{indexCP2}) is
generically fractional. Nevertheless, it will turn out to be the one
appropriate to our K-theory construction.

Next we turn to the construction of the tachyon field
(\ref{tachyoncomplex}). For this, it is useful to recall the mapping
onto the $\su$ spins $j_\pm=j_\pm(n,m)$ defined in
(\ref{jpmSU2spins}). The shifts along vertices induced by the arrows
of the symmetric quiver in these variables take the simple forms
\bea
j_+(n+1,m+3)\=j_+(n,m)+\mbox{$\frac12$} \qquad & \mbox{and} &
\qquad j_+(n-1,m+3)\=j_+(n,m) \ , \nonumber\\[4pt]
j_-(n-1,m+3)\=j_-(n,m)-\mbox{$\frac12$} \qquad & \mbox{and} &
\qquad j_-(n+1,m+3)\=j_-(n,m) \ .
\label{SU2spinshifts}\eea
The vertex redefinition (\ref{jpmSU2spins}) thus orients the symmetric
quiver diagram onto a rectangular quiver of the same type as those
which arise for the symmetric space $\CP^1\times\CP^1$. This will
enable us to adapt some of the constructions of~\cite{LPS2} to the
present case.

Using the Dirac operator analysis above, we decompose the vertex set
$\quiver_0(k,l)$ into the disjoint subsets
\beq
\quiver_0^+\=\big\{(n,m)\in\quiver_0(k,l)~\big|~m^2>n^2\big\}
\qquad \mbox{and} \qquad
\quiver_0^-\=\big\{(n,m)\in\quiver_0(k,l)~\big|~m^2\leq n^2\big\}
\label{symquiv0pm}\eeq
corresponding respectively to positive and negative values of the
index (\ref{indexCP2}). Using the $\su$ spin variables in
(\ref{mnqjpm}), these subsets further split into disjoint unions
$\quiver_0^\pm=\quiver_0^{\pm +}\sqcup\quiver_0^{\pm -}$ with
\bea
\quiver_0^{+\pm}&=&\big\{(n,m)\in\quiver_0^+~\big|~
2j_\pm-j_\mp<\pm\,\mbox{$\frac{k-l}2$}\big\} \ , \nonumber\\[4pt]
\quiver_0^{-\pm}&=&\big\{(n,m)\in\quiver_0^-~\big|~
j_\pm-2j_\mp\leq\pm\,\mbox{$\frac{k-l}2$} \quad \mbox{and} \quad
j_\pm-j_\mp\geq\pm\,\mbox{$\frac{k-l}3$}\big\} \ .
\label{symquiv0pmsplits}\eea
Using these decompositions we can define a bi-grading of the fibre
space (\ref{Vklmodulegrading}) by
\beq
\underline{V}_{\,\pm}\=\underline{V}_{\,\pm+}\,\oplus\,
\underline{V}_{\,\pm-} \qquad \mbox{with} \quad
\underline{V}_{\,\pm\bullet}\=\bigoplus_{(n,m)\in\quiver_0^{\pm
\bullet}}\,\underline{V}_{\,n,m}\,\otimes\,\underline{(n,m)} \ .
\label{Vklsymbigrading}\eeq

Given the operators (\ref{phisymansatz}), we define morphisms in
(\ref{oddgradedconnsubsp}) by
\beq
\mphi^\pm:=\sum_{(n,m)\in\quiver_0(k,l)}\,\phi_{n,m}^\pm\otimes
\bigg(~\sum_{q\in\{-n+2j\}_{j=0}^n}\,\left(\big|\npmoverqmo\,,\,m+3
\big\rangle\big\langle\noverq\,,\,m\big|
+\big|\npmoverqpo\,,\,m+3\big\rangle\big\langle\noverq\,,\,
m\big|\right)\,\bigg) \ .
\label{phipmgrconn}\eeq
Using (\ref{SU2spinshifts}) along with finite-dimensionality of the
path algebra associated to the given symmetric quiver, we find the
generic nilpotency conditions
\bea
\big(\mphi^+\big)^i~\neq~0 \ , \quad i\=1,\dots,k \qquad 
&\mbox{and}& \qquad \big(\mphi^+\big)^{k+1}\=0 \ , \nonumber\\[4pt]
\big(\mphi^-\big)^j~\neq~0 \ , \quad j\=1,\dots,l \qquad 
&\mbox{and}& \qquad \big(\mphi^-\big)^{l+1}\=0 \ .
\label{symnilpotent}\eea
The operators (\ref{phipmgrconn}) are thus naturally associated with
the zero-form components of a $\Z_{k+1}\times\Z_{l+1}$-graded
connection. From the holomorphic relations (\ref{symCklhol}) it
follows that they obey the commutation relation
\beq
\big[\mphi^+\,,\,\mphi^-\big]=0 \ ,
\label{mphipmcommrel}\eeq
and therefore generate a $p$-dimensional representation of the
two-dimensional abelian algebra $\mathfrak{u}$. In addition, from the
non-holomorphic relations (\ref{symCklnonhol}) one finds the
commutativity condition
\beq
\big[\mphi^+\,,\,\mphi^-\,{}^\dag\,\big]=0
\label{nonholmphipmrel}\eeq
along with hermitean conjugates.

For any positive integer $s$, we use (\ref{partialisom}) and
(\ref{phisymansatz}) to derive the identities
\bea
\big(\mphi^\pm\big)^s&=&\sum_{(n,m)\in\quiver_0(k,l)}\,
T_{n\pm s,m+3s}\,T^\dag_{n,m}\otimes\bigg(~
\sum_{q\in\{-n+2j\}_{j=0}^n}\, \Big(\big|
{\stackrel{\scriptstyle n\pm s}{\scriptstyle q-s}}\,,\,m+3s
\big\rangle\big\langle{\stackrel{\scriptstyle n}{\scriptstyle q}}
\,,\,m\big| \label{mphisymspower} \\ && \hspace{3cm}
+\,s\,\sum_{j=1}^{s-1}\,\big| 
{\stackrel{\scriptstyle n\pm s}{\scriptstyle q-s+2j}}\,,\,m+3s
\big\rangle\big\langle{\stackrel{\scriptstyle n}{\scriptstyle q}}
\,,\,m\big|+\big|
{\stackrel{\scriptstyle n\pm s}{\scriptstyle q+s}}\,,\,m+3s
\big\rangle\big\langle{\stackrel{\scriptstyle n}{\scriptstyle q}}
\,,\,m\big|\Big)\,\bigg) \ . \nonumber
\eea
Using these formulas we now introduce the operators
\beq
\mmu^+~:=~\big(\mphi^+\big)^{[\frac k2]+1} \qquad \mbox{and}
\qquad \mmu^-~:=~\big(\mphi^-\big)^{[\frac l2]+1}
\label{mmusympmdefs}\eeq
where $[x]$ denotes the integer part of $x\in\R$. With respect to the
$\Z_2\times\Z_2$-grading in (\ref{Vklsymbigrading}), using
(\ref{symnilpotent}), (\ref{mphisymspower}) and (\ref{SU2spinshifts})
one can straightforwardly show that they are odd holomorphic maps
\bea
\mmu^+\,:\, \underline{V}_{\,++}\otimes\Hcal~\longrightarrow~
\underline{V}_{\,--}\otimes\Hcal \qquad &\mbox{with}& \quad
\big(\mmu^+\big)^2\=0 \ , \nonumber\\[4pt]
\mmu^-\,:\,\underline{V}_{\,-+}\otimes\Hcal~\longrightarrow~
\underline{V}_{\,+-}\otimes\Hcal \qquad &\mbox{with}& \quad
\big(\mmu^-\big)^2\=0 \ .
\label{mmusymoddmaps}\eea
The operators (\ref{mmusympmdefs}) thus produce the desired bi-complex
of noncommutative tachyon fields between branes and antibranes.

It is instructive at this stage to look at the limiting case
$l=0$. Then the quiver collapses to a holomorphic chain with $k+1$
vertices. One has $j_-=0$, $n=2j_+$ and $m=3n-2k$. In this case
$m-n=2(n-k)\leq0$ at each vertex $v=(n,m)$, while
$m+n=2(2n-k)$. The branes are located at vertices $v=(n,m)$ with
$n=0,1,\dots,\big[\frac k2\big]-1$ along the chain, while the
antibranes have $n=\big[\frac k2\big],\big[\frac
k2\big]+1,\dots,k$. The tachyon field $\mmu^+$ shifts $m+n$ by
$4\big[\frac k2\big]+4$ and hence takes branes into antibranes. A
completely analogous picture holds for $k=0$, whereby $j_+=0$,
$n=2j_-$ and $m=-3n+2l$, with the tachyon field $\mmu^-$ taking
antibranes into branes. In this case $m+n\geq0$ while $m-n$ is
positive for $0<n<\big[\frac l2\big]$ and negative for $\big[\frac
l2\big]\leq n\leq l$. These chains are qualitatively the same as the
brane-antibrane systems associated to the symmetric space
$\CP^1$~\cite{PS1}. However, the symmetric quiver diagram associated
to a generic module $\underline{C}^{k,l}$ is not simply the product of
two chains as in the $\CP^1\times\CP^1$ case~\cite{LPS2}.

In contrast to the quiver gauge theories associated to the symmetric
space $\CP^1$, the tachyon fields (\ref{mmusymoddmaps}) do not map
between branes of equal and opposite K-theory charges
(\ref{indexCP2}). On isotopical components, their non-trivial kernels
and cokernels are given by the finite-dimensional vector spaces
\bea
\ker\big(\mmu_{n,m}^+\big)\=\Imm\big(P_{n,m}\big) \qquad &
\mbox{and} & \qquad 
\ker\big(\mmu_{n,m}^+{}^\dag\big)\=\Imm\big(P_{n+[\frac k2]+1,
m+3[\frac k2]+3}\big) \ , \nonumber \\[4pt]
\ker\big(\mmu_{n,m}^-\big)\=\Imm\big(P_{n,m}\big) \qquad &
\mbox{and} & \qquad 
\ker\big(\mmu_{n,m}^-{}^\dag\big)\=\Imm\big(P_{n-[\frac l2]-1,
m+3[\frac l2]+3}\big) \ .
\label{musymkercoker}\eea
Denote by $\mmu_{\alpha\beta}^\pm$, $\alpha,\beta=\pm$, the
restrictions of the operators (\ref{mmusympmdefs}) to
$\underline{V}_{\,\alpha\beta}$. Following~\cite{LPS2}, with respect
to the $\Z_2$-grading (\ref{Vklmodulegrading}) and a suitable basis
for the vector space $\underline{V}^{k,l}$, we define the operator
\beq
\mbf T:=\mmu^+_{++}\oplus\mmu^-_{+-}{}^\dag \ .
\label{mbfTsymdef}\eeq
It is an odd map (\ref{tachyoncomplex}) which produces the appropriate 
two-term complex representing the brane-antibrane system with tachyon
field (\ref{mbfTsymdef}).

To incorporate the twistings by the instanton and monopole bundles, we
use the ABS construction to extend the tachyon field
(\ref{mbfTsymdef}) to the operator
\beq
\mbf\Tcal:=\mbf T\otimes\Idd\,:\,\underline{\Ccal}_{\,+}~
\longrightarrow~\underline{\Ccal}_{\,-} \ ,
\label{symtachyonABS}\eeq
where
\beq
\underline{\Ccal}_{\,\pm}~:=~\bigoplus_{(n,m)\in\quiver_0^\pm}\,\big(\,
\underline{V}_{\,n,m}\otimes\Hcal\big)\,\otimes\,
\underline{C}^{k_{n,m},l_{n,m}}_{\,\pm} \qquad \mbox{with} \quad
\underline{C}^{k_{n,m},l_{n,m}}_{\,\pm}\=\ker\big(\Dirac_{n,m}^\pm\big)
\ .
\label{symCcalmodule}\eeq
From (\ref{indexCP2}) and (\ref{musymkercoker}) it follows that the
K-theory charge represented as the index of the tachyon field
(\ref{symtachyonABS}) is given by
\bea
\Qcal&:=&\ind(\mbf\Tcal) \= \dim\ker(\mbf\Tcal)-\dim\ker\big(
\mbf\Tcal^\dag\,\big) \label{symKtheorycharge} \\[4pt]
&=&\sum_{(n,m)\in\quiver_0^{++}}\,\mbox{$\frac18$}\,(n+1)\,\big(
m^2-n\,(n+2)-1\big) \nonumber\\ && \times\,\Big[\big(
N_{n,m}+N_{n+[\frac k2]-[\frac l2],m+3[\frac l2]+3[\frac k2]+6}\big)-
\big(N_{n+[\frac k2]+1,m+3[\frac k2]+3}+
N_{n-[\frac l2],m+3[\frac l2]+3}\big)\Big] \ , \nonumber
\eea
where we have used the commutativity relations (\ref{mphipmcommrel})
and (\ref{nonholmphipmrel}). This charge is related to the instanton
charge (\ref{CP2Qfinal}) through
\beq
Q=4\Qcal+\mbox{$\frac12$}\,N \ ,
\label{topKtheorysymrel}\eeq
where $N$ is the noncommutative vortex number (\ref{NsymD0}).

\bigskip

\noindent
{\bf Non-symmetric $\mbf{\underline{C}^{k,l}}$ quiver charges. \ } The
representation ring of the maximal torus $T$ is a product
$\rep(T)=\rep(\uo)\otimes\rep(\uo)$, which can be identified with the
formal Laurent polynomial ring
\beq
\rep(T)\cong\Z\big[\Lcal_{(1)}\,,\,\Lcal_{(1)}^\vee\,\big]\otimes
\Z\big[\Lcal_{(2)}\,,\,\Lcal_{(2)}^\vee\,\big]
\label{repringT}\eeq
in the two monopole line bundles $\Lcal_{(i)}\to Q_3$. The (reduced)
K-theory of the space $Q_3$ is generated by $\Lcal_{(i)}$,
$\Lcal_{(1)}\otimes\Lcal_{(i)}$ and
$\Lcal_{(1)}\otimes\Lcal_{(1)}\otimes\Lcal_{(2)}$ with $i=1,2$. For
the Dirac index in (\ref{repringT}), we will now show that the index
of the spin$^c$ Dirac operator on $Q_3$ in the background monopole
fields corresponding to the irreducible representation
$\underline{(q,m)_n}$ of $T$ is given by
\beq
\ind(\Dirac_{q,m})=\mbox{$\frac18$}\,q\,\big(m^2-q^2\big) \ .
\label{indDiracqm}\eeq
The Dirac spectrum in this case was also computed in~\cite{Dolan}
from the natural spin$^c$ connection on $Q_3$ with torsion, by using
exactly the same technique as in the symmetric case above.

The chiral case $\ker(\Dirac_{q,m}^-)=\{0\}$ corresponds to background
gauge field configurations on $Q_3$ with $q^2\geq m^2$, for which
$\ker(\Dirac_{q,m}^+)$ is isomorphic to the $G$-module
$\underline{C}^{k,l}$ having $|q|=k+l$, $|m|=k-l$. The corresponding
dimension (\ref{dimkl}) coincides with minus the index
(\ref{indDiracqm}) after shifting $q\to q\pm2$. The antichiral case
$\ker(\Dirac_{q,m}^+)=\{0\}$ corresponds to $q^2\leq m^2$, for which
the Dirac kernel $\ker(\Dirac_{q,m}^-)$ is isomorphic to the $\sut$
representation $\underline{C}^{k,l}$ with $|q|=k$, $|m|=k+2l$. The
corresponding dimension (\ref{dimkl}) agrees with (\ref{indDiracqm})
after the shifts $q\to q\pm1$, $m\to m\pm3$. In both cases the shifts
account for the contributions of the intrinsic spin$^c$ fermion to the
$\uo$ monopole charges.

We correspondingly decompose the vertex set $\quiver_0(k,l)$ into
disjoint subsets
\beq
\quiver_0^+\=\big\{(q,m)_n\in\quiver_0(k,l)~\big|~q^2>m^2\big\}
\qquad \mbox{and} \qquad
\quiver_0^-\=\big\{(q,m)_n\in\quiver_0(k,l)~\big|~q^2\leq m^2\big\} \
.
\label{nonsymQ0pm}\eeq
Using the $U$-spin and $V$-spin charge variables (\ref{qUdefs}), we
further decompose these subsets into $\quiver_0^\pm=\quiver_0^{\pm
  +}\sqcup\quiver_0^{\pm -}$ with
\bea
\quiver_0^{++}\=\big\{(q,m)_n\in\quiver_0^+~\big|~q_U<0\big\}
\quad &\mbox{and}& \quad
\quiver_0^{+-}\=\big\{(q,m)_n\in\quiver_0^+~\big|~q_V<0\big\} \ ,
\label{nonsymQ0pmpm} \\[4pt]
\quiver_0^{-+}\=\big\{(q,m)_n\in\quiver_0^-~\big|~q_U\geq0 \ , \ q>0
\big\} \quad &\mbox{and}& \quad
\quiver_0^{--}\=\big\{(q,m)_n\in\quiver_0^-~\big|~q_V\geq0 \ , \ q\leq
0 \big\} \ . \nonumber
\eea
The corresponding bi-grading of the fibre space
(\ref{Vklmodulegrading}) is then given by
\beq
\underline{V}_{\,\pm}\=\underline{V}_{\,\pm+}\,\oplus\,
\underline{V}_{\,\pm-} \qquad \mbox{with} \quad
\underline{V}_{\,\pm\bullet}\=\smash{\bigoplus_{(q,m)_n\in\quiver_0^{\pm
\bullet}}}\,{}^n\underline{V}_{\,q,m}\,\otimes\,\underline{(q,m)_n} \ .
\label{Vklnonsymbigrading}\eeq

Given the operators (\ref{nonsymphiansatz}), we define morphisms in
(\ref{oddgradedconnsubsp}) by
\bea
\mphi^\pm&:=&\sum_{(q,m)_n\in\quiver_0(k,l)}\,\left(
{}^{n-1}T_{q\pm1,m+3}\,{}^nT_{q,m}{}^\dag\otimes
\big|\nmoverqpmo\,,\,m+3
\big\rangle\big\langle\noverq\,,\,m\big| \right. \nonumber\\ &&
\hspace{3cm} \left.
+\,{}^{n+1}T_{q\pm1,m+3}\,{}^nT_{q,m}{}^\dag\otimes
\big|\npoverqpmo\,,\,m+3\big\rangle\big\langle\noverq\,,\,
m\big|\right) \ , \nonumber\\[4pt]
\mphi^0&:=&\sum_{(q,m)_n\in\quiver_0(k,l)}\,{}^nT_{q+2,m}\,
{}^nT_{q,m}{}^\dag\otimes\big|\noverqpt\,,\,m
\big\rangle\big\langle\noverq\,,\,m\big| \ .
\label{mphinonsymdefs}\eea
From the quadratic holomorphic relations
(\ref{nonsymCklquad1})--(\ref{nonsymCklquad3}) it follows that these
$p\times p$ matrix-valued operators satisfy the commutativity
equations
\beq
\big[\mphi^+\,,\,\mphi^-\,\big]\=0 \qquad \mbox{and} \qquad
\big[\mphi^\pm\,,\,\mphi^0\,\big]\=0 \ ,
\label{mphiholquadrels}\eeq
while from the non-holomorphic relations (\ref{quadnonholrels}) one
finds
\beq
\big[\mphi^+\,,\,\mphi^-\,^\dag\,\big]\=0 \qquad \mbox{and} \qquad
\big[\mphi^\pm\,,\,\mphi^0\,^\dag\,\big]\=0 \ ,
\label{mphinonholquadrels}\eeq
plus hermitean conjugates. From the linear holomorphic relations
(\ref{linholweak}) one has 
\beq
\mphi^+\=\mphi^-\,\mphi^0 \=\mphi^0\,\mphi^- \ ,
\label{mphihollinrels}\eeq
while from (\ref{linnonholrels}) it follows that
\beq
\mphi^-\=\mphi^+\,\mphi^0\,^\dag\=\mphi^0\,^\dag\,\mphi^+
\qquad \mbox{and} \qquad
\mphi^0\=\mphi^-\,^\dag\,\mphi^+\=\mphi^+\,\mphi^-\,^\dag \ ,
\label{mphinonhollinrels}\eeq
along with hermitean conjugates.

From the isospin range (\ref{qrange}) it follows that the operators
$\mphi^0$ obey the generic nilpotency conditions
\beq
\big(\mphi^0\big)^i~\neq~0 \ , \quad i\=1,\dots,k+l+1 \qquad 
\mbox{and} \qquad \big(\mphi^0\big)^{k+l+2}\=0 \ .
\label{mphi0nil}\eeq
From the relations (\ref{mphiholquadrels})--(\ref{mphinonhollinrels})
it then also follows that
\beq
\big(\mphi^\pm\big)^i~\neq~0 \ , \quad i\=1,\dots,k+l+1 \qquad 
\mbox{and} \qquad \big(\mphi^\pm\big)^{k+l+2}\=0 \ .
\label{mphipmnil}\eeq
Generic non-vanishing powers of the operators (\ref{mphinonsymdefs})
are readily computed with the results 
\bea
\big(\mphi^\pm\big)^s&=&\sum_{(q,m)_n\in\quiver_0(k,l)}\,
\Big({}^{n-s}T_{q\pm s,m+3s}\,{}^nT_{q,m}{}^\dag\otimes\big|
{\stackrel{\scriptstyle n- s}{\scriptstyle q\pm s}}\,,\,m+3s
\big\rangle\big\langle{\stackrel{\scriptstyle n}{\scriptstyle q}}
\,,\,m\big|  \nonumber \\ && \hspace{3cm}
+\,s\,\sum_{j=1}^{s-1}\,{}^{n-s+2j}T_{q\pm s,m+3s}\,
{}^nT_{q,m}{}^\dag\otimes\big| 
{\stackrel{\scriptstyle n-s+2j}{\scriptstyle q\pm s}}\,,\,m+3s
\big\rangle\big\langle{\stackrel{\scriptstyle n}{\scriptstyle q}}
\,,\,m\big|\nonumber \\ && \hspace{3cm}
+\,{}^{n+s}T_{q\pm s,m+3s}\,{}^nT_{q,m}{}^\dag\otimes\big|
{\stackrel{\scriptstyle n+ s}{\scriptstyle q\pm s}}\,,\,m+3s
\big\rangle\big\langle{\stackrel{\scriptstyle n}{\scriptstyle q}}
\,,\,m\big|\Big) \ , \nonumber\\[4pt]
\big(\mphi^0\big)^s&=&\sum_{(q,m)_n\in\quiver_0(k,l)}\,
{}^nT_{q+2s,m}\,{}^nT_{q,m}{}^\dag\otimes\big|
{\stackrel{\scriptstyle n}{\scriptstyle q+2s}}\,,\,m
\big\rangle\big\langle{\stackrel{\scriptstyle n}{\scriptstyle q}}
\,,\,m\big| \ ,
\label{mphinonsymspower}\eea
for any positive integer $s$.

From (\ref{mphihollinrels}) and (\ref{mphinonhollinrels}) it
follows that only two of the operators in (\ref{mphinonsymdefs}) are
independent. We will work with $\mphi^\pm$ and use the second relation
in (\ref{mphinonhollinrels}) to generate $\mphi^0$. The operators
$\mphi^\pm$ thus form the odd zero-form components of a
$\Z_{k+l+2}\times\Z_{k+l+2}$-graded connection. Using
(\ref{mphinonsymspower}) we then define the bi-tachyon fields
\beq
\mmu^\pm:=\big(\mphi^\pm\big)^{[\frac{k+l+2}2+1]+1} \ .
\label{mmupmnonsymdef}\eeq
Similarly to (\ref{mmusymoddmaps}), they are odd holomorphic maps
\bea
\mmu^+\,:\, \underline{V}_{\,++}\otimes\Hcal~\longrightarrow~
\underline{V}_{\,--}\otimes\Hcal \qquad &\mbox{with}& \quad
\big(\mmu^+\big)^2\=0 \ , \nonumber\\[4pt]
\mmu^-\,:\,\underline{V}_{\,+-}\otimes\Hcal~\longrightarrow~
\underline{V}_{\,-+}\otimes\Hcal \qquad &\mbox{with}& \quad
\big(\mmu^-\big)^2\=0
\label{mmunonsymoddmaps}\eea
whose isotopical components have non-trivial kernels and cokernels
given by
\beq
\ker\big({}^n\mmu^\pm_{q,m}\big)\=\Imm\big({}^nP_{q,m}\big)
\qquad \mbox{and} \qquad
\ker\big({}^n\mmu^\pm_{q,m}{}^\dag\big)\=\bigoplus_{j=0}^s\,
\Imm\big({}^{n-s+2j}P_{q\pm s,m+3s}\big) \ ,
\label{mmunonsymker}\eeq
with $s:=\big[\frac{k+l+2}2\big]+1$. Note that the cokernel in
(\ref{mmunonsymker}) naturally takes into account the degeneracies of
weight vectors, i.e. the multiple arrows in the quiver diagram.

The tachyon field
\beq
\mbf T:=\mmu^+_{++}\oplus\mmu^-_{+-}
\label{mbfTnonsymdef}\eeq
then yields the requisite two-term complex (\ref{tachyoncomplex}). We
extend it as in (\ref{symtachyonABS}) using the noncommutative ABS spaces
where
\beq
\underline{\Ccal}_{\,\pm}~:=~\bigoplus_{(q,m)_n\in\quiver_0^\pm}\,\big(\,
{}^n\underline{V}_{\,q,m}\otimes\Hcal\big)\,\otimes\,
\underline{C}^{k_{q,m},l_{q,m}}_{\,\pm} \qquad \mbox{with} \quad
\underline{C}^{k_{q,m},l_{q,m}}_{\,\pm}\=\ker\big(\Dirac_{q,m}^\pm\big)
\ .
\label{nonsymCcalmodule}\eeq
Using (\ref{indDiracqm}), (\ref{mphiholquadrels}),
(\ref{mphinonholquadrels}) and (\ref{mmunonsymker}) we compute the
index $\Qcal$ of the tachyon field (\ref{mbfTnonsymdef}) as before to
get
\beq
\Qcal=\sum_{(q,m)_n\in\quiver_0^{++}}\,\mbox{$\frac18$}\,q\,
\big(q^2-m^2\big)\,\Big[\big({}^nN_{q,m}+
{}^nN_{q,m+6s}\big)-\big({}^nN_{q+s,m+3s}+{}^nN_{q-s,m+3s}\big)\Big] \
.
\label{Qcalnonsym}\eeq
This charge is related to the instanton charge (\ref{Q3Qfinal}) as
\beq
Q=4\Qcal \ .
\label{QQcalnonsymrel}\eeq

\subsection{Euler-Ringel characters\label{ERform}}

The category of representations of the quiver with relations
$(\quiver(k,l),\rel(k,l))$ provides a complete framework for
understanding our instanton solutions. It gives a more detailed
picture of the dynamics, particularly of how the original
configuration on $X$ folds itself into branes and antibranes on
$M_{2d}$ within the category of quiver modules. Following~\cite{LPS2},
we start from the instanton module (\ref{Tvquivmodule}) over
$(\quiver(k,l),\rel(k,l))$ and its projective Ringel resolution. Since
there are no relations among our relations, this leads to the exact
sequence
\bea
0~\longrightarrow~\bigoplus_{v\in\quiver_0(k,l)}~
\bigoplus_{\rho\in\rel(k,l)}\,\underline{\Pcal}_{\,\rho(v)}\otimes
\ker\big(T_v^\dag\big)&\longrightarrow&\bigoplus_{v\in\quiver_0(k,l)}~
\bigoplus_{\Phi\in\quiver_1(k,l)}\,\underline{\Pcal}_{\,\Phi(v)}
\otimes\ker\big(T_v^\dag\big)~\longrightarrow \nonumber\\ 
&\longrightarrow&
\bigoplus_{v\in\quiver_0(k,l)}\,\underline{\Pcal}_{\,v}\otimes
\ker\big(T_v^\dag\big)~\longrightarrow~\underline{T}~
\longrightarrow~0
\label{Ringelresolution}\eea
where $\underline{\Pcal}_{\,v}$ is the quiver representation defined
as the subspace of the path algebra generated by all paths which start
from vertex $v\in\quiver_0(k,l)$. The first sum in
(\ref{Ringelresolution}) runs through the holomorphic relations of the
quiver which are indexed by paths starting at vertex $v$ and ending at
vertex $\rho(v)$.

Let
\beq
\underline{W}=\bigoplus_{v\in\quiver_0(k,l)}\,\underline{W}_{\,v}
\label{Wrepgen}\eeq
be the canonical representation of $(\quiver(k,l),\rel(k,l))$
determined by the ``folding'' of K-theory charges in the
equivariant ABS construction. We regard it as an element of the
representation ring of the quiver. It will be determined explicitly
below in terms of the Dirac kernels
$\underline{C}^{k_v,l_v}=\ker(\Dirac_v)$ of Section~\ref{Kcharge}
above. The module $\underline{W}$ represents the coupling of K-theory
charges on $G/H$ to the instanton modules $\underline{T}$.

Applying the covariant functor $\Hom(-\,,\,\underline{W}\,)$ to the
projective resolution (\ref{Ringelresolution}) yields the complex
\bea
0&\longrightarrow&\Hom\big(\,\underline{T}\,,\,\underline{W}\,\big)~
\longrightarrow~\bigoplus_{v\in\quiver_0(k,l)}\,\Hom\big(
\ker(T_v^\dag)\,,\,\underline{W}_{\,v}\big)~\longrightarrow
\nonumber\\ &\longrightarrow&\bigoplus_{v\in\quiver_0(k,l)}~
\bigoplus_{\Phi\in\quiver_1(k,l)}\,\Hom\big(
\ker(T_v^\dag)\,,\,\underline{W}_{\,\Phi(v)}\big)~\longrightarrow
\nonumber\\ &\longrightarrow&
\bigoplus_{v\in\quiver_0(k,l)}~\bigoplus_{\rho\in\rel(k,l)}\,\Hom\big(
\ker(T_v^\dag)\,,\,\underline{W}_{\,\rho(v)}\big)~\longrightarrow~
\Ext^2\big(\,\underline{T}\,,\,
\underline{W}\,\big)~\longrightarrow~0 \ .
\label{Ringelcomplex}\eea
We define $\Ext^p(\,\underline{T}\,,\,\underline{W}\,)$ to be the
$p$-th cohomology group of this complex. For $p=0$, the group
$\Ext^0(\,\underline{T}\,,\,\underline{W}\,)=
\Hom(\,\underline{T}\,,\,\underline{W}\,)$ corresponds to the vertices
$\quiver_0(k,l)$ and classifies partial gauge symmetries of the
instanton system. The group
$\Ext^1(\,\underline{T}\,,\,\underline{W}\,)=
\Ext(\,\underline{T}\,,\,\underline{W}\,)$ corresponds to the arrows
$\quiver_1(k,l)$ and classifies deformations of
$\underline{T}\,\oplus\,\underline{W}$ which describe bound states of
the constituent D-branes arising from partial gauge symmetries. The
group $\Ext^2(\,\underline{T}\,,\,\underline{W}\,)$ corresponds to the
relations~$\rel(k,l)$.

We now compute the relative Euler-Ringel form on the
representation ring of the quiver. Using (\ref{Ringelcomplex}) we find
\bea
\chi\big(\,\underline{T}\,,\,\underline{W}\,\big)&:=&
\sum_{p\geq0}\,(-1)^p~\dim\,\Ext^p\big(\,\underline{T}\,,\,
\underline{W}\,\big) \nonumber\\[4pt] &=&
\sum_{v\in\quiver_0(k,l)}\,\Big(\dim\,\Hom\big(
\ker(T_v^\dag)\,,\,\underline{W}_{\,v}\big)+\sum_{\rho\in\rel(k,l)}\,
\dim\,\Hom\big(\ker(T_v^\dag)\,,\,\underline{W}_{\,\rho(v)}\big)
\nonumber\\ &&\qquad\qquad\qquad -\,\sum_{\Phi\in\quiver_1(k,l)}\,
\dim\,\Hom\big(\ker(T_v^\dag)\,,\,\underline{W}_{\,\Phi(v)}\big)
\Big) \nonumber\\[4pt] &=& \sum_{v\in\quiver_0(k,l)}\,N_v\,\Big(
w_v+\sum_{\rho\in\rel(k,l)}\,w_{\rho(v)}-
\sum_{\Phi\in\quiver_1(k,l)}\,w_{\Phi(v)}\Big) \ ,
\label{Eulerformgen}\eea
where $w_v:=\dim(\,\underline{W}_{\,v})$. We will define the virtual
representation (\ref{Wrepgen}) such that the virtual dimensions $w_v$,
$v\in\quiver_0(k,l)$ obey the linear inhomogeneous recursion relations
\beq
w_v+\sum_{\rho\in\rel(k,l)}\,w_{\rho(v)}-
\sum_{\Phi\in\quiver_1(k,l)}\,w_{\Phi(v)}=\ind(\Dirac_v) \ ,
\label{wrecrelgen}\eeq
together with vanishing conditions at the boundaries of the pertinent
quiver. Then the character (\ref{Eulerformgen}) coincides with the topological
charges computed previously. Let us now turn to the explicit
constructions of the modules $\underline{W}$.

\bigskip

\noindent
{\bf Symmetric $\mbf{\underline{C}^{k,l}}$ quiver charges. \ } The
symmetric quiver diagram has only one holomorphic relation $\rho$
given by (\ref{holrelCkl}), which takes a vertex $v=(n,m)$ to
$\rho(v)=(n,m+6)$, and two arrows $\Phi^\pm$ taking $v=(n,m)$ to
$\Phi^\pm(v)=(n\pm1,m+3)$. We thus need to solve the four-term linear
inhomogeneous recursion relation
\beq
w_{n,m}+w_{n,m+6}-w_{n+1,m+3}-w_{n-1,m+3}=\ind(\Dirac_{n,m}) \ ,
\label{Sym4termrecrel}\eeq
where the index is given by (\ref{indexCP2}). By introducing the $\su$
spin variables $i:=-2j_+$ and $\alpha:=2j_-$, using
(\ref{SU2spinshifts}) one finds that the left-hand side of
(\ref{Sym4termrecrel}) is the same as that which arises for the
symmetric $\CP^1\times\CP^1$ quiver~\cite{LPS2}. Rewriting the
solution of that case in terms of our variables, one has
\beq
w_{n,m}=\sum_{n'=0}^{n-2}~
\sum_{\stackrel{\scriptstyle j=0}{
\scriptstyle m'=-2(k-l)+3j}}^{2(j_+-j_-)}
\,\ind(\Dirac_{n',m'}) \ .
\label{Symwnmsol}\eeq

These numbers correspond to the virtual dimensions of the $\sut$
representations
\beq
\underline{W}_{\,n,m}=\bigoplus_{n'=0}^{n-2}~
\bigoplus_{\stackrel{\scriptstyle j=0}{
\scriptstyle m'=-2(k-l)+3j}}^{2(j_+-j_-)}
\,\underline{\ind}(\Dirac_{n',m'}) \ .
\label{SymWnmsol}\eeq
This produces a non-decreasing sequence of representations
$\{\,\underline{W}_{\,n,m}\}$ as we move along the quiver of
constituent D-branes, such that the $G$-module $\underline{W}_{\,n,m}$
gives extensions of the instanton and monopole fields carried by the
elementary brane state at vertex $v=(n,m)\in\quiver_0(k,l)$. Thus all
in all, aside from a few details arising from the nonabelian nature of
the holonomy group $H=\su\times\uo$ in this case, the physics of the
$\CP^2$ quiver gauge theory is qualitatively similar to that of the
$\CP^1\times\CP^1$ quiver gauge theory.

\bigskip

\noindent
{\bf Non-symmetric $\mbf{\underline{C}^{k,l}}$ quiver charges. \ } The
non-symmetric quiver diagram has several holomorphic relations. From
(\ref{Ckllinrels}) we obtain a linear relation $\rho\in\rel(k,l)$
taking vertex $v=(q,m)_n$ into $\rho(v)=(q+1,m+3)_{n\pm1}$. From
(\ref{Cklquadrels}) we obtain the three respective quadratic holomorphic
relations $\rho^{+-}$, $\rho^{\pm0}$ taking $v=(q,m)_n$ into
$\rho^{+-}(v)=(q,m+6)_{n\pm1}$, $\rho^{+0}(v)=(q+3,m+3)_{n\pm1}$ and
$\rho^{-0}(v)=(q+1,m+3)_{n\pm1}$. There are three arrows $\Phi^0$,
$\Phi^\pm$ taking $v$ into $\Phi^0(v)=(q+2,m)_n$,
$\Phi^+(v)=(q+1,m+3)_{n\pm1}$ and $\Phi^-(v)=(q-1,m+3)_{n\pm1}$.
Since the index (\ref{indDiracqm}) in this case is independent of the
degeneracy label $n$, we will suppose that the same is true of the
virtual dimensions ${}^nw_{q,m}=w_{q,m}$. Then (\ref{wrecrelgen})
simplifies to the six-term recursion relation
\beq
w_{q,m}+w_{q+1,m+3}+w_{q,m+6}+w_{q+3,m+3}-w_{q+2,m}-w_{q-1,m+3}=
\ind(\Dirac_{q,m}) \ .
\label{Nonsym6termrecrel}\eeq
We see therefore that the number of independent terms in
(\ref{wrecrelgen}) is equal to the dimension $d_H$ of the coset space $G/H$.

We have not succeeded in finding an instructive and compact explicit
solution to the system 
(\ref{Nonsym6termrecrel}). Nevertheless, we may prove the existence of
a unique solution as follows. Let us recall the $U$-spin and $V$-spin
electric charge eigenvalues $u:=Y_U=-\frac12\,(q+\frac m3)$ and
$v:=Y_V=\frac12\,(q-\frac m3)$ from Section~\ref{Finitesols}. Setting
$w_{q(u,v),m(u,v)}=:a_{u,v}$, we may rewrite the recurrence relations
(\ref{Nonsym6termrecrel}) in the form
\beq
a_{u,v}=a_{u,v-1}+a_{u-1,v+1}-a_{u-1,v}-a_{u-1,v-1}-a_{u-2,v+1}+
b_{u,v}
\label{auvrecrel}\eeq
where
\beq
b_{u,v}=\mbox{$\frac12$}\,(v-u)\,(u+2v)\,(2u+v) \ .
\label{buvcharge}\eeq
It is now straightforward to see that (\ref{auvrecrel}) satisfies the
hypotheses of Theorem~5 in~\cite{B-MP}, from which we deduce the
existence of a unique solution $a_{u,v}$ to (\ref{auvrecrel}). The
sequence $a_{u,v}$ can in this way be evaluated inductively, or
alternatively via the kernel method which yields some solution of the
recursion relation. Several properties of this solution can be deduced
from the results of~\cite{B-MP}. It describes the moduli of the
folding of branes and antibranes in this case. Note that the index
(\ref{buvcharge}) changes sign when one interchanges $U$-spin and
$V$-spin electric charges.

\bigskip

\section*{Acknowledgments}

\noindent
We thank B.~Dolan, D.~O'Connor and B.~Schroers for helpful
discussions. This work was supported in part by the Deutsche
Forschungsgemeinschaft~(DFG). O.L. is grateful for hospitality to the
Theory Division at CERN, where part of this work was done.
The work of R.J.S. was supported in part
by the EU-RTN Network Grant MRTN-CT-2004-005104.

\bigskip

\end{document}